\documentclass[%
preprint,
superscriptaddress,longbibliography,
 amsmath,amssymb,
 aps,prd,
]{revtex4-1}

\usepackage{xcolor}
\usepackage{comment}
\usepackage{multirow}
\usepackage{braket}
\usepackage{relsize}
\usepackage{rotating}
\usepackage{yfonts}
\usepackage{eufrak}
\usepackage{graphicx}
\usepackage{dcolumn}
\usepackage{bm}
\usepackage{epstopdf}
\definecolor{light}{rgb}{0.5, 0.5, 0.5}
\def\light#1{{\color{light}#1}}
\newcommand{\para}{\mbox{$\mathsmaller{\Pi}$}}
\newcommand{\veq}{\mathrel{\rotatebox{90}{$=$}}}

\begin{document}
\preprint{APS/123-QED}

\title{Adjustable reach in a network centrality based on current flows}

\author{Aleks J. Gurfinkel}%
 \email{To whom correspondence should be addressed. E-mail: alexander.gurfinkel@gmail.com.} 
 \affiliation{%
 Department of Physics, Florida State University, Tallahassee, FL 32306-4350, USA\\
}%

\author{Per Arne Rikvold}%

\affiliation{%
 Department of Physics, Florida State University, Tallahassee, FL 32306-4350, USA\\
}%
\affiliation{%
PoreLab, NJORD Centre, Department of Physics, University of Oslo, P.O. Box 1048 Blindern, 0316 Oslo, Norway
}%

\date{\today}

\begin{abstract}

Centrality, which quantifies the ``importance'' of individual nodes,
is among the most essential concepts in modern network theory.
Most prominent centrality measures can be expressed as an aggregation of
influence flows between pairs of nodes.
 As there are
many ways  in which influence can be defined, many different centrality measures 
are in use.  
Parametrized centralities allow further flexibility and utility 
by tuning the centrality calculation to the regime most appropriate for a given purpose and  network. 
Here, we identify two categories of centrality parameters.
{\it Reach} parameters  control the attenuation of influence flows between distant nodes. 
{\it Grasp} parameters  control the centrality's tendency to send influence flows along multiple,
often nongeodesic paths.
Combining these categories with Borgatti's centrality types [S. P. Borgatti, Social Networks $\mathbf{27}$, 55-71 (2005)], we arrive at a novel classification system for
{\it parametrized } centralities. Using this classification, we identify the notable absence of any centrality measures that are radial, reach parametrized, and based on acyclic, conserved  flows of influence. We therefore introduce the {\it ground-current centrality}, which is a measure of precisely this type. Because of its unique position in  
the taxonomy, the ground-current centrality differs significantly from similar centralities. We demonstrate that, compared to other conserved-flow centralities, it has a simpler mathematical description. Compared to other  reach-parametrized centralities,  it robustly preserves an intuitive rank ordering across a wide range of network architectures, capturing aspects of both the closeness and betweenness centralities.  We also show that it produces a consistent distribution of centrality values among the nodes, neither trivially equally spread (delocalization), nor overly focused on a few nodes (localization).
Other reach-parametrized centralities exhibit both of these behaviors on regular networks and hub networks, respectively. We compare the properties of the ground-current centrality with several other reach-parametrized centralities on four artificial networks and seven real-world networks.
\end{abstract}

\pacs{Valid PACS appear here}
\maketitle

\section{Introduction \label{sec:intro}}

 Centrality measures are  prescriptions for assigning importance values to nodes in complex networks, and the power of the concept stems from the flexibility of characterizing importance in different ways. As such, centralities can be applied everywhere from Internet search results (Google's PageRank \cite{page1999pagerank}) to identifying important structures in neuron networks \cite{joyce2010new}. Centrality is one of the most basic and widely studied concepts in network theory.

Recently, we summarized how many prominent centrality measures arise from the aggregation of ``influences'' flowing between pairs of nodes  \cite{gurfinkel2020absorbing}. These influences are encoded in the entries of a centrality matrix $\mathbf{M}$, whose specification is equivalent to that of the overall measure.  As we demonstrate here, these pair influences can be revealing measurements in their own right (see Sec.~\ref{sec:sstar}).
Centrality results are also useful beyond identifying influential nodes and influence flows between node pairs.  Often, researchers posses  quantitative information about individual nodes---information which  is {\it external} to the specification of the network structure. A centrality that approximately reproduces these data can reveal
principles inherent in the structure of the network. 
In \cite{xu2014architecture}, we investigated the architecture of the
Florida electric power grid along these lines. A strong correlation was revealed between  the known generating capacities of power plants and the values of a centrality based on 
Estrada's communicability  
\cite{estrada2008communicability,estrada2009communicability}, here referred to as the {\it communicability centrality}.
Quantification of such correlations between node attributes and network structure requires 
centrality measures with a built-in tuning parameter.

The communicability centrality has a parameter that controls the (graph) distance over which nodes can 
influence each other. Such parameters can reveal the length scale over which the network is optimized.   Since there are many ways for a centrality to limit the distance that influence can spread, we introduce the {\it reach-parametrized}  category to describe  centralities with parameters that have this effect. We will discuss how the reach-parametrized category includes the well-known PageRank \cite{page1999pagerank}, Katz \cite{katz1953new}, and $\alpha$ \cite{ghosh2011parameterized} centralities.  The reach-parametrized category is not exhaustive. In \cite{gurfinkel2020absorbing}, we introduced the {\it conditional walker-flow centralities}, which include parameters that interpolate these centralities between older, well-known measures. The conditional walker-flow measures belong to a distinct category:  {\it grasp-parametrized } centralities. These centralities' parameters also attenuate influence, but in a way different from reach parameters. While reach parameters control how far centrality influence can spread,  grasp parameters control how many alternative paths influence can follow.

In addition to reach and grasp   parametrization, here we further classify parametrized centralities according to the conceptual dimensions introduced by Borgatti \cite{borgatti2005centrality,borgatti2006graph}. Referencing this classification system, we  show that there is a notable absence of centrality measures that are radial,  reach parametrized, and based on acyclic, conserved flows of influence. To fill this void, we introduce a new centrality, the {\it ground-current centrality}. There, influence is modeled by the flow of electrical current from the source node to all possible end nodes, from which the current  flows to ground. (The method is fully described in Sec.~\ref{sec:gcc}.) The physics of current flow naturally satisfies the conservation and acyclicity criteria, while variable resistances to ground naturally limit the spread of currents (and hence influences), thus representing a reach parameter.

Conservation and acyclicity enable the ground-current centrality to perform differently from other reach-parametrized centralities in several ways.
Most importantly, we demonstrate that, compared to other  reach-parametrized centralities, the ground-current centrality robustly preserves an intuitive rank ordering across a  range of simple network topologies. Here we take the closeness centrality (specifically, its harmonic variant \cite{Dekker2005,Rochat2009,NEWM10}) to provide the paradigmatic intuitive centrality ranking for simple networks, since it places greater importance on nodes that are close to many others. 

However, the ground-current centrality can also reproduce aspects of betweenness: when the reach is high, it is highly sensitive to network bottlenecks, assigning them high centrality rank, whereas other  reach-parametrized centrality measures almost completely ignore bottlenecks in certain situations. We further show that, on hub networks, the ground-current centrality does not lead to excessive localization. This is a phenomenon \cite{martin2014localization} whereby the majority of the net centrality is assigned to a small fraction of nodes. On the other hand, in regular networks the ground-current centrality does not lead to excessive {\it de}localization: the assignment of nearly the same centrality value to every node. Other measures, such as the Katz and communicability centralities, exhibit both these behaviors. Recently, it has also been proposed to construct centrality measures from diffusion dynamics \cite{arnaudon2020scale}. 

The remainder of this paper is organized as follows. In Sec.~\ref{sec:centralities}, we present a classification system for {\it parametrized} centrality measures, discussing in detail the distinction between reach and grasp parameters.  
 In Sec.~\ref{sec:gcc} we define the ground-current centrality. In Sec.~\ref{sec:results}, we discuss the   properties of the ground-current centrality relative to other similarly classified measures. To that end, we perform a numerical study of the centralities' detailed performance on a variety of networks.  These include  four artificial networks designed to highlight a particular network property, as well as seven real-world networks.  In Sec.~\ref{sec:conc}, we conclude that the special properties of the ground-current centrality stem from its unique position as a radial reach-parametrized centrality based on acyclic, conserved flows. 

\section{Reach and grasp parameters for network centralities \label{sec:centralities}}
In this section, we present a wide-ranging classification of parametrized centrality measures, which includes the most prominent measures in the literature. We find that a simple and reasonable combination of centrality characteristics has not yet been studied, which motivates us to introduce a new measure, the ground-current centrality, to which we devote Secs. \ref{sec:gcc}-\ref{sec:conc}.

\subsection{Notation and conventions}

The  $N\times N$ adjacency matrix is denoted $\mathbf{A}$. Here we consider both weighted and unweighted adjacency matrices. The graph distance between nodes $i$ and $j$ is denoted $d_{i j}$. In the case of weighted networks, we may instead use $D_{i j}$, which  is the length of the shortest edge path from node $i$ to $j$, where the length of a given edge $(a,b)$ is $(\mathbf{A}_{ab})^{-1}$ \cite{gurfinkel2020absorbing}.

The most commonly studied centrality measures  can be found in, \em e.g.\/\em, Ch.7 of \cite{NEWM10},  and many can be written \cite{borgatti2006graph} in the  matrix  form: 
\begin{flalign}
\phantom{\text{}}&&
c_i=  \alpha \sum_{j} \mathbf{M}_{i j},&&\text{}\label{eq:centrality} \end{flalign}
where $c_i$ is the centrality of node $i$, and the sum is over the $N$ nodes in the network. We focus on centralities with a single parameter $\para$, so $\mathbf{M}=\mathbf{M}(\para)$. The matrix elements $\mathbf{M}_{i j}$ of the $N\times N$ centrality matrix $\mathbf{M}$ encode the level of influence that node $j$ exerts on node $i$, and the final centrality is the sum of such influences.
In this paper we denote column (row) vectors as kets (bras). The  normalization factor $\alpha$  ensures that $\braket{c\,|1}=1$, where $\ket{1}$ is the column vector with all elements equal to one \footnote{In \cite{gurfinkel2020absorbing}, we chose to present {\it unnormalized } centrality results  to better track centrality behavior across a range of parameter values.}. The normalization factor is different for every centrality measure, and for each choice of parameter value, so $\alpha=\alpha(\para)$. To maintain readability, we will omit the $\para$ dependence of $\alpha$ and $\mathbf{M}$, and we will not specify which centrality $\alpha$ normalizes when it is clear from the context. 

The {\it degree centrality} (DEG) is one of the simplest and most commonly studied network measures. It can be put into the above form by setting $\mathbf{M}^\mathrm{DEG}$ equal to  $\mathbf{A}$ so that $c^\mathrm{DEG}_i=\alpha \sum_j A_{ij}=\alpha k_i$. In this paper, we consider (potentially) weighted,   symmetric adjacency matrices. The $k_i$ are, thus,  (potentially) weighted degrees, and there is no distinction between indegree and outdegree.

A very important centrality that { cannot} be expressed in the above form is the {\it closeness centrality} (CLO): $c^\mathrm{CLO}_i = (\sum_j{D_{i j}})^{-1}$. In \cite{gurfinkel2020absorbing}, we therefore used  the {\it harmonic closeness centrality} (HCC) \cite{Dekker2005,Rochat2009,NEWM10}, which {\it can} be written in matrix form as: 
\begin{equation}
\mathbf{M}^\mathrm{HCC}_{i j} = {D_{i j}}^{-1}. 
\end{equation}

A useful modification of Eq.~(\ref{eq:centrality}) involves subtracting  the diagonal of the centrality matrix $\mathbf{M}$:
\begin{flalign}
\phantom{\text{}}&&\widetilde{c}_i^{ \mathrm{\hspace{.15em}}}= \widetilde{\alpha}  \sum_{j} \widetilde{\mathbf{M}}_{i j}^{ \mathrm{\hspace{.15em}}} = \widetilde{\alpha}\sum_{j} (\mathbf{M}-\pmb{\mathrm{Diag}}(\mathbf{M}))_{i j},
&&\text{}
\label{eq:exocentcentrality}
\end{flalign}
where $\widetilde{\alpha}$ is the new normalization factor.
This modified form $\widetilde{\mathbf{M}}$ simply prevents self influence,  and we thus refer to $\widetilde{c}$ as the {\it exogenous} centrality. 

Above, we have used $\pmb{\mathrm{Diag}}(\mathbf{M})$ to indicate the modified form of matrix $\mathbf{M}$ that has all nondiagonal entries set to zero. In the following, we will also use the symbol $\pmb{\mathrm{Diag}}(\ket{v})$ to indicate the diagonal matrix with the elements of the vector $\ket{v}$ appearing on the diagonal.

\subsection{Reach-parametrized centralities}
A centrality parameter $\para$ is a {\it reach parameter } if changing it tends to attenuate the influence flow $\mathbf{M}_{ij}$ between pairs of nodes $i$ and $j$ separated by large graph distances $d_{ij}$. For weighted networks, it is possible to instead use the weighted graph distance $D_{ij}$.

Three prominent reach-parametrized centralities with similar definitions are the \em PageRank \em (PRC), \em Katz \em (KC),  and \em $\alpha$  \em centralities.  The first two of these can be defined \cite{noteoninverseparam,katz1953new,page1999pagerank}, respectively, by 
\begin{equation}\label{eq:prc}
\mathbf{M}^\mathrm{PRC}= [\pmb{\mathbb{I}} - \para_\mathrm{PRC}^{-1} \mathbf{A} \hspace{.2em} \pmb{\mathrm{Diag}}(\ket{k^{-1}}) ]^{-1}, 
\end{equation}
and
\begin{equation}\label{eq:katz}
\mathbf{M}^\mathrm{KC}= (\pmb{\mathbb{I}}-  \para_\mathrm{KC}^{-1}\mathbf{A})^{-1},
\end{equation}
where $\ket{k^{-1}}_i = k_i^{-1}$, the identity matrix is $\pmb{\mathbb{I}}$, and where we have employed the matrix inverse. (For the PRC, we have used a simplified definition that works for the {\it symmetric} adjacency matrices considered in this paper.) The $\alpha$ centrality is a variation of the Katz centrality, involving another parameter \cite{ghosh2011parameterized}. Here, we focus on the KC.

 The fact that the parameters $\para$ control the network distance over which influence can spread is seen from the series expansion for the
  Katz centrality: 
\begin{equation}\label{eq:katzexpand}
\mathbf{M}^\mathrm{KC} = \pmb{\mathbb{I}} + \para_\mathrm{KC}^{-1} \mathbf{A} +\para_\mathrm{KC}^{-2} \mathbf{A}^2 +\mbox{$\mathsmaller{\Pi}$}_\mathrm{KC}^{-3} \mathbf{A}^3+\cdots.
\end{equation}
 Since, in general, $(\mathbf{A}^l)_{ij}$ is equal to the number of walks of length $l$ from node $i$ to node $j$, one can see that larger values of $\para_\mathrm{KC}$ tend to suppress the influence of longer walks. The case of the PageRank centrality is similar, except that each term in the series expansion describes a single random walk, rather than counting the total number of walks. This is because the value of $[\mathbf{A} \hspace{.2em} \pmb{\mathrm{Diag}}(\ket{k^{-1}})]^l_{ij}$ is the probability of a walker starting on node $j$ being on $i$ after $l$ steps \cite{random_walk_note}. Thus $\para_\mathrm{PRC}$ controls the length of walks in the same way as $\para_\mathrm{KC}$. Incidentally, the series expansion in Eq.~(\ref{eq:katzexpand}) makes clear that the Katz centrality will diverge at some small value $\para_\mathrm{KC}$---the same is true for the PageRank centrality at $\para_\mathrm{PRC}=1$.  In what follows, we restrict $\para_\mathrm{KC}$ and $\para_\mathrm{PRC}$ to the range where Eq.~(\ref{eq:katzexpand}) and the corresponding series expansion for the PageRank are convergent.

The series form of Katz centrality above suggests a class of reach-parametrized centralities based on power series in the adjacency matrix (with the PageRank case similar). These take the form $\mathbf{M}(\para)=\sum_{l=0}^\infty f(l) \mathbf{A}^l \para^{-l}$, where the Katz centrality sets all factors $f(l)$ to one.
This choice, however, is not ideal because the series does not converge when $\para_\mathrm{KC}$ is smaller than  the largest eigenvalue of $\mathbf{A}$ ($\lambda_1$, with corresponding eigenvector $\ket{\psi_1}$). In the general case, for small $\para$, the higher-order terms are dominated by 
\begin{equation}\label{eq:matlimit}
f(l) (\lambda_1/\para)^l \ket{\psi_1}\bra{\psi_1}.
\end{equation}

\noindent For $\mathbf{M}$ to converge for all $\para$, $1/f(l)$ must grow super-exponentially in $l$. 
A reasonable choice, inspired both by the Estrada communicability metric \cite{estrada2008communicability} and by the desire to make contact with statistical physics, is to choose the factors $f(l)=(l!)^{-1}$. This formula, which defines the \em communicability centrality \em (COM) in terms of the matrix exponential function,  means that 

\begin{equation}
\label{eq:communicability}
\mathbf{M}^\mathrm{COM}(\para_T) = \exp (\mathbf{A} / \para_T)
=
 \pmb{\mathbb{I}} + \frac{\para_\mathrm{T}^{-1} \mathbf{A}}{1!} +\frac{\para_\mathrm{T}^{-2} \mathbf{A}^2}{2!} +\frac{\mbox{$\mathsmaller{\Pi}$}_\mathrm{T}^{-3} \mathbf{A}^3}{3!}+\cdots,
\end{equation} 
where we have  introduced the ``temperature"  parameter  $\para_T$. (This is very similar to the \em total communicability \em studied in \cite{benzi2013total}.)
 In past work \cite{gurfinkel2015centrality}, we compared the communicability centrality to several other centrality measures prominent in the literature, finding that it gives the best match to the generating capacities in the Florida power grid.

The communicability and Katz centralities have several satisfying properties, especially in their exogenous forms $\widetilde{\mathbf{M}}^{ \mathrm{COM}}$ and $\widetilde{\mathbf{M}}^\mathrm{KC}$. From the series expansions, it is easy to see that the degree centrality is recovered in the low reach limits ($\para_T \to \infty$ and $\para_\mathrm{KC} \to \infty$). In fact, in these limits we obtain $\widetilde{\mathbf{M}}^\mathrm{COM}=\widetilde{\mathbf{M}}^\mathrm{KC}=\mathbf{A}$. In the high reach limits ($\para_T \to 0 $ and $\para_\mathrm{KC} \to \lambda_1$), the largest eigenvalue dominates as in Eq.~(\ref{eq:matlimit}), so the centralities reduce to the well-known eigenvector centrality \cite{NEWM10}.  For large, fully connected networks, the exogenous forms $\widetilde{\mathbf{M}}$ give very similar results. 
These centralities also satisfy two very reasonable conditions on assigning influence between nodes $i$ and $j$: (1) the existence of many walks leads to more influence due to the presence of the term $(\mathbf{A}^l)_{ij}$, but (2) long walks are suppressed due to the weights $\para^{-l}$. (PageRank satisfies very similar conditions.)

Though several individual examples of reach-parametrized centralities are  well-known in the field of network science, we believe that we are the first to identify  reach-parametrized as a distinct category of centrality measures. We emphasize that, for every reach-parametrized centrality in this paper, we have defined the parameters $\para$ such that small $\para$ results in high reach, while large $\para$ results in low reach.

\subsection{Grasp-parameterized centralities}

\begin{figure*}
\includegraphics[scale=1.2, trim={0 0cm 0cm 0cm},clip]{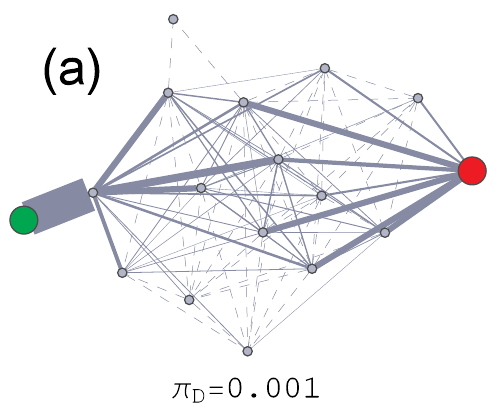}
\includegraphics[scale=1.2, trim={0 0cm 0cm 0cm},clip]{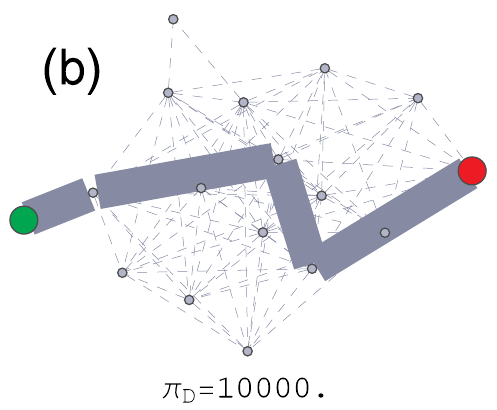}
\caption{\label{fig:grasp_demo} {High and low grasp centralities.} The figures depict the current of random walkers used to calculate the conditional current betweenness and the conditional resistance closeness from \cite{gurfinkel2020absorbing}. This is demonstrated on the (weighted) kangaroo interaction network from \cite{kangadata,grant1973dominance}. Line thickness is proportional to current magnitude.  A unit current flows from the source node (large, green) to the target node (large, red). Dashed lines indicate negligible current ($<.01$ units). (a) At high grasp  (low $\para_D$),  the current takes advantage of many parallel paths. (b) At low grasp (high $\para_D$), the current follows only the shortest weighted path from the source to the target. }
\end{figure*}

A centrality parameter $\para$ is a {\it grasp parameter } if it tends to attenuate the influence of indirect paths between two nodes in a weighted graph. As illustrated in Fig.~\ref{fig:grasp_demo}, when the centrality parameter is set to  high grasp, the measure takes into account many parallel paths between the nodes, while when the centrality parameter is set to low grasp, the measure only considers the shortest path between the two nodes. This is distinct from the behavior of reach parameters because the two nodes can be an arbitrary (weighted) distance apart. Thus, reach-parametrized and grasp-parametrized are distinct centrality categories.

 In \cite{gurfinkel2020absorbing}, we introduced the grasp-parametrized centrality category, as well as two grasp-parametrized measures, based on absorbing random walks: the {\it conditional current betweenness } [$\mathbf{M}^\mathrm{CBT}(\para_D)$] and the {\it conditional resistance closeness} [$\mathbf{M}^\mathrm{RCC}(\para_D)$]. Collectively, these are the {\it conditional walker-flow} centralities, parametrized by the ``walker death parameter'' $\para_D$.  The  conditional current betweenness interpolates from  betweenness, at low grasp, to Newman's {\it random-walk betweenness } \cite{newman2005measure}, at high grasp. Similarly, the conditional resistance closeness interpolates from the harmonic closeness, at low grasp, to the harmonic form of the Stephenson--Zelen {\it information centrality} \cite{stephenson1989rethinking} (also known as the current-flow closeness \cite{brandes2005centrality} and the resistance closeness \cite{gurfinkel2020absorbing}), at high grasp.

\subsection{Classification of parametrized centralities}
\label{sec:difficulties}

There is a proliferation of centrality measures in the network-science literature. Even in the case of parametrized centrality measures, which have not yet been studied extensively, there are sufficiently many measures to require an organizing principle. Here, we build on  the typologies introduced by Borgatti in  \cite{borgatti2005centrality,borgatti2006graph}. There,  centralities are situated along the conceptual dimensions of Summary Type, Walk Position, and 
Walk Type. Each of these is described below. In Table~\ref{tab:classification}, all of the parametrized centralities discussed in this paper are classified according to Walk Position (columns) and Walk Type (rows).

\subsubsection{Summary Type: how influences are aggregated}
The difference between the standard (row-sum) centrality $\mathbf{M}$ and exogenous centrality $\widetilde{\mathbf{M}}$ lies in what Borgatti calls Summary Type, which dictates the way influences are aggregated, not the fundamental nature of the centrality.  Another possible variation is the diagonal centrality $\overline{\mathbf{M}}=\pmb{\mathrm{Diag}}(\mathbf{M})$. Estrada's {\it subgraph centrality } \cite{estrada2005subgraph} is equivalent to $\overline{\mathbf{M}}^\mathrm{COM}$ at $\para_T=1$.

\subsubsection{Walk Position: radial and medial centralities}

Though the conditional current betweenness and conditional resistance closeness are parametrized by the same ``walker death" process, they are very different measures. In Borgatti's typology, the first of these is a {\it medial } centrality while the latter is {\it radial}. This means that the former assigns importance to a node based on the walks passing through it, while the latter assigns importance based only on the walks that start on the node. The classic examples of medial and radial centrality are {\it betweenness } and {\it closeness}, respectively, and we have seen that the conditional walker-flow centralities reduce to these at low grasp. The columns in Table~\ref{tab:classification} group the parametrized centralities discussed in this paper into radial and medial categories.

\subsubsection{Walk Type: reach, grasp, conserved flows, duplicating flows, cyclic flows, and acyclic flows}
The Walk Type conceptual dimension describes the characteristics of the walks through which influence is spread. For example, influence might be restricted to geodesic paths, or to walks of a certain length. It is clear, then, that the categories of reach-parametrized and grasp-parametrized  centrality represent differences in Walk Type.  

A further distinction within the Walk Type  is described in \cite{borgatti2005centrality}, which compares conserved flow processes ({\it e.g.}, the movement of physical objects) to duplicating flow processes ({\it e.g.}, the spread of gossip). The conditional current betweenness and conditional resistance closeness are both calculated using the conserved current created by a single random walk, so they are conserved-flow centralities. On the other hand, the Katz, PageRank, and Communicability centralities rely on infinite summations, as in Eqs.~(\ref{eq:katzexpand}) and (\ref{eq:communicability}), aggregating influence from an infinite number of walks. These are thus duplicating-flow centralities. 

Another important subcategory within the Walk Type dimension is cyclicity. (Borgatti addresses cyclicity within his ``trajectory dimension''.) The Katz, PageRank, and Communicability are cyclic: the spread of influence within these centralities is free to form cycles, potentially even recrossing the same edge over and over. Thus, for all the measures considered here,  cyclic centralities are  based on duplicating flows, while acyclic centralities are based on conserved flows.  However, in general, cyclicity and duplication are independent of each other. 

The rows in Table~\ref{tab:classification} group the parametrized centralities discussed in this paper into  Walk Type categories.

\subsubsection{Disfavored centrality combinations} \label{sec:reachgrasp}

Generally, reach parametrization is not compatible with medial measures like betweenness, since every pair of source and target nodes is considered equally, no matter how far apart they may be. This is why there are no well-known measures in the light-font areas of the right column of Table~\ref{tab:classification}. However, any reach-parametrized  relationship (such the entries in the matrix  $\mathbf{M}^\mathrm{COM}$) may be used to weight pairs of nodes, allowing betweenness-like measures to use reach parameters. (This modification would also allow the simultaneous use of reach and grasp parameters.) These areas of the table are marked with stars to indicate that these centrality combinations are achievable, though they have not been studied extensively to our knowledge. 

Centralities that are both  duplicating and grasp parametrized are also disfavored. It is difficult to control the grasp of duplicating-flow centralities since, by the nature of duplicating influence, they generally cannot restrict influence to geodesic paths. 
However, an exception to this rule is found---for the {\it medial} parameter type---in the form of the communicability betweenness \cite{estrada2009communicability}, and similarly constructed centralities. They rely on a mathematical technique for converting radial reach-parametrized centralities into medial grasp centralities. This is described in Appendix \ref{app:cmb}.  We are not aware of any similar techniques for arriving at {\it radial}, duplicating, grasp centralities, which is why this area of  Table~\ref{tab:classification} remains empty.

\subsubsection{A new radial reach-parametrized centrality based on acyclic, conserved flows}
\label{sec:newcent}

Aside from the disfavored centrality combinations described above, there is one location in Table~\ref{tab:classification} (indicated with bold stars) that has, to our knowledge, not yet been studied.  The Katz centrality, which is radial and reach-parametrized, is one of the oldest measures in the network science literature, and the PageRank, of the same type, is one of the most prominent. It is striking, therefore, that there is no well-known  {\it conserved-flow} centrality of this type, given the importance of conserved flows in both theoretical and practical domains. Therefore, in Sec.~\ref{sec:gcc}, we introduce the {\it ground-current  centrality}, which is of the radial, reach-parametrized, and conserved-flow type. It is also acyclic, whereas the duplicating radial reach measures are cyclic. In Sec.~\ref{sec:results}, we  show that the use of acyclic, conserved flows leads the ground-current centrality to some notable differences from the other measures in the radial, reach-parametrized category. 

\begin{table}[h]
\caption{\label{tab:classification}{\it Classification of parametrized centralities.} Centrality measures are classified according to Borgatti's \cite{borgatti2005centrality,borgatti2006graph} Walk Position (columns) and Walk Type (rows).  Conditional current betweenness subsumes betweenness and random walk betweenness, while conditional resistance closeness subsumes closeness and information centrality \cite{gurfinkel2020absorbing}. Reference \cite{avrachenkov2013alpha} describes the beta current-flow centrality, whose derivation is similar to that in Sec. \ref{sec:gccf}. The positions in the table depicted with a light font represent disfavored centrality types, discussed in Sec.~\ref{sec:reachgrasp}. The starred  entries represent centralities introduced in this paper, filling in ``blanks'' within the table. The ground-current centrality is the main result of this paper.}
\begin{ruledtabular}
\begin{tabular}{cllll}

                                      & \multicolumn{2}{l}{Radial}                                & \multicolumn{2}{l}{Medial}                                 \\ \hline
\multirow{3}{5.5em}{\vspace{-3em} \begin{minipage}{5.5em} \vspace{-1em} Acyclic \\ \vspace{-1em}conserved \\ \vspace{-1em} flow\end{minipage}}            & Grasp:  &cond. resistance closeness \hspace{1em} & Grasp: & cond. current betweenness, and \cite{avrachenkov2013alpha}\\ 
                                      & Reach: & $\textbf{*}$\textbf{ground current (Sec.~\ref{sec:gcc}-\ref{sec:results})}$\textbf{*}$                          & \light{Reach:}& \light{$*${see Sec.~\ref{sec:conc}}$*$} \\ \hline
\multirow{2}{5.5em}{\vspace{-1em} \begin{minipage}{5.5em} \vspace{-1em} Cyclic \\ \vspace{-1em}duplicating \\ \vspace{-1em} flow\end{minipage}} &\light{Grasp:}&\light{none (see Sec.~\ref{sec:reachgrasp} )} & Grasp: & communicability betweenness \\ 
                                      & Reach: & Katz, PageRank,  communicability \hspace{2em}                           &  \light{Reach:}& \light{$*${see Sec.~\ref{sec:reachgrasp}}$*$}                                                                                        \\ 
\end{tabular}
\end{ruledtabular}
\end{table}

\section{The ground-current centrality \label{sec:gcc}}

\subsection{Generalizing the resistance-closeness centrality}
\label{sec:gcc_first}

This paper is concerned with developing a conserved-flow   centrality measure that features a reach parameter, tuning the distance that influences can spread across the network. To estimate the node centralities in network $\mathcal{N}$, we focus our model on the electrical current  flows in the  resistor network derived from $\mathcal{N}$ (or equivalently, random walkers \cite{Doyle06randomwalks} on $\mathcal{N}$). In this interpretation, an element of $\mathcal{N}$'s adjacency matrix $\mathbf{A}_{ij}$ is  taken to be the conductance (inverse resistance) of the direct electrical connection between nodes $i$ and $j$ \footnote{The interpretation of non-zero adjacency matrix elements as conductances in a resistor network makes sense for {\it affinity-weighted}  networks \cite{newman2005measure}, as well as multigraphs where $\mathbf{A}_{ij}$ stands for the number of parallel edges between $i$ and $j$. When an adjacency matrix element stands for, {\it e.g.}, a distance rather than an affinity, it is more appropriate for it to be interpreted as a resistance. In either case, when $\mathbf{A}_{ij}=0$, there is no edge between $i$ and $j$. }. By using current flow to spread influence, we guarantee that the resulting centrality will be both conserved and acyclic.

It is not possible to explicitly limit the reach of current  (and hence influence) by increasing the resistance along all edges, or changing the strength of voltage sources. Since network current flow is a linear theory, any introduction of a multiplicative constant $m$ on either (1) all voltage sources, or (2) all resistances, will only scale the resulting currents by $m$. And since centrality vectors are normalized by the factor $\alpha$ in Eq.~(\ref{eq:centrality}), any multiplicative constants do not affect the final centrality assignments. This provides motivation to build a parametrization around resistors {\it external } to the equivalent resistor network.

\begin{figure*}
\includegraphics[scale=0.85, trim={0 1cm 0cm 0cm},clip]{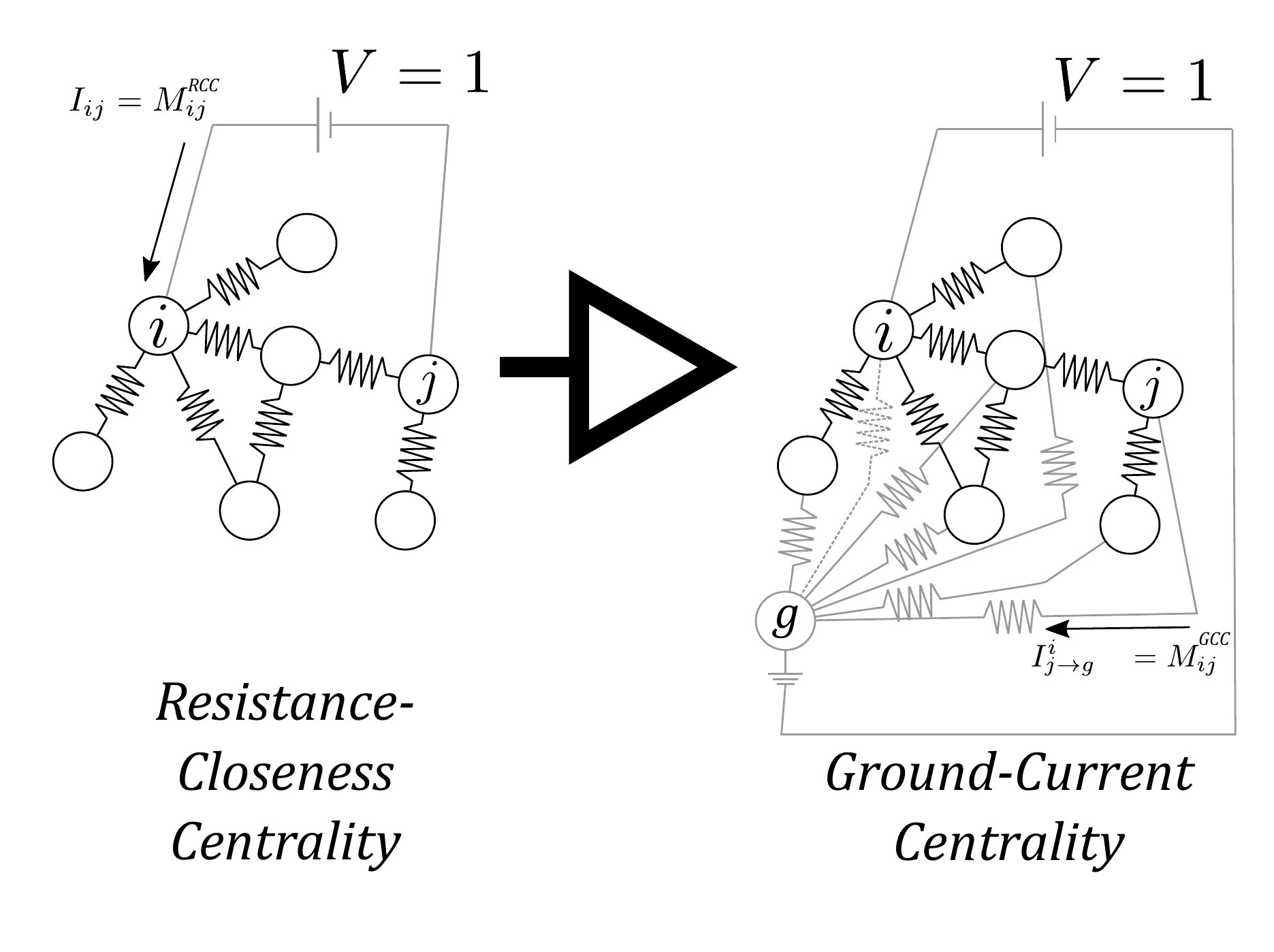}
\caption{\label{fig:gcc} The ground-current centrality (right) as a multinode generalization of the resistance-closeness centrality (left). The ground-current centrality of a node $i$ is given as a function of the finite ground conductances (shown in light gray), by the currents flowing from that node to the ground node $g$ when a unit voltage is introduced between node $i$ and $g$.  The exogenous ground-current centrality $\widetilde{\mathbf{M}}^\mathrm{GCC}_{ij}$ is equivalent to the removal of the (dotted) connection between $i$ and and $g$. Ignoring the voltage sources, the left side of the figure illustrates the resistor-network interpretation of the network $\mathcal{N}$, while the right side illustrates the modified network $\mathcal{N}_g$}.
\end{figure*}

We now present a new centrality, which is a generalization of (but not a parametrization of) the resistance-closeness centrality (RCC) studied in \cite{gurfinkel2020absorbing}. There, $\mathbf{M}^\mathrm{RCC}_{ij}$ is equal to the inverse of the effective resistance $R^{\mathrm{eff}}_{ij}$, which is the current resulting  from connecting a 1-Volt battery between $i$ and $j$ in $\mathcal{N}$, as seen in Fig. \ref{fig:gcc}(left). Without affecting the results, we may set the absolute potential scale by connecting $j$ to the ground node $g$ with a resistance-less wire; the current then returns to the battery through the ground node. Extrapolating the measure to multiple nodes is achieved simply by connecting \em all \em nodes directly to ground. The currents $ I^i_{j \rightarrow g}$ from each $j$ to ground (when the voltage source is on $i$) are straightforwardly interpreted as the contribution of $j$ to the centrality value of $i$; that is, the ground currents are just the $\mathbf{M}_{ij}$. 
Thus, we name the new measure the \em ground-current centrality \em (GCC).  In summary:
\begin{equation}
\label{eq:gccbasic}
\mathbf{M}^\mathrm{GCC}_{ij} = I^i_{j \rightarrow g}\qquad  \mbox{(unit voltage source between $i$ and ground)}
\end{equation}
(In what follows, we will often omit the superscripts in $\mathbf{M}^\mathrm{GCC}$ and $c^\mathrm{GCC}$ when it is clear from the context that we are referring to the ground-current centrality.)

This centrality measure represents a transition from the resistance distance, a node-node relation, to a  node-network relation; this process is illustrated in Fig. \ref{fig:gcc}. The ground-current centrality also represents a complementary approach to our previous work in \cite{gurfinkel2020absorbing}. There, the conditional walker-flow centralities employ the portion of the current that does not eventually reach ground. Here, the entirety of the current eventually reaches ground, and the centrality is based on the magnitudes of the ground currents.

If all the nodes were directly connected to ground with zero resistance, then they would all be at the same potential. This would mean that no current could flow between them, leading to a centrality insensitive to the details of the network structure. To prevent this behavior, we introduce the ground-conductance vector $\ket{C}$, where $C_j$ is the finite conductance of the edge connecting $j$ to ground. 
The node potentials are now $V_j = I^i_{j \rightarrow g} /C_j = \mathbf{M}_{i j}/ C_j$---in general they are all different. Since the network $\mathcal{N}$ has $N$ nodes, adding $g$  and its adjacent edges creates a  $(N+1)$-node network. This new network, called $\mathcal{N}_{g}(\ket{C})$, is illustrated on the right side of Fig. \ref{fig:gcc}. Note that one of the edges between $g$ and $i$ (indicated by the battery symbol in the circuit diagram) represents voltage boundary conditions, and  is therefore not included in $\mathcal{N}_{g}(\ket{C})$.

\subsection{The ground-current centrality formula \label{sec:gccf}}

We now derive a compact formula for the ground-current centrality. The foundational relation for resistor networks \cite{NEWM10}---as applied to $\mathcal{N}_g (\ket{C})$---is 
\begin{equation}\label{eq:networkvoltage} \ket{I^\mathrm{in}} = \mathbf{L}^g\ket{V}.\end{equation}
Here, $\ket{V}$ is the vector of node voltages, and $\mathbf{L}^g$ is the $(N+1)\times(N+1)$ Laplacian matrix of $\mathcal{N}_{g}(\ket{C})$.  The $j$th element of the vector $\ket{I^\mathrm{in}}$ is equal to the current entering ($I^\mathrm{in}_j>0$) or leaving ($I^\mathrm{in}_j<0$) the network at node $j$. In the present case, illustrated in Fig.~\ref{fig:gcc}(right), $I^\mathrm{in}_j=0$ when $j$ is not $i$ or $g$.

Because $\mathbf{L}^g \ket{1}=0$, Eq.~(\ref{eq:networkvoltage}) cannot be inverted as is. A standard solution \cite{newman2005measure} is to remove one node from the network, leading to the invertible $N\times N$ {\it reduced } Laplacian $\mathbf{L}^\mathrm{red}$. This specifies the gauge in which the removed node is at  zero potential (see Appendix \ref{app:reduced}). 

We choose to remove node $g$, appropriately setting its potential to zero. Proceeding similarly to the derivation in \cite{avrachenkov2013alpha}, removing $g$ leads to the reduced Laplacian    $\mathbf{L}^\mathrm{red}= \mathbf{L} +\pmb{\mathrm{Diag}}(\ket{C})$. Here $\mathbf{L}$ is the standard Laplacian of the $N$-node network $\mathcal{N}$: $\mathbf{L}=\pmb{\mathrm{Diag}}(\ket{k})-\mathbf{A}$, where $\ket{k}$ is the {\it weighted } degree vector. Therefore, inverting the reduced version of  Eq.~(\ref{eq:networkvoltage}) leads to
\begin{equation}
V_j=  \left[\mathbf{L} +\pmb{\mathrm{Diag}}(\ket{C})\,\right]^{-1}_{i j} I^\mathrm{in}_i,
\end{equation}
where we used the fact that $\mathbf{L}$ is symmetric. Recall that $i$ is the index of the node connected to the battery [see fig.~2(b)], while $j$ can stand for any node, including $i$.

From the requirement that $V_i=1$,  we have $I^\mathrm{in}_i= 1/\left[\mathbf{L} +\pmb{\mathrm{Diag}}(\ket{C})\,\right]^{-1}_{i i} $. The current $\mathbf{M}_{ij}$ from $j$ to $g$ is just $V_j C_j$.  And because all the current entering the network at $i$ must also leave the
 network at $g$, $\sum_j \mathbf{M}_{ij} =\sum_j I^i_{j \rightarrow g}= I^\mathrm{in}_i$, so $I^\mathrm{in}_i$ is equal to the  centrality $c_i$ of node $i$.   Assembling these results, we arrive at a generalized formula for the ground-current centrality:
\begin{eqnarray}\label{eq:gccf}
c_i =&\;1/& \left[\mathbf{L}+\pmb{\mathrm{Diag}}(\ket{C}) \,\right]_{i i}^{-1}\nonumber \\
\mathbf{M}_{i j}= &  c_i & [\mathbf{L}+\pmb{\mathrm{Diag}}(\ket{C}) \,]_{i j}^{-1} C_j .
\end{eqnarray}
 Every row of $\mathbf{M}$ corresponds to a different experimental situation, where the voltage boundary conditions are changed by connecting a different node $i$ to the 1-Volt battery.   In matrix form, this can be expressed as $\mathbf{M}= \{\pmb{\mathrm{Diag}}([\mathbf{L}^\mathrm{red}]^{-1})\}^{-1}   [\mathbf{L}^\mathrm{red}]^{-1}  \pmb{\mathrm{Diag}}(\ket{C}) .$

For notational convenience, in this section we use the {\it unnormalized} form of the centrality. 
 It can be easily verified that $\sum_j [\mathbf{L}+\pmb{\mathrm{Diag}}(\ket{C}) \,]_{i j}^{-1} C_j=1$. This leads to $\sum_j \mathbf{M}_{ij}=c_i$, which is the unnormalized form of Eq.~(\ref{eq:centrality}).
 
 We note that, unlike for other centralities, the elements of the ground-current  centrality matrix $\mathbf{M}^\mathrm{GCC}$ do not need to be calculated to find the $c_i$---in fact, the reverse is true. Nonetheless, the  $\mathbf{M}^\mathrm{GCC}_{ij}$ are informative in their own right, since they encode the influence of node $j$ on $i$'s centrality. Here, they will be useful for analyzing test cases that show how the ground-current centrality differs from similar measures; see Sec.~\ref{sec:results}.

The vector $\ket{C}$ in Eq.~(\ref{eq:gccf})  can be used to tune the relative importance of nodes in the network. For example, in a power-grid network, we may set $C_i=0$ when $i$ is a generator, thereby ensuring that the centrality only rewards connections to loads.  However, the simplest case, as in \cite{avrachenkov2013alpha}, is to set all ground conductances  to the same value $\para_C$, meaning that $\pmb{\mathrm{Diag}}(\ket{C}) = \para_C \pmb{\mathbb{I}}$, for identity matrix $\pmb{\mathbb{I}}$. This leads us  to the final parametrized form of our centrality:
\begin{equation}\label{gccfinal}
\left.\begin{aligned}
 c_i(\para_C) &=& 1/&(\mathbf{L} +\para_C \pmb{\mathbb{I}}\,)_{i i}^{-1} \; \qquad \\
\mathbf{M}_{i j}(\para_C) &=&I^i_{j \rightarrow g}=\hspace{1.2em} c_i& (\mathbf{L} +\para_C  \pmb{\mathbb{I}}  \,)_{i j}^{-1}  \; \para_C \, \  \qquad  
\end{aligned}
\right\}
\qquad \text{ground-current centrality} 
\end{equation}

We emphasize that, unlike  other network measures based on current flows, it is not necessary to perform a summation to obtain the centrality $c_i$ of  node $i$. In what follows, we use the normalized form of the ground-current centrality matrix $\mathbf{M}$, as per Eq.~(\ref{eq:centrality}).

\subsection{Properties and limits of the ground-current centrality }
\label{ssc:gccp}

\begin{table}
\caption{\label{tab:gcc}Ground-Current Centrality High/Low $\para_C$ Limits. The ground-current centrality is formulated in Eq.~(\ref{gccfinal}). The limits for the generalized ground-current centrality [Eq.~(\ref{eq:gccf})] are in square brackets. In the generalized version,  $\para_C$ is not defined, and the limits  should be interpreted as high and low values of $\braket{C|1}$. }
\begin{ruledtabular}
\begin{tabular}{c | c | c | c}
Measure& Symbol &High $\para_C$ & Low $\para_C$ \\ \hline
Ground-Current Centrality &$\mathbf{M}^\mathrm{GCC}_{ij}$&$\delta_{ij}\para_C$& $\para_C$\\ 
&&$\big[\;\delta_{ij}C_i\;\big]$ &$\big[\;C_j\;\big]$  \\
Exogenous Ground-Current Centrality &$\widetilde{\mathbf{M}}^\mathrm{GCC}_{ij}$&$\mathbf{A}_{ij}$& $(1-\delta_{ij})\para_C$\\
&&$\big[\;\mathbf{A}_{ij}\;\big]$ &$\big[\;(1-\delta_{ij})C_j\;\big] $
\end{tabular}
\end{ruledtabular}

\end{table}

We have argued that the ground-current centrality has a naturally arising parameter $\para_C$. Though $\para_C$ was necessary to force the centrality to interact with the network structure, it is easy to see that this parameter also has the  effect of tuning the centrality's reach. 
Consider the $\para_C\rightarrow\infty$ limit. When $\para_C$ is large,  the vast majority of the current leaving the battery at node $i$ follows the very high-conductance edge directly to ground, rather than following the relatively low-conductance edges  leading to other locations in the network. A node can thus only influence itself, and $\mathbf{M}$ becomes diagonal. This can also be seen from setting $j=i$ in the second line of  Eq.~(\ref{gccfinal}), whereby $c_i \approx \mathbf{M}_{i i} = I_{i\rightarrow g}$  for large $\para_C$. Thus the reach is low when $\para_C$ is high.

The behavior in the low-$\para_C$ limit is easy to understand through physical properties of  resistor networks:
As $\para_C\rightarrow0$ the effective resistance to ground approaches infinity, leading to very small currents in the network; therefore all node potentials approach the value 1 because the potential drop between adjacent network nodes becomes tiny. Therefore all ground currents are identical: $\mathbf{M}_{ij}=(1 -V_g) \para_C=\para_C$, since $V_g=0$.  Nodes at large graph distances from $i$ are not penalized by the centrality. This means that $c_i=N \para_C$ for all $i$. When $\para_C$ is low, the reach is  high and the network looks the same from every node.

\begin{figure*}
\includegraphics[scale=1.2, trim={0 0cm 0cm 0cm},clip]{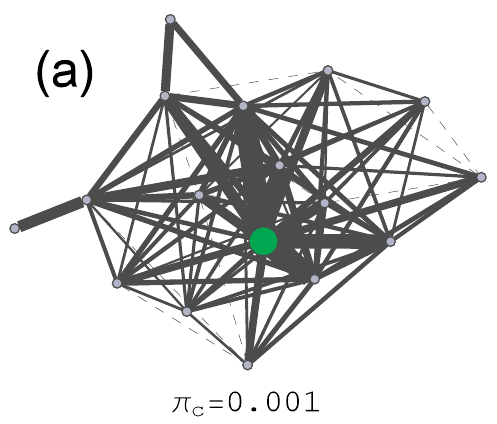}
\includegraphics[scale=1.2, trim={0 0cm 0cm 0cm},clip]{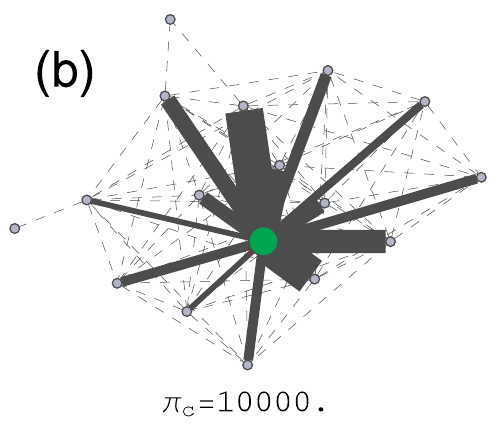}
\caption{\label{fig:reach_demo} {High and low reach in the exogenous ground-current centrality.}  This is demonstrated on the (weighted) kangaroo interaction network from
 \cite{kangadata,grant1973dominance}. Compare the grasp behavior of the conditional current betweenness in Fig.~\ref{fig:grasp_demo}. Line thickness for edges $(k,l)$ indicates the product of the normalization factor $\tilde{\alpha}$ from
  Eq.~(\ref{eq:exocentcentrality}) and the edge current  magnitude $I^i_{k\rightarrow l}$, where the current flow results from a unit potential difference between the source node $i$ (large, green) and the ground node $g$ (not pictured). For readability, the line thickness is proportional to the {square root} of $\tilde{\alpha} I^i_{k\rightarrow l}$.
   Dashed lines indicate negligible current: $\tilde{\alpha} I^i_{k\rightarrow l}<.0001$.  All connections to ground have conductance $\para_C$ and,
   because this is the {\it exogenous } centrality variant $(\widetilde{\mathbf{M}})$, every node other than the source node is connected to ground.  Node $j$'s final contribution to $i$'s centrality is $\tilde{\alpha}\hspace{.2em} \widetilde{\mathbf{M}}^\mathrm{GCC}_{ij} = \tilde{\alpha} I^i_{j\rightarrow g}$. (a) At high reach (low $\para_C$), the current spreads out to every node. Though the  currents $I^i_{k\rightarrow l}$  are very small at this parameter value, the normalization factor results in nonnegligible influences $\tilde{\alpha}\hspace{.2em} \widetilde{\mathbf{M}}^\mathrm{GCC}_{ij}$.  In accordance with Table~\ref{tab:gcc}, the current to ground is the same at every node. (b) At low reach (high $\para_C$), the current  only flows along edges adjacent to the source, weighted by the edge
      conductance---see Table \ref{tab:gcc}.  }
\end{figure*}

\begin{figure*}
\includegraphics[scale=1.9, trim={3.0cm 2.2cm 2.8cm 2.2cm},clip]{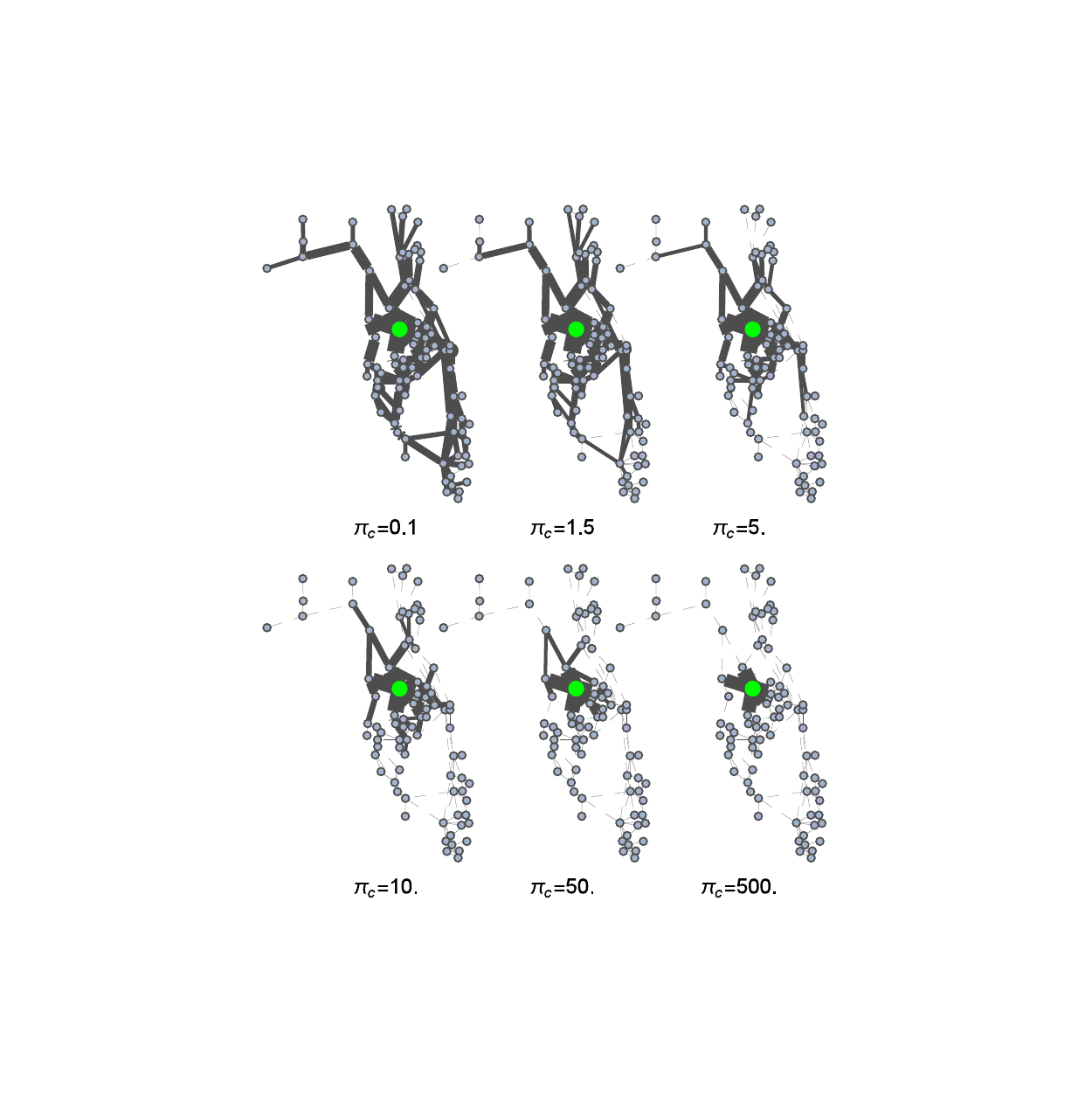}
\caption{\label{fig:reach_intermediate} {Intermediate reach in the exogenous ground-current centrality.} See the caption to Fig.~\ref{fig:reach_demo} for explanatory details.  The reach is demonstrated on the (weighted) Florida power-grid network from \cite{dale,xu2014architecture}.
 In this version of the network, the weights are readable from the figure: they are inversely proportional to the Euclidean distance between nodes. When the reach is high ($\para_C$ is low), the  currents spread to nodes at large weighted distance from the voltage source. In this regime (e.g., at $\para_C=0.1$), the amount of current flowing to ground from each node is approximately identical. As the reach decreases ($\para_C$ increases to, e.g., $5.0$), the ground-currents are no longer identical. The currents along edges far from the voltage source are diminished and, at very low reach (e.g., $\para_C=500$), only currents to the voltage source's nearest neighbors remain.   }
\end{figure*}

It is also useful to consider the exogenous ground-current centrality $\widetilde{\mathbf{M}}^\mathrm{GCC}$. Referencing Fig.~\ref{fig:gcc}, this amounts to the removal of the dotted connection to ground.  This variant  can recover the adjacency matrix for large values of $\para_C$---much like the communicability centrality recovers the adjacency matrix for large values of $\para_T$. 
Detailed calculations for the limiting forms of the two variants of ground-current centrality for \em arbitrary \em $\ket{C}$ vectors are presented in Appendix \ref{app:gcc}. We summarize the limits in Table \ref{tab:gcc}. 

The behavior of the ground-current centrality at intermediate values of $\para_C$ is intermediate to the behavior at the limits. As $\para_C$ decreases from $\infty$, pairs of nodes ($i,j$) separated by larger  weighted graph distances $D_{ij}$ start to receive non-negligible ground current $\mathbf{M}_{ij}$. This means that the {\it reach} of the centrality increases as $\para_C$ decreases and, therefore, $\para_C$ is a reach parameter. Finally, as $\para_C$ approaches $0$, {\it all } pairs produce the same value of  $\mathbf{M}_{ij}$, regardless of the distance between $i$ and $j$---reach is maximized. (The centrality {\it at } $\para_C=0$ is undefined, however, since there is no ground-current flow  in that situation.)

 Increasing the reach by decreasing $\para_C$ allows longer network paths to be explored, which leads to more parallel paths to the same destination. Therefore, tuning reach in this case also necessarily tunes grasp, but this is a secondary effect.

The reach behavior of the exogenous ground-current centrality at   high reach (low $\para_C$) and  low reach (high $\para_C$) is illustrated in Fig.~\ref{fig:reach_demo}. The intermediate reach behavior is illustrated in Fig.~\ref{fig:reach_intermediate}.
The figures also clearly illustrate the ground-current centrality's status as a {\it radial } measure: influence spreads outward from the node $i$. Further, the centrality of every node is derived from a single conserved current flow in a resistor network. These key properties of the ground-current centrality are reflected in its position in Table~\ref{tab:classification}.

\section{Unique features of the ground-current centrality}\label{sec:results}

Because of its unique position in the taxonomy presented in Table~\ref{tab:classification}, the ground-current centrality differs significantly from similar centralities. In Sec.~\ref{sec:adv_conserved} we compare it to other conserved flow centralities (top row in Table I), while in Sec.~\ref{sec:adv_reg} we compare it to other radial reach-parametrized centralities (left column in the table).

\subsection{Differences from other conserved-flow centralities}
\label{sec:adv_conserved}

Referencing the final expressions in Eqs.~(\ref{eq:gccf}) and (\ref{gccfinal}), we consider the differences between the ground-current centrality and other current-based centrality measures previously considered (the first two rows in Table \ref{tab:classification}). Of course, the most important difference is that the ground-current centrality is the only one of these that can control  reach, which is in many ways a more intuitive type of parametrization than grasp. Further, the other methods' centrality matrices do not reduce to the adjacency matrix at any parameter value---this is a consequence of these centralities not using a reach parameter, and thus being unable to restrict influence to nearest neighbors.

The ground-current  centrality is also mathematically simpler than the alternatives.  The  closeness and betweenness centralities rely on algorithms (Dijkstra's algorithm and the method described by Brandes in \cite{brandes2001faster}, respectively), while the ground-current  centrality  has a closed-form solution. The resistance closeness and the current betweenness rely on the calculation of currents using the pseudoinverse or the inverse of a reduced Laplacian matrix. On the other hand, the ground-current  centrality  uses an ordinary matrix inverse and the ordinary Laplacian $\mathbf{L}$. This is convenient for formula manipulations such as those in Appendix \ref{app:gcc}. Further, the conditional forms \cite{gurfinkel2020absorbing} of the resistance closeness and current betweenness require the calculation of current on every edge, while the ground-current  centrality  only calculates currents that correspond to elements of $\mathbf{M}^\mathrm{GCC}$. In fact, even this is unnecessary:  Eqs.~(\ref{eq:gccf}) and (\ref{gccfinal}) show that the final centralities can be found from the diagonal of the inverted matrix, without summing over $\mathbf{M}_{ij}$.

Finally, we emphasize that the ground-current centrality is significantly simpler conceptually than the alternative measures. All of these involve solving  a current (or walker) flow problem between  pairs of nodes and   aggregating all such pairs to calculate the final centrality. The ground-current  centrality, however, requires only a single current-flow problem for every node whose centrality we wish to calculate.

\subsection{Differences from other radial reach-parametrized centralities}
\label{sec:adv_reg}


In Table \ref{tab:classification}, the ground-current centrality is the only radial reach-parametrized centrality that is based on an acyclic, conserved flow. As a result, it differs significantly from the  Katz, PageRank, and communicability centralities. Especially at high reach, these alternative centralities lead to unintuitive centrality rankings on simple example networks. The reason is that the cyclic flows employed by these centralities are forced to retrace their steps when the reach is high, while the ground-current  centrality's conserved current flow never does so because current flow is acyclic. 

\begin{table}[]
\caption{\em Summary of real-world example networks\/\em. Networks have $N$ nodes and $M$ edges. The density of a network is defined as the number of edges divided by the number of possible edges: $M/(0.5 N (N-1))$.  }
\label{tab:nets}
\setlength\tabcolsep {.35em}
\begin{tabular}{clccccc}
Network               & Refs. & \textit{N} & \textit{M}& Density & Weights                                                           \\ \hline
\textit{C. elegans} Neuronal Network & \cite{Choe-2004-connectivity}& 277 & 1918 & 0.05 &Unweighted\\

Weighted Florida Power Grid   &     \cite{xu2014architecture,dale}      & 84         & 137   &0.04     & Continuous   \\
Unweighted Florida Power Grid &     \cite{xu2014architecture,dale}      & 84         & 137 &0.04       & Unweighted        \\
Italian Power Grid & \cite{hama10} & 127 & 169 & 0.02 & Unweighted \\
Vole Trapping & \cite{nr-aaai15, voles} & 118 & 283 & 0.04 & Integer\\
Kangaroo Group             &      \cite{kangadata,grant1973dominance}     & 17         & 91    & 0.67     & Integer                       \\
 Benchmark Circuit & \cite{milo2004superfamilies} &  512 & 819 & 0.006 & Unweighted \\
\end{tabular}
\end{table}

We  compare the behavior of the radial reach-parametrized centralities on  line networks, subdivided star networks, Cayley trees modified to become regular networks, and a lattice network with a weighted bottleneck.  In these simply structured networks,  the nodes' centrality rankings are intuitive. Here we take the closeness centrality  to provide the paradigmatic intuitive centrality ranking, since it  assigns greater importance to nodes that are close to many others. In our simply structured example networks, such nodes are easy to identify by eye. In addition to the simple example networks, we analyze numerical data from seven real-world example networks, summarized in Table \ref{tab:nets}. Because we compare node-node centrality flows (elements of $\mathbf{M}$) as well as final centralities (elements of $\ket{c}$), we rely specifically on the harmonic \cite{Dekker2005,Rochat2009,NEWM10} closeness centrality: $\mathbf{M}^\mathrm{HCC}=D_{i j}^{-1}$. 

Of the parametrized centralities considered here, only the ground-current centrality can reproduce the intuitive ordering in all the simply structured example networks. Furthermore, in the case of networks with bottlenecks---including real-world networks---the ground-current centrality reproduces aspects of the betweenness centrality, as well as the harmonic closeness. In the case of regular networks, the ground-current centrality does not result in nearly identical centrality values, as do several of the alternative measures. Conversely, in the case of real-world  networks with hubs, we show that the ground-current centrality assigns centrality weight more equitably than the communicability centrality, while still giving the most weight to the hub.

In this section we use the exogenous form ($\widetilde{\mathbf{M}}$) of the discussed centralities,  since
only the exogenous forms of the communicability, Katz, and ground-current centralities reduce to degree centrality at low reach ($\mathbf{M}$ reduces to $\mathbf{A}$). 
 Furthermore, only the exogenous communicability centrality leads to nontrivial results in the case of regular networks (see Sec.~\ref{sec:reg}). However, the results for the full ground-current centrality $\mathbf{M}^\mathrm{GCC}$ are very similar to those for  $\widetilde{\mathbf{M}}^\mathrm{GCC}$.
We also limit the discussion to  {\it normalized} centralities, introducing the normalization factor $\widetilde{\alpha}$ into Eq.~(\ref{gccfinal}) so that $\widetilde{\alpha} \sum_{ij}\widetilde{\mathbf{M}}=1$. Without normalization, centrality values for the communicability ($\widetilde{\mathbf{M}}^\mathrm{COM}$)  become unmanageably large at high reach, while ground-current centrality values ($\widetilde{\mathbf{M}}^\mathrm{GCC}$) go to zero in the same regime.

\subsubsection{Line Networks}\label{sec:lines}
Consider the unweighted network of $N$ nodes arranged in a straight line, so that the two end nodes have degree 1, while the middle $N-2$ nodes have degree 2. Here, the harmonic closeness centrality   specifies a centrality ranking that grows with proximity to the center of the line. Indeed, this intuitive ordering is reproduced by almost all the centrality measures under consideration, and across all parameter values  (except those extremal values where all centralities are equal). The PageRank is the only centrality that does not reproduce this ordering. 

The PageRank  places the degree 1 nodes in the lowest centrality rank, but the rankings from there on out are the {\it reverse} of those of the harmonic closeness, so that the node at the center of the line has the second-lowest rank. This unintuitive ordering occurs at all nonextremal parameter values. More generally, the PageRank has properties that make it unsuitable as a reach-parametrized centrality. As the parameter goes to zero, the reach technically increases. However, at this parameter value, the random walk behind the PageRank is allowed to take many steps, including steps that retrace its own path. Thus the walk approaches its stationary distribution, which is proportional to the degree of nodes \cite{NEWM10}. The result is the paradoxical situation where increasing the PageRank's reach tends to make it more like the degree centrality, which is inherently low-reach. We believe that this behavior leads to the unintuitive ordering on the line network.

Originally, the PageRank centrality was developed to rank websites, which form {\it directed} networks of hyperlinks. Our simple test case suggests that the PageRank is not well suited to {\it undirected} networks.

\subsubsection{Subdivided Star Networks}
\label{sec:sstar}
\begin{figure}[h]
\includegraphics[scale=0.7, trim={0 0cm 0cm 0cm},clip]{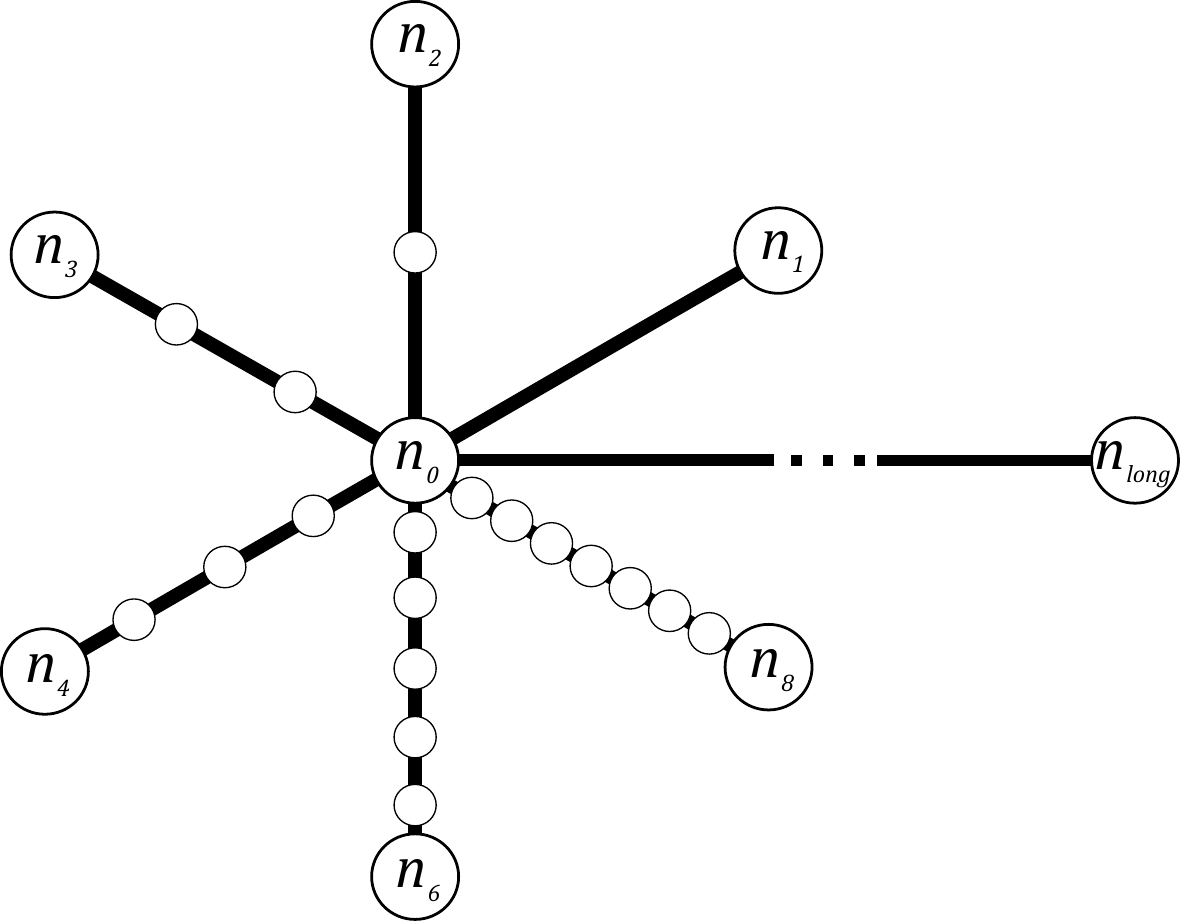}
\caption{\label{fig:sstar} The subdivided star network $\mathcal{S}_{\{1,2,3,4,6,8\}}$. We only compare the centralities of the large, labeled nodes. However, all 26 nodes are accounted for in the adjacency matrix. The node labels indicate the number of edges in the ``spoke'' terminated by that node, {\it e.g.}, one must traverse 4 edges to move from $n_0$ to $n_4$.    The weights are chosen to make the total weighted graph distance along the spoke equal to unity: Here, $D_{n_0 n_1}=D_{n_0 n_2}=D_{n_0 n_3}=D_{n_0 n_4}=D_{n_0 n_6}=D_{n_0 n_8}=1$. All edges within a spoke have the same length, and thus the same weight. (The edge weights are inversely proportional to the Euclidean distances in the figure). For example, the 6 edges  between $n_0$ and $n_6$ have weight 6. Because edge weights are inverse to weighted edge distances,  $6 \times (1/6)=D_{n_0 n_6}=1$. There is one exception to the previous rules: a long edge $(n_0,n_\mathrm{long})$, where $d_{n_0 n_\mathrm{long}}=1$ and $D_{n_0 n_\mathrm{long}}=1000$. }
\end{figure}

We now introduce a simple class of {weighted} networks that also have intuitive centrality matrix values based on the harmonic closeness.    These {\it subdivided star networks\/} $\mathcal{S}_{\{d\}}$ comprise  a series of ``spokes'' emanating from the hub node $n_0$. Each spoke consists of a chain of edges. The network is specified precisely by  ${\{d\}}$, the list of {\it unweighted} distances along the spokes. The edge weights are chosen to make the {\it weighted} distance ($D$) along each spoke equal to unity.  See the caption to Fig.~\ref{fig:sstar} for further details and  an illustration for ${\{d\}}={\{1,2,3,4,6,8\}}$. We also intend to compare the behavior of a node very distant from $n_0$. To do this, we connect a final node $n_\mathrm{long}$ directly to $n_0$,  setting $D_{n_0 n_\mathrm{long}}=1000$.

We are only concerned with influence flows  between the hub node and the nodes at the ends of the spokes. We choose $\mathcal{S}_{\{1,2,5,10,18,30\}}$ as a representative example network, on which we compare the influence values $\widetilde{\mathbf{M}}$ for different centralities. Specifically, we consider $\widetilde{\mathbf{M}}_{n_0 i^\mathrm{p}}$, for peripheral nodes $i^\mathrm{p} \in {\{n_\mathrm{long},n_1,n_2,n_5,n_{10} \ldots\}}$.  All the  nodes $i^\mathrm{p}$ (except $n_\mathrm{long}$) are the same weighted distance from $n_0$, but their unweighted distances $d_{n_0 i^\mathrm{P}}$ are all different. As a result, the ordering of matrix elements in the {\it unweighted} harmonic closeness (HCC) is clear: $\widetilde{\mathbf{M}}^\mathrm{HCC}_{n_0 i^\mathrm{p}}$ goes down for $i^\mathrm{p}$ on ``longer'' spokes, while these matrix elements are all the same for the {\it weighted} HCC. On the other hand, the { unweighted} HCC matrix elements $\widetilde{\mathbf{M}}^\mathrm{HCC}_{n_0 n_\mathrm{long}}$ and  $\widetilde{\mathbf{M}}^\mathrm{HCC}_{n_0 n_1}$ are the same, while in the weighted case $\widetilde{\mathbf{M}}^\mathrm{HCC}_{n_0 n_\mathrm{long}}$ is much smaller than any other $\widetilde{\mathbf{M}}^\mathrm{HCC}_{n_0 i^\mathrm{P}}$ . (Note that we use the harmonic closeness, because the standard closeness does not specify matrix elements.)  Of all the parametrized centralities considered here, the ground-current centrality is the only one that matches the intuitive ordering of both weighted and unweighted HCC.

Note that we are not comparing the final centrality values $c_{i^\mathrm{p}}$ of the peripheral nodes, but rather the matrix elements $\widetilde{\mathbf{M}}_{n_0 i^\mathrm{p}}$, since these values are not inflated by the presence of nodes along the spokes \footnote{In the final calculation, the many nonperipheral nodes (unlabeled in Fig.~\ref{fig:sstar}) account for the majority of the contribution to $i^\mathrm{p}$'s centrality. This means that peripheral nodes on ``long'' spokes will have larger centrality, just because they are near many nonperipheral nodes. As a result, the  $c_{i^\mathrm{p}}$ will have an ordering that places greater importance on nodes a farther unweighted distance from the hub node $n_0$.  All of the centralities under discussion reproduce this expected $c_{i^\mathrm{p}}$ ordering. We focus on the matrix elements $\tilde{\mathbf{M}}_{n_0 i^\mathrm{p}}$ rather than the final centrality $c_i$ because they are a direct measurement of the influence between two nodes, and as such are more sensitive to the chosen centrality method.  }.

\begin{figure}[h]
\includegraphics[scale=.75, trim={0 .5cm 0cm 0cm},clip]{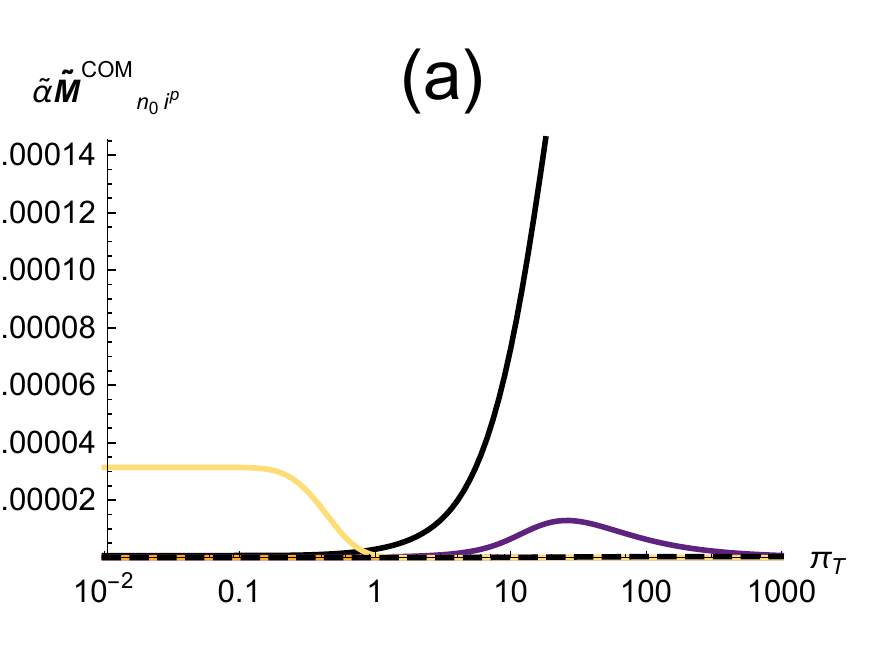}
\includegraphics[scale=.9, trim={0 0cm 0cm 0cm},clip]{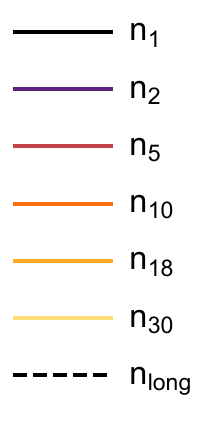}
\includegraphics[scale=.75, trim={0cm .5cm 0cm 0cm},clip]{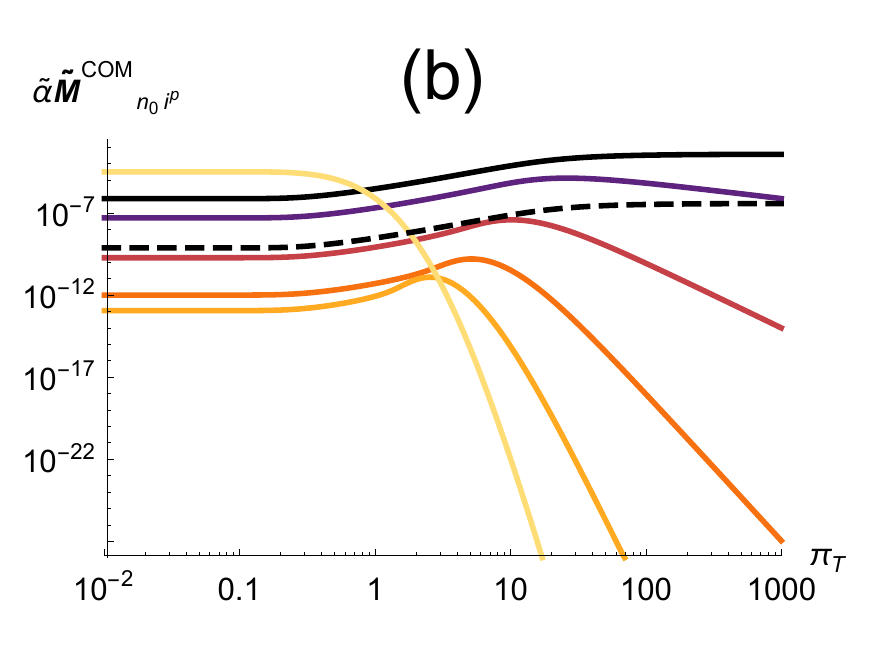}
\caption{
\label{fig:exp_star} 
	{Selected values of $\widetilde{\alpha} \widetilde{\mathbf{M}}^\mathrm{COM}$ for the $\mathcal{S}_{\{1,2,5,10,18,30\}}$ network}. The same data are plotted on (a) log-linear and (b) log-log scales. Note that the normalization factor $\widetilde{\alpha}$ depends on $\para_T$. Without $\widetilde{\alpha}$, the $\widetilde{\mathbf{M}}$ values become unmanageably large. The Katz centrality is qualitatively similar, but with convergence failure at high reach (low values of $\para_T$).  At high reach, the COM fails to reproduce the intuitive HCC ranking of the $\widetilde{\mathbf{M}}_{n_0 i^\mathrm{p}}$.
}
\end{figure}

Figure \ref{fig:exp_star} depicts  the  communicability centrality $\widetilde{\mathbf{M}}^\mathrm{COM}_{n_0 i^\mathrm{p}}$. Though the rank ordering for all $i^\mathrm{P}$ except $n_\mathrm{long}$ matches HCC at low reach (high $\para_T$), the levels begin to cross as  the reach is increased, and at high reach ($\para_T\to0$) $\widetilde{\mathbf{M}}^\mathrm{COM}_{n_0 n_{30}}$ becomes the highest, though in HCC it is the lowest. This matrix element alone deviates from the ordering established at low-reach (high $\para$). The reason, to be discussed in Sec.~\ref{sec:conc}, is  the duplicating nature of the communicability centrality.  Another issue is that the  ranking of centrality element $\widetilde{\mathbf{M}}^\mathrm{COM}_{n_0 n_\mathrm{long}}$ does not appreciably change as the parameter is decreased (reach is increased).

 Furthermore, the addition of spokes to the network can affect the rank ordering of the other $i^\mathrm{p}$. For example, while the figure shows that $\widetilde{\mathbf{M}}^\mathrm{COM}_{n_0 n_5}>\widetilde{\mathbf{M}}^\mathrm{COM}_{n_0 n_{10}}$ for the network $\mathcal{S}_{\{1,2,5,10,18,30\}}$, this is not the case for the  network $\mathcal{S}_{\{1,2,5,10\}}$, even though they only differ by the addition of two spokes. In the smaller network $\widetilde{\mathbf{M}}^\mathrm{COM}_{n_0 n_{10}}$ is the largest at high reach (low $\para_T$) (and in general the largest $\widetilde{\mathbf{M}}^\mathrm{COM}_{n_0 i^\mathrm{p}}$ at high reach occurs for the $i^\mathrm{p}$ with the largest value of $d_{n_0 i^\mathrm{p}}$ in the network ).  The ground-current centrality is not susceptible to such reshuffling upon the addition of spokes because  different spokes are electrically independent when $n_0$ is the network's voltage source, as in the calculation of $\widetilde{\mathbf{M}}^\mathrm{GCC}_{n_0 i^\mathrm{p}}$ (see Sec.~\ref{sec:gcc_first}).

The Katz centrality on the $\mathcal{S}_{\{1,2,5,10,18,30\}}$ network is qualitatively similar to the communicability centrality in Fig.~\ref{fig:exp_star}, reproducing the features discussed above. As with the communicability, $\widetilde{\mathbf{M}}^\mathrm{KC}_{n_0 n_{30}}$ begins to overtake the other values of $\widetilde{\mathbf{M}}^\mathrm{KC}_{n_0 i^\mathrm{p}}$ as $\para_T$ is reduced. However, the convergence fails before it can overtake $\widetilde{\mathbf{M}}^\mathrm{KC}_{n_0 n_5}$.

The PageRank centrality reproduces HCC's $\widetilde{\mathbf{M}}_{n_0 i}$ ranking  for all nodes except $i^\mathrm{P}=n_\mathrm{long}$. In fact, $\widetilde{\mathbf{M}}^\mathrm{PRC}_{n_0 n_1}=\widetilde{\mathbf{M}}^\mathrm{PRC}_{n_0 n_\mathrm{long}}$ for all values of $\para_{PRC}$---therefore $\widetilde{\mathbf{M}}^\mathrm{PRC}_{n_0 n_\mathrm{long}}$ is consistently tied for the highest rank. This happens because the random walker beginning on either $n_1$ or $n_\mathrm{long}$ {\it must} traverse the edge to $n_0$, regardless of the weight of that edge.  The result does not seem reasonable, because the connection from $n_0$ to $n_\mathrm{long}$ is meant to carry very little influence.

\begin{figure}[h]
\includegraphics[scale=.75, trim={0 .5cm 0cm 0cm},clip]{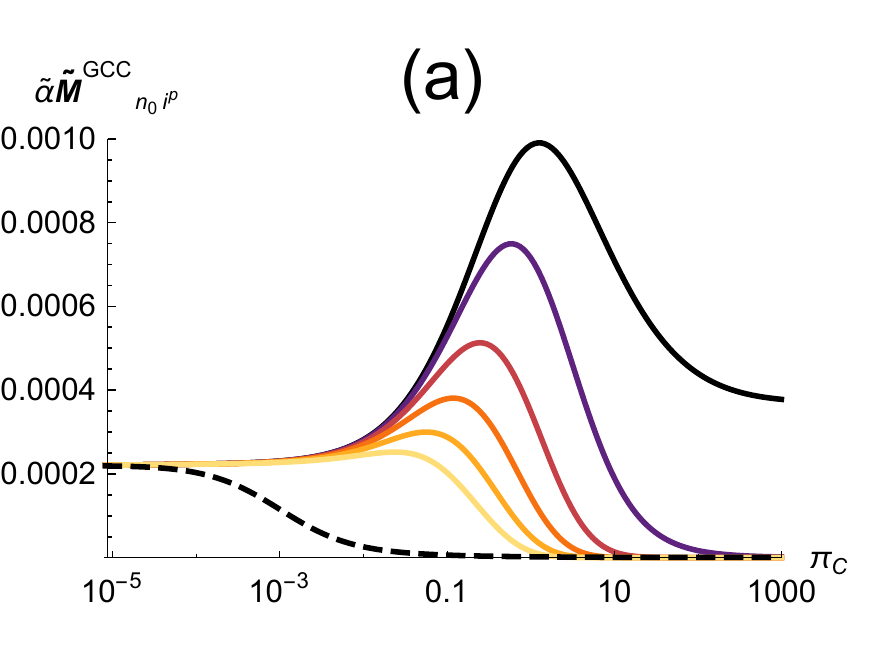}
\includegraphics[scale=.9, trim={0 0cm 0cm 0cm},clip]{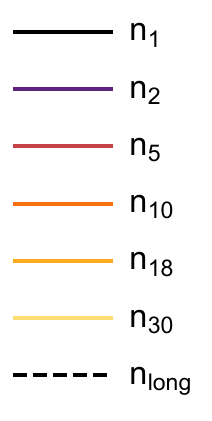}
\includegraphics[scale=.75, trim={0cm .5cm 0cm 0cm},clip]{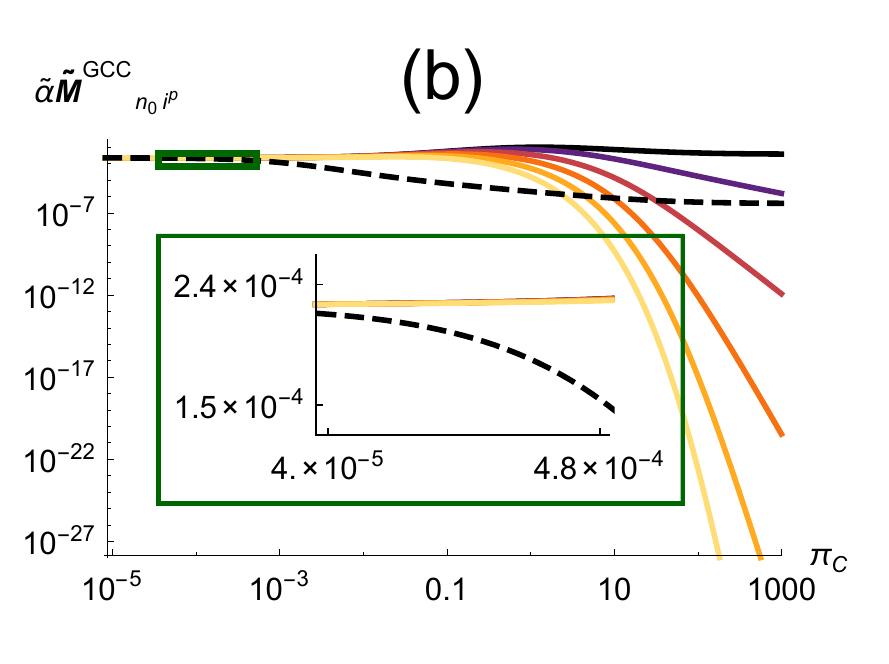}
\caption{\label{fig:gcc_star} { Selected values of $ \widetilde{\alpha} \widetilde{\mathbf{M}}^\mathrm{GCC}$ for the $\mathcal{S}_{\{1,2,5,10,18,30\}}$ network}. The same data are plotted on (a) log-linear and (b) log-log scales.  Note that the normalization factor $\widetilde{\alpha}$ depends on $\para_c$. Without $\widetilde{\alpha}$, $\widetilde{\mathbf{M}}$ values go to zero at small $\para_c$. The inset shows the detailed behavior of the curves at high reach (low $\para_C$), where the values for all peripheral nodes $i^\mathrm{p}$ become indistinguishable well before $\widetilde{\mathbf{M}}_{n_0 n_\mathrm{long}}$ acheives the same value. At all parameter values, the ground-current  centrality  reproduces the HCC's intuitive centrality ordering on the $\widetilde{\mathbf{M}}_{n_0 i}$.
}
\end{figure}

Figure \ref{fig:gcc_star} shows that, for the ground-current centrality, the ordering of the  $\widetilde{\mathbf{M}}^\mathrm{GCC}_{n_0 i^\mathrm{p}}$ matches the HCC ordering at all parameter values for all $i^\mathrm{p}$ except $n_\mathrm{long}$. In addition, the $\widetilde{\mathbf{M}}^\mathrm{GCC}_{n_0 n_\mathrm{long}}$ is  ranked lowest  at high reach (low $\para_C$), matching the weighted HCC. While this matrix element is not ranked lowest at low reach (high $\para_C$),  Fig. \ref{fig:gcc_star}(a) shows that it does not amount to a significant centrality contribution at those parameter values.  The inset of Fig. \ref{fig:gcc_star}(b)  shows that, while all $\widetilde{\mathbf{M}}_{n_0 i^\mathrm{p}}$ values eventually converge as $\para_C\to 0$, those for the peripheral nodes $i^\mathrm{p}$ other than $n_\mathrm{long}$ converge at much higher $\para_C$. This behavior is reasonable, given that $D_{n_0 i^\mathrm{p}}=1$ for all $i^\mathrm{p}$ other than $n_\mathrm{long}$, and that $D_{n_0 n_\mathrm{long}}=1000$.

\subsubsection{Regular Networks}
\label{sec:reg}
We have seen that (the exogenous forms of) several centralities under discussion reduce to degree centrality at low reach (high $\para$). In a sense, then, lower parameter values (higher reach) are perturbations on the degree centrality. Therefore, it becomes reasonable to factor out the contribution of  nearest-neighbor influence to probe each centrality method's unique characteristics. Testing on a $k$-regular network, where every node has degree $k$, accomplishes this goal.

For $k$-regular networks, the communicability, Katz, and PageRank---but not ground-current---centralities are always trivial, with every node's centrality value equal to $1/N$. More generally, this result obtains for any $\mathbf{M}$ that can be written as a power series in 
the adjacency matrix: $\mathbf{M}(\mathbf{A})= a_0 \pmb{\mathbb{I}}  + a_1 \mathbf{A} + a_2 \mathbf{A}^2 + \cdots$.  This is because $\mathbf{A}\ket{1}=k\ket{1}$, and so $\mathbf{M}(\mathbf{A}) \ket{1}$ is proportional to $\ket{1}$ as well. Applying the normalization factor from Eq.~(\ref{eq:centrality}) results in $\ket{c}= \alpha\mathbf{M}(\mathbf{A})\ket{1} = (1/N)\ket{1}$.

Equations (\ref{eq:communicability}) and (\ref{eq:katzexpand}), respectively, show that the communicability and Katz centralities display this degeneracy.  Equation (\ref{eq:prc}) for the PageRank centrality shows the same, noting that, for regular graphs the factor of $\pmb{\mathrm{Diag}}(\ket{k^{-1}})$ becomes a scalar. Indeed, in the case of regular graphs, the PageRank becomes identical to the Katz centrality with $\para^\mathrm{KC}= k \para^\mathrm{PRC}$.

It is still possible to achieve nontrivial results by removing the diagonal of $\mathbf{M}$, {\it i.e.}, using the {\it exogenous} forms of these centralities, given by $\widetilde{\mathbf{M}}$.
(On the other hand, the diagonal forms $\overline{\mathbf{M}}$ tend to produce the inverse 
 centrality 
 ranking, because $\mathbf{M}\ket{1} =\widetilde{\mathbf{M}}\ket{1}+\overline{\mathbf{M}}\ket{1}$. ) Nonetheless, the centrality values are still nearly identical, because the diagonal does not account for a large fraction of the final centrality weight. In general, the ground-current centrality results in nontrivial and more varied centrality values for both $\mathbf{M}$ and $\widetilde{\mathbf{M}}$. 
 
\begin{figure}
\includegraphics[scale=1.0, trim={1cm 0cm 0cm 0cm},clip]{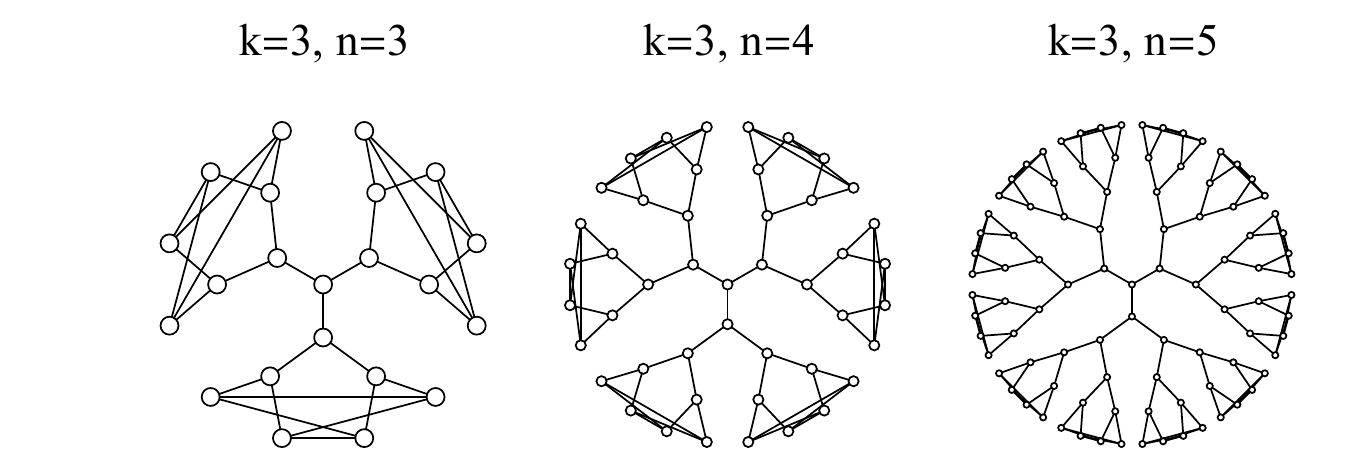}
\caption{\label{fig:cayley_closed_example} 
Closed Cayley trees with degree $k$ and $n$ generations.
 }
\end{figure}

 As a test case, we consider the modified Cayley trees depicted in Fig.~\ref{fig:cayley_closed_example}. The (unmodified) Cayley tree is an acyclic nearly regular network, defined by two parameters: $k$ and $n$.  The first of these is the degree of  every interior (i.e., nonleaf) node, while the second is the number of generations grown out from the central generation-0 node.  For $m\ge1$, the $m$th generation contains $k (k-1)^{m-1}$ nodes. Cayley trees have the special property that it is intuitive which nodes are more central than others: the lower the generation, the higher the centrality, in accordance with the harmonic closeness, HCC. This is because, as can be seen in Fig. \ref{fig:cayley_closed_example}, lower-generation nodes are  closer to the center, while higher-generation nodes are more peripheral. To arrive at the modified Cayley tree, we add edges to every leaf node, resulting in a $k$-regular graph.  
 
The new edges are added in such a way as to keep the leaf nodes on the network's periphery and the lower-generation nodes closer to the center. This
``tree closure" method, described below, can be employed for all odd values of $k$. However, here
we report centrality results only for $k=3$ and $n=7$, since results are qualitatively similar for other values of $k$ and $n$.  To ``close'' a $k=3$ Cayley tree, every leaf node $i$ makes two additional connections. The closest leaf node to $i$, which lies  graph distance
$d=2$ away, is skipped. Then $i$ is connected to the next-closest two
leaf nodes, a graph distance $d=4$ away. This produces a symmetric
network, where every node at a given generation is equivalent. The HCC ordering is unaffected by the addition of these edges.

\begin{figure}
\includegraphics[scale=.8, trim={0 0cm 0cm 0cm},clip]{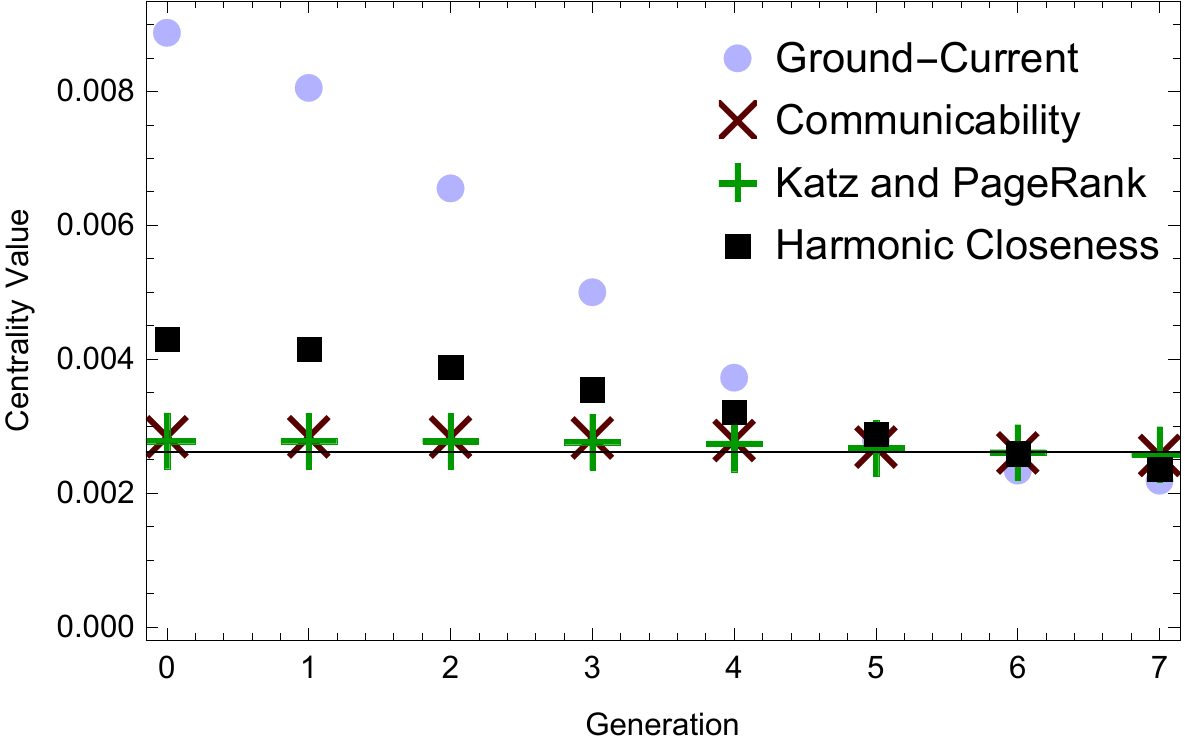}
\caption{\label{fig:cayley_results}  
Exogenous centrality values for the closed Cayley tree with $k=3$ and $n=7$. In this network, all nodes at a given generation are equivalent, so there are only 8 unique data points.  The parameter values for each centrality are chosen to give the largest possible spread in the centrality values of the generations (ground-current: $\para_C=0.010$, communicability: $\para_T=0.202$, PageRank: $\para_\mathrm{PRC}=1.055$). As discussed in the text, the Katz and PageRank centralities are identical on this network. The communicability values are similar but not equal to the Katz values. The horizontal line indicates the value of $1/N$, which coincides with the normalized degree centrality values on this network. 
}

\end{figure}

All the centralities under discussion reproduce the HCC centrality hierarchy: lower generation nodes have higher centralities. However,  the centralities other than the ground-current  centrality  are nearly trivial. In Fig.~\ref{fig:cayley_results}, we plot the centralities for the parameters  that produce the largest range between the centrality values of the 0th and $n$th generation nodes.
 For consistency, we have used the exogenous ($\widetilde{\mathbf{M}}$) forms of every centrality. However, the full ($\mathbf{M}$) ground-current centrality is very similar. The full form of the other centralities leads to the trivial result of centrality values of $1/N$ for every node, illustrated by the horizontal line in the figure. However, for the other centralities, even the exogenous form does not produce much deviation from $1/N$.  
 
The analysis presented here  also leads to similar results when applied to square-lattice segments, made into regular networks by the addition of multiedges along the periphery. Based on these considerations, we propose that the ground-current centrality as a reasonable choice for discriminating central and noncentral network structure in regular graphs. This may also be true for nearly-regular graphs, such as the street networks of cities that have gridlike layouts.

\subsubsection{Networks with bottlenecks}
\label{Networks with bottlenecks}\label{sec:bnecks}

The ground-current  centrality  is the only radial reach  centrality in Table \ref{tab:classification} that is based entirely on a single acyclic, conserved flow. As a result, it is more sensitive to bottlenecks than the other centralities. 

{\it Lattices with a Bottleneck }

We have argued that, of all the reach-parametrized centralities considered here, the ground-current centrality is the only centrality that reproduces intuitive centrality orderings on a range of networks. To this end, we have showed that it matches the centrality rankings specified by the harmonic closeness. In this section, we further show that the ground-current centrality also captures intuitive aspects of the betweenness centrality when applied to networks with bottlenecks. 

\begin{figure}
\includegraphics[scale=1.0, trim={0cm 0cm 0cm 0cm},clip]{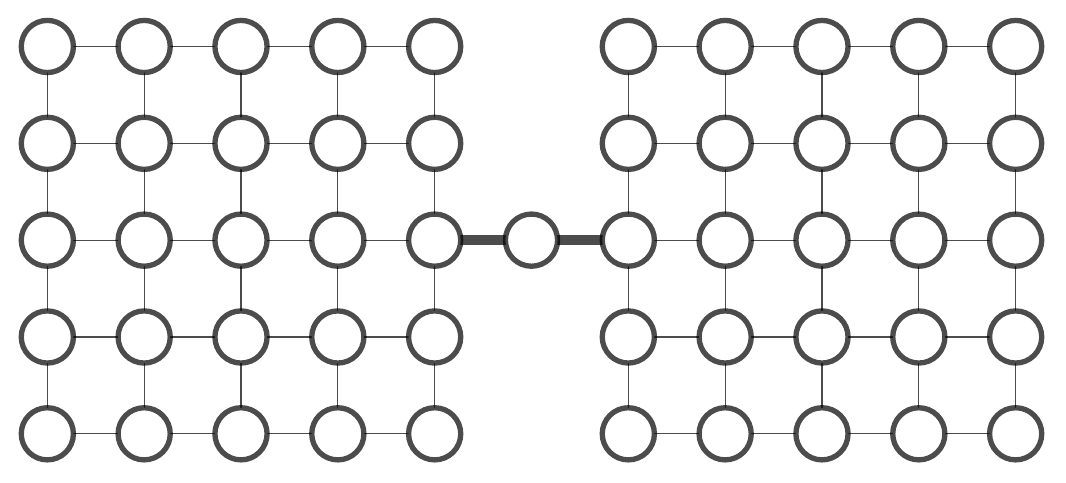}
\caption{\label{fig:bottleneck_network} The weighted bottleneck network with length 5: $\mathcal{B}(L=5)$. All but two of the edges have unit lengths. The two edges forming the bottleneck have lengths of 10. They are depicted as thick lines in the figure. }
\end{figure}

\begin{figure}
\includegraphics[scale=1, trim={0cm 0cm 0cm 0cm},clip]{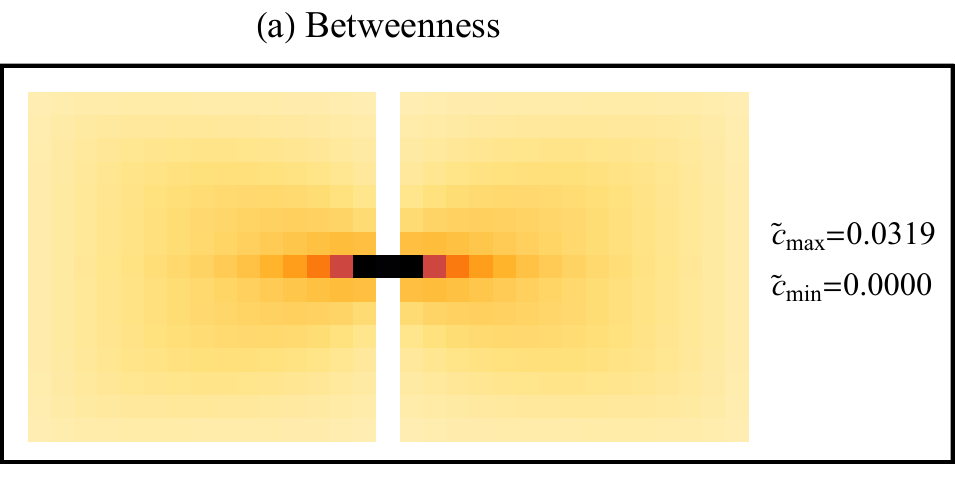}
\includegraphics[scale=1, trim={0cm 0cm 0cm 0cm},clip]{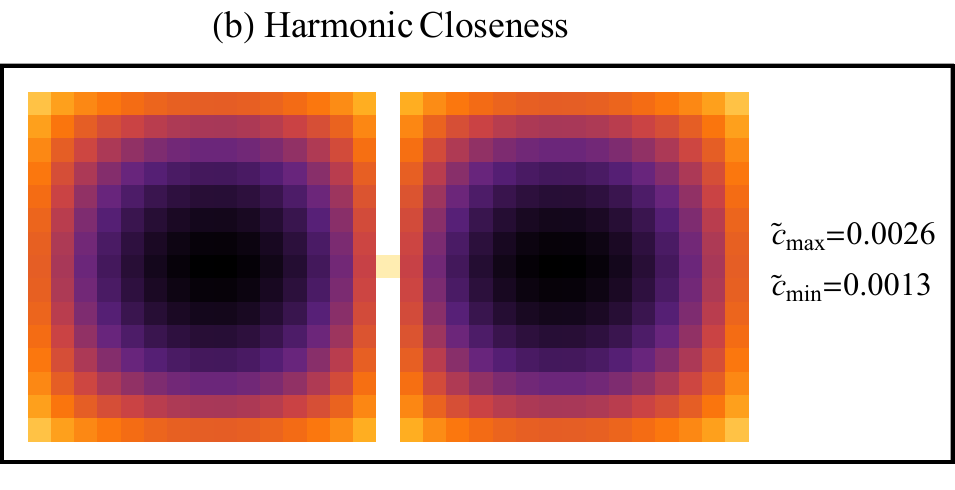}
\includegraphics[scale=1, trim={0cm 0cm .4cm 0cm},clip]{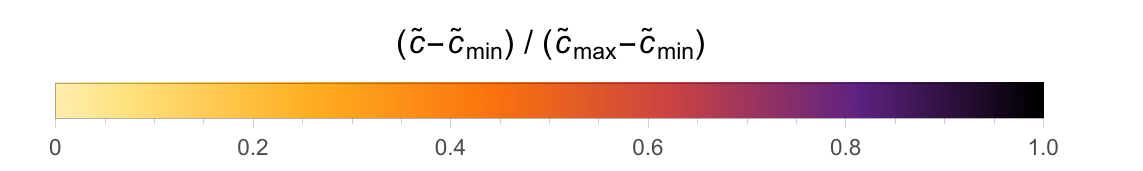}
\caption{\label{fig:bottleneck_network_besclos} Normalized (a) betweenness and (b) closeness results on $\mathcal{B}(L=15)$. Each non-white pixel corresponds to a node of $\mathcal{B}(L=15)$.  For readability, the color scale is chosen such that the maximum centrality value (at given $\para$) is black and the minimum nearly white. The (normalized) centrality values corresponding to these colors are reported for every $\para$. A completely white region in the subfigures indicates a lack of network nodes in that location. }
\end{figure}

To show that the ground-current centrality independently captures aspects of harmonic closeness and betweenness, we construct a network to which those two centralities assign very different centrality rankings. Consider the weighted bottleneck network  $\mathcal{B}(L=5)$ depicted in Fig.~\ref{fig:bottleneck_network}. It consists of two $L\times L$  square sublattices, connected by a single node bottleneck node. All edges have unit length, except for the two edges incident on the bottleneck node, which have length 10. The weighting of these edges helps distinguish the harmonic closeness and the (weighted \cite{brandes2001faster}) betweenness on this network, as shown in Fig.~\ref{fig:bottleneck_network_besclos}.

The addition of the bottleneck node significantly changes the structure of the network by increasing the number of nodes reachable from the peripheral regions of the two sublattices. It is remarkable then, that the Communicability, Katz, and PageRank centralities are largely insensitive to the bottleneck node's inclusion.

\begin{figure}
\includegraphics[scale=1.0, trim={0cm 0cm 0cm 0cm},clip]{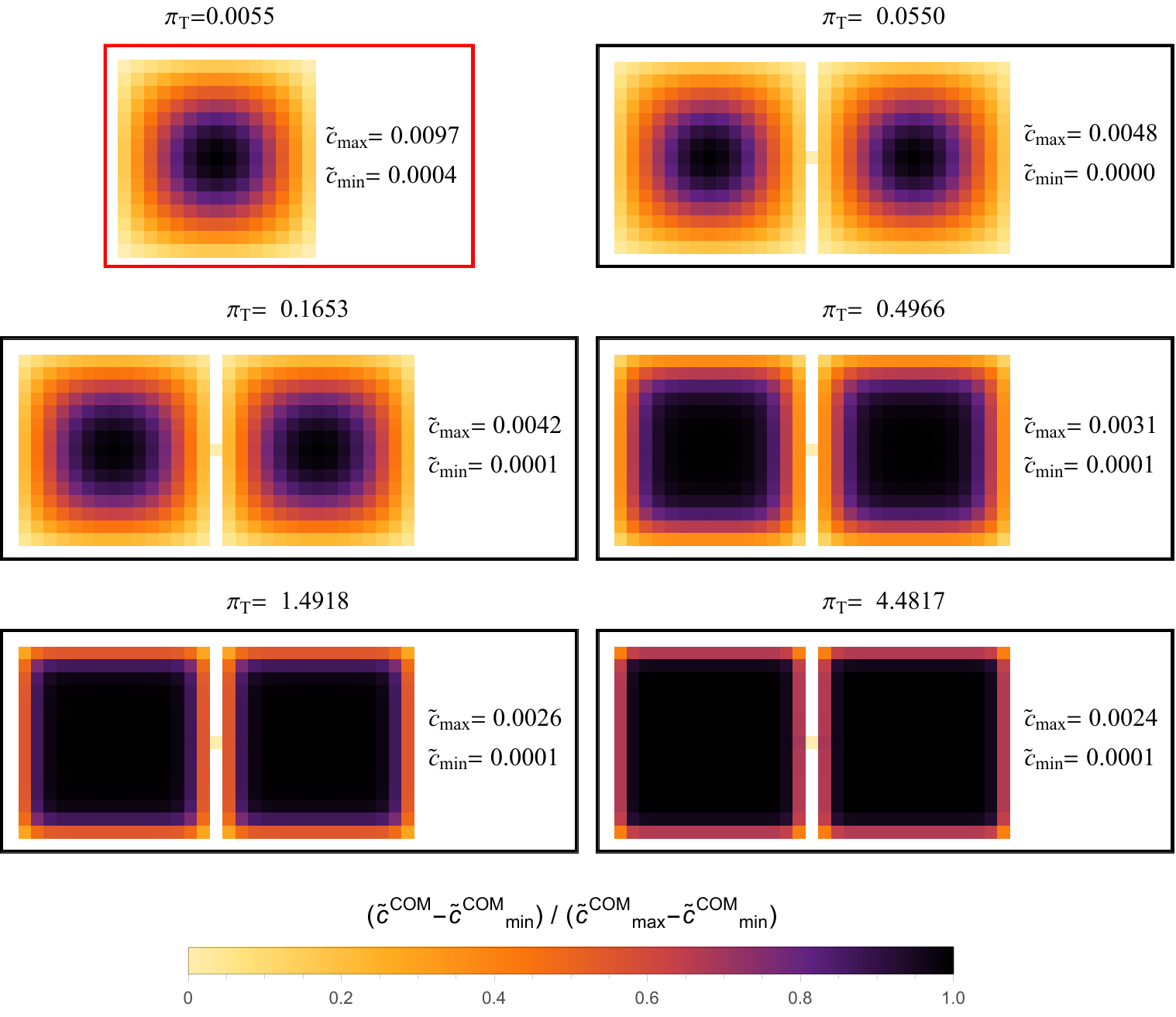}
\caption{\label{fig:exp_contraction} Communicability centrality on $\mathcal{B}(L=15)$.
The full range of parameters is shown, in the sense that increasing/decreasing their values does not alter the image. The parameters are equally spaced on a log scale. For comparison, the red-bordered subfigure illustrates the centrality on the isolated $L=15$ lattice, using a maximum reach $\para_T$ value for {that} network, so that decreasing $\para_T$  does not alter the image.  See the caption to Fig.~\ref{fig:bottleneck_network_besclos} for details.
}
\end{figure}

Consider Fig.~\ref{fig:exp_contraction}, which depicts the exogenous communicability centrality values $\widetilde{c}^\mathrm{COM}$ on $\mathcal{B}(L=15)$ on a range of $\para_T$ values. (The results in this section also hold for other values of $L$.) The full range of parameters is shown, in that increasing/decreasing the parameter values does not alter the image. The bottom-right portion of the figure confirms that the exogenous centrality is proportional to the degree centrality at low reach (high $\para$): all nonperipheral nodes have the identical, highest centrality rank. As the the reach is increased ($\para$ decreased), the region of high centrality rank shrinks towards the middle of each sublattice, largely insensitive to the presence of the bottleneck node. The top-left portion of the figure shows the high reach (low $\para$) centrality values of the isolated $L=15$ lattice---its centrality ranks are almost indistinguishable from the sublattices of $\mathcal{B}(L=15)$. The Katz centrality behaves similarly, and so is not pictured.

\begin{figure}
\includegraphics[scale=1.0, trim={0cm 0cm 0cm 0cm},clip]{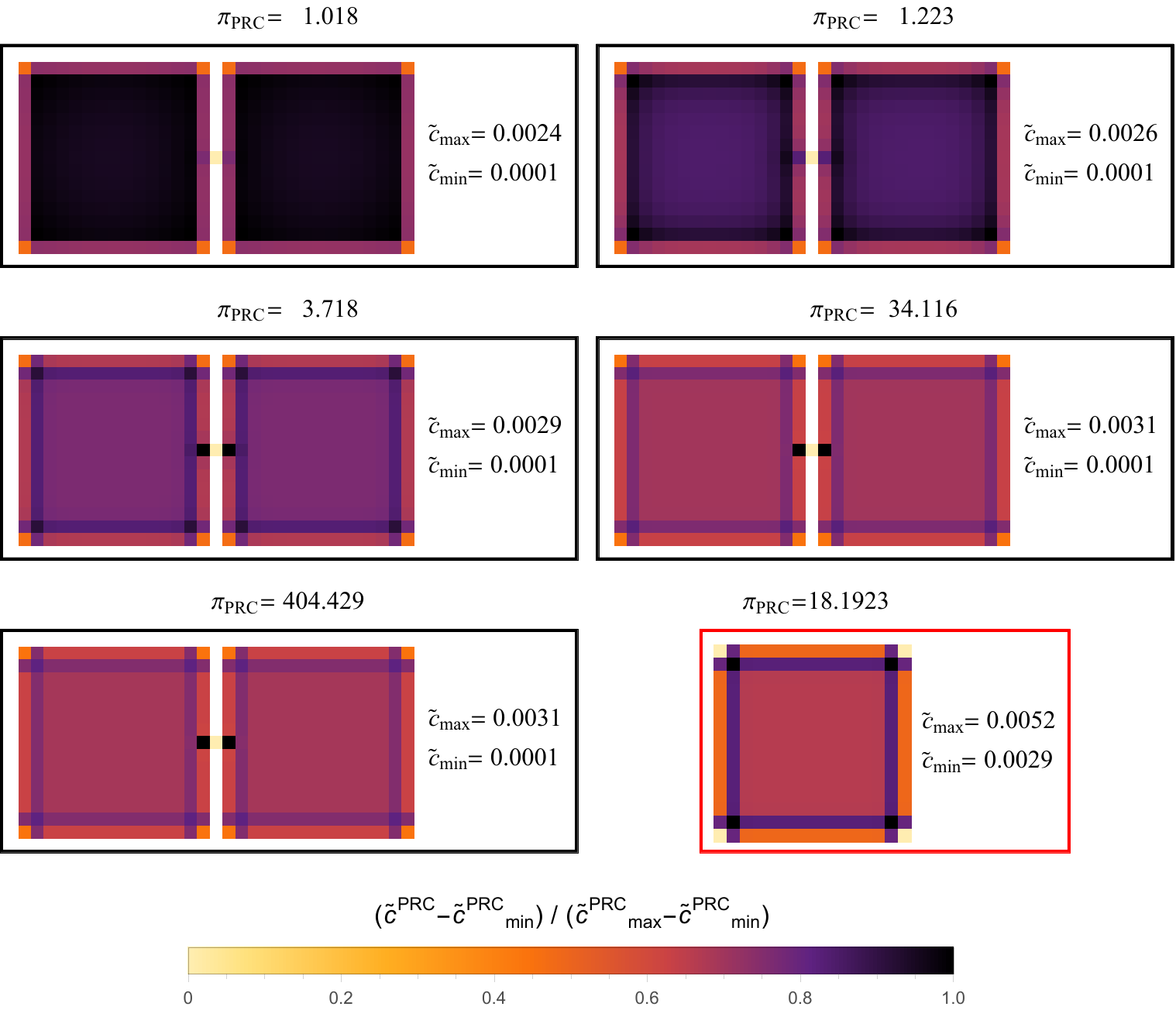}
\caption{\label{fig:pagerank_contraction} PageRank centrality on $\mathcal{B}(L=15)$. See the caption to Fig.~\ref{fig:bottleneck_network_besclos} for details. Here, the red-bordered subfigure illustrates the centrality on the isolated $L=15$ lattice at very low reach. }
\end{figure}

\begin{figure}
\includegraphics[scale=1.0, trim={0cm 0cm 0cm 0cm},clip]{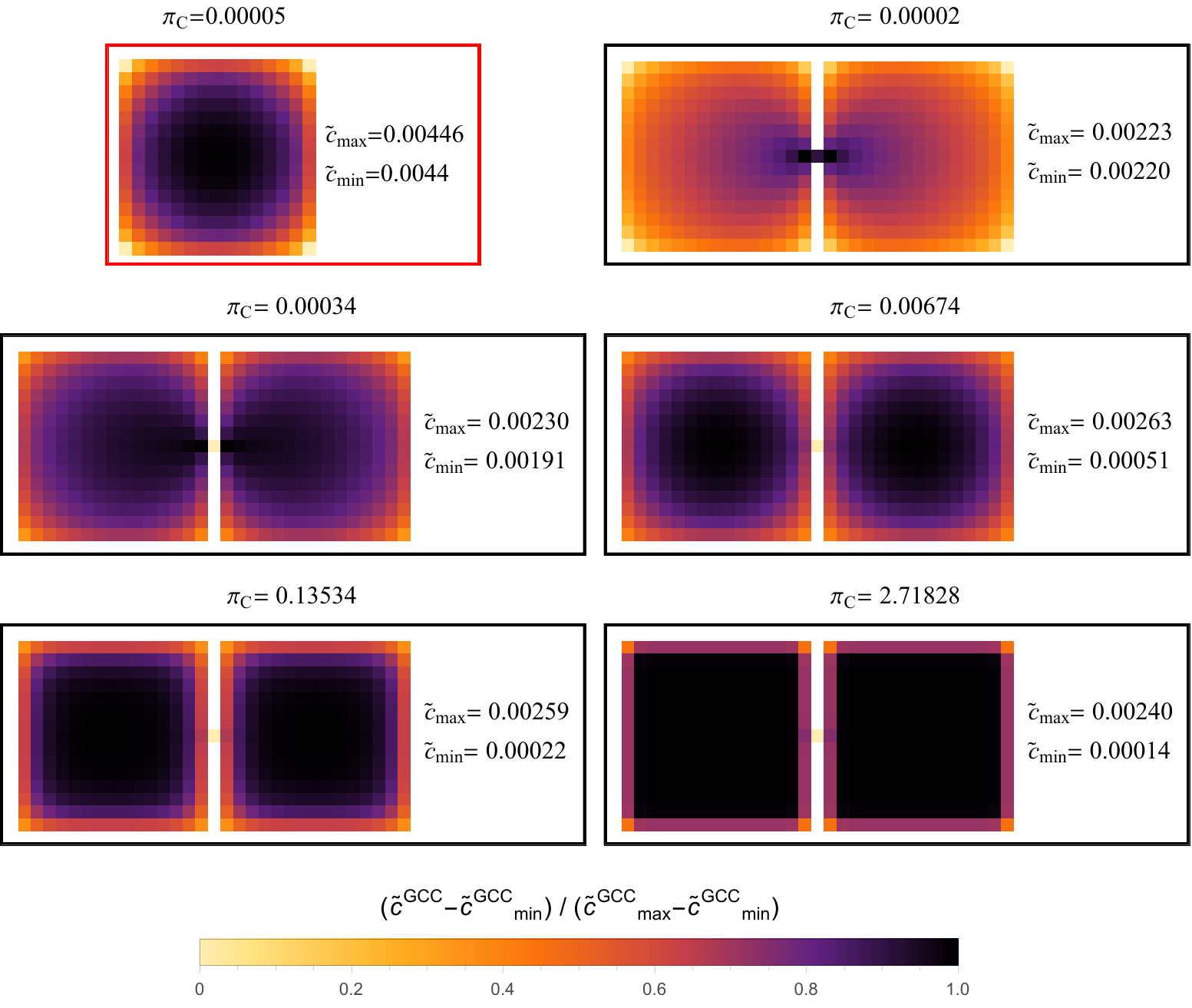}
\caption{\label{fig:gcc_contraction} Ground-current centrality on $\mathcal{B}(L=15)$. See the caption to Fig.~\ref{fig:bottleneck_network_besclos} for details.}
\end{figure}

The PageRank is also insensitive to the bottleneck, as seen in Fig.~\ref{fig:pagerank_contraction}. There, the top-left portion shows that the PageRank reduces to degree centrality at high reach (low $\para$), unlike the communicability, Katz, and ground-current  centralities. As the reach is decreased ($\para$ increased), the centrality ranks remain largely symmetric {\it within} each sublattice, regardless of proximity to the bottleneck node. The bottom-right portion of the figure shows that the resulting pattern is very similar to that produced by PageRank on an isolated $L=15$ lattice.

In contrast, the high-reach ground-current centrality is highly sensitive to the presence of the network's bottleneck, as shown in Fig.~\ref{fig:gcc_contraction}. At intermediate reach ($\para_C=0.00674$), the centrality ranks within the sublattices are very similar to those of the isolated lattice at high reach ($\para_C=0.00002$), shown in the figure's top-left. The rankings are also similar to the harmonic closeness centrality of Fig.~\ref{fig:bottleneck_network_besclos}(b). While increasing the reach (lowering $\para_C$) does not change the centrality pattern in the isolated lattice, it has a large effect on the  weighted bottleneck network. The figure shows that the region of high centrality contracts tightly around the bottleneck as $\para_C \to 0$, creating a pattern much more similar to the betweenness centrality of Fig.~\ref{fig:bottleneck_network_besclos}(a).

{\it Bottlenecks in Real Networks}

\begin{figure}
\includegraphics[scale=1.3, trim={0cm 0cm 0cm 0cm},clip]{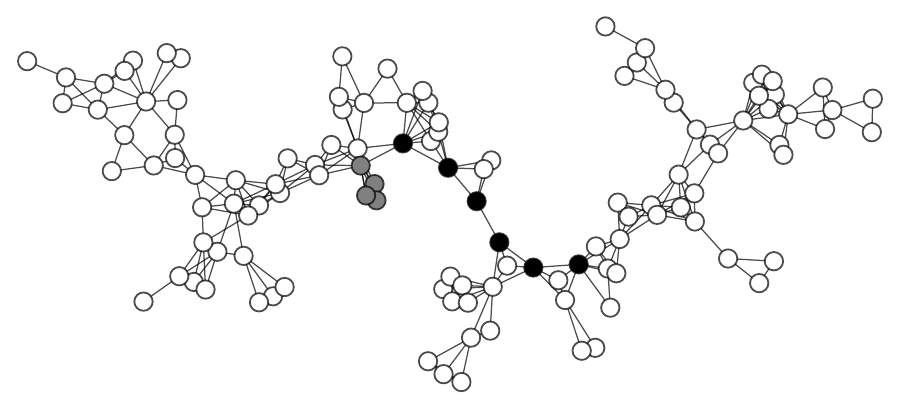}
\caption{\label{fig:voles} Vole trapping network \cite{nr-aaai15,voles}. The black nodes are those in the top 5\% of betweenness rank. The gray nodes are, at high reach, those in the top  5\% exogenous communicability centrality  (equivalently, eigenvector centrality) rank. They are the two nodes with the highest weighted degree and some of their high-degree neighbors.}
\end{figure}

The ground-current centrality's sensitivity to bottlenecks at high reach is also present in real networks. Here, we use high-betweenness nodes as a proxy for bottleneck structures.
We compare the betweenness and the communicability, PageRank, and ground-current centralities as applied to seven example networks, including the previously discussed kangaroo network, Florida power grid network, and weighted bottleneck network $\mathcal{B}(L=15)$. We also analyze the Italian power grid previously studied in \cite{hama10}. The unweighted {\it C. elegans} network \cite{Choe-2004-connectivity} consists of 277 nodes corresponding to the majority of the nematode worm's neurons. The nematode is well studied in network theory \cite{watts1998collective, newman2002assortative} and neuroscience \cite{yan2017network} because it has one of the simplest neural structures of any organism. Here we analyze only the undirected version of this network. Finally, we analyze the largest connected component of the vole trapping network from \cite{nr-aaai15, voles}, depicted in Fig.~\ref{fig:voles}. The  network's 118 nodes represent voles, while its 283 edges link voles that were caught in the same trap during a particular trapping session, where the integer edge weights correspond to the number of times they were trapped together. This network is different from the other real networks under consideration because it has high betweenness nodes that do not also have high degree.

\begin{figure}
\hspace{0cm}
\includegraphics[scale=.838, trim={0cm 0cm 0cm 0cm},clip]{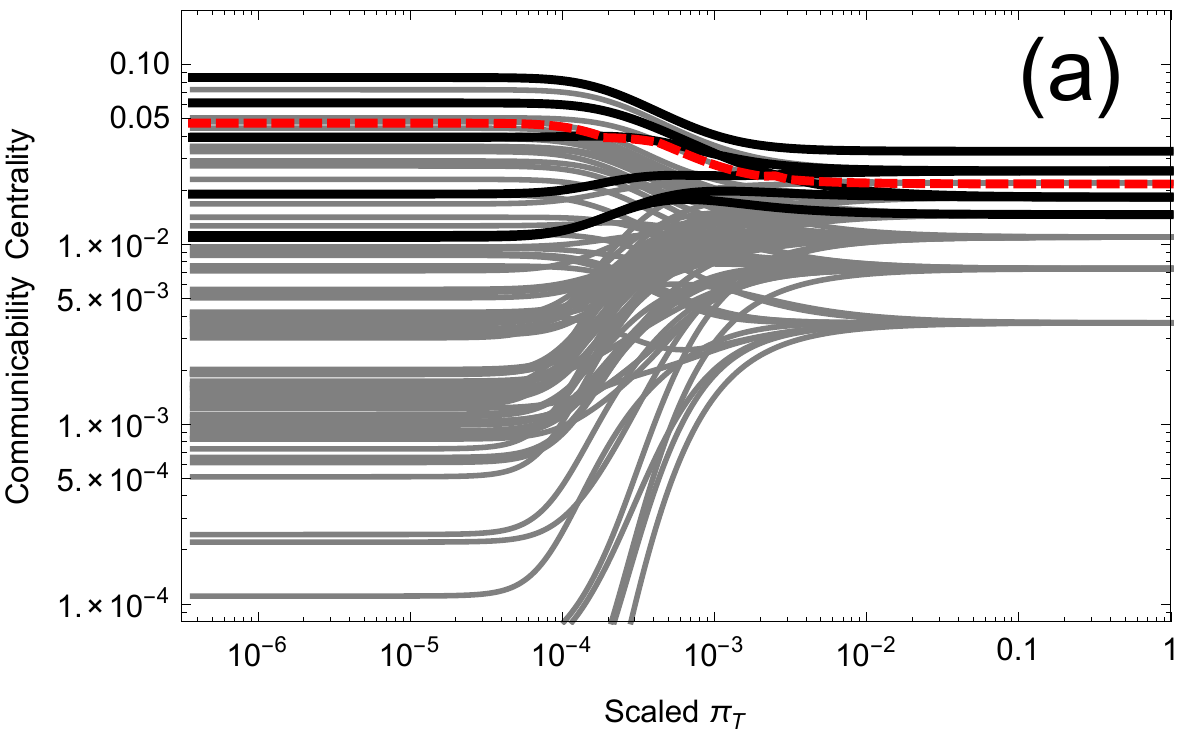}\hspace{2cm}
\includegraphics[scale=.855, trim={0cm 0cm 0cm 0cm},clip]{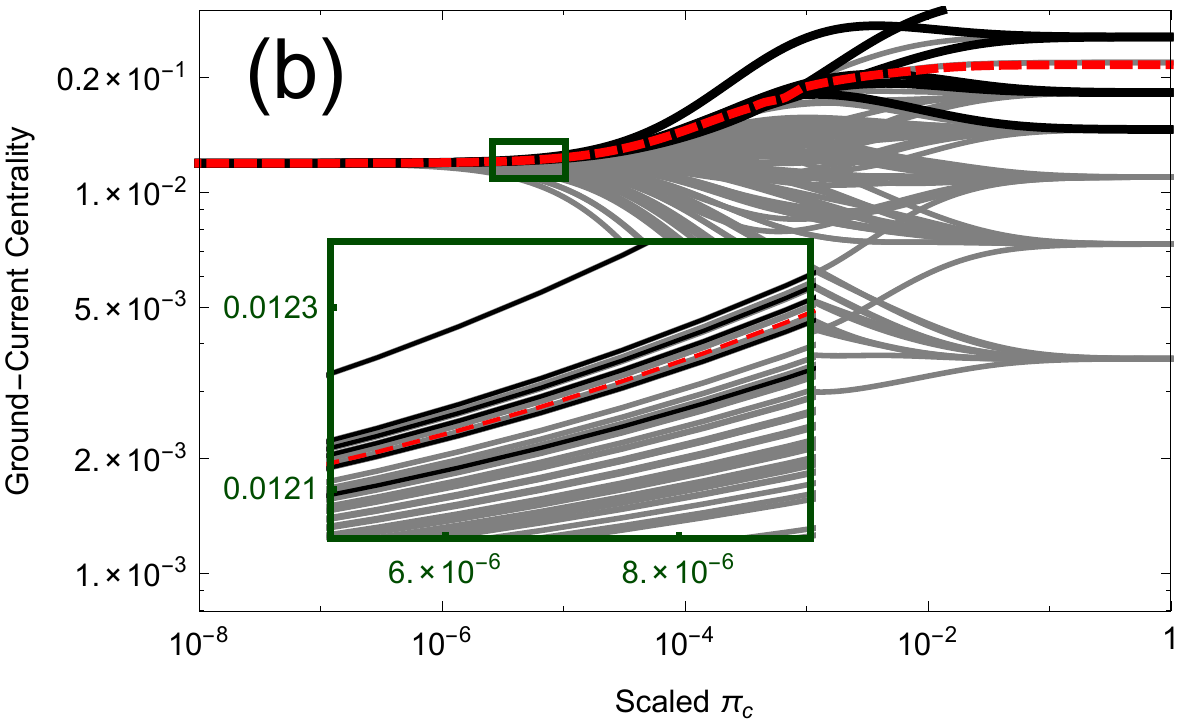}
\caption[(a) Exogenous communicability centrality on the unweighted Florida power-grid network. (b) Exogenous ground-current centrality on the unweighted Florida power-grid network]{\label{fig:unweightedFlorida_comm}{(a) Exogenous communicability centrality on the unweighted Florida power-grid network. (b) Exogenous ground-current centrality on the unweighted Florida power-grid network}. The black curves correspond to nodes in the top 5\% of betweenness rank. All the curves above the dashed red line correspond to nodes in the top 5\% of (a) exogenous communicability  and (b) exogenous ground-current centrality  rank.  See the text for details.}

\end{figure}

\begin{figure}
\hspace{.7cm}
\includegraphics[scale=.8, trim={0cm 0cm 0cm 0cm},clip]{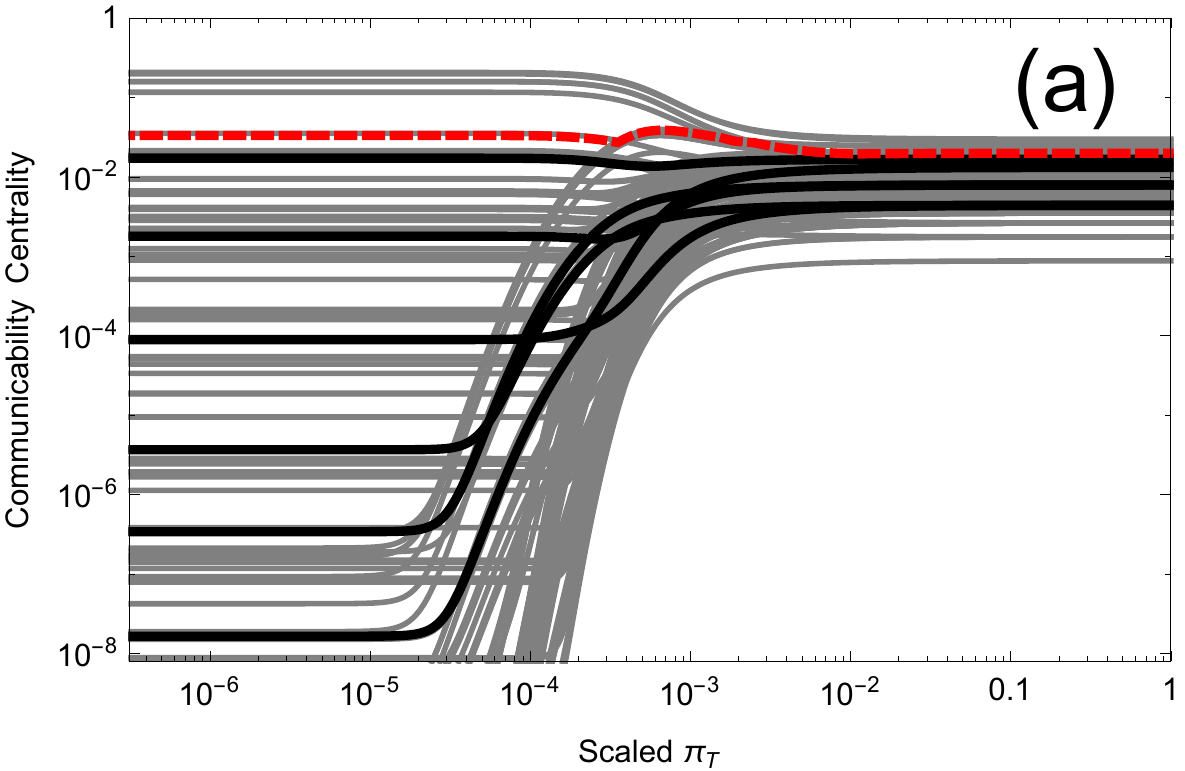}\hspace{1cm}
\includegraphics[scale=.855, trim={0cm 0cm 0cm 0cm},clip]{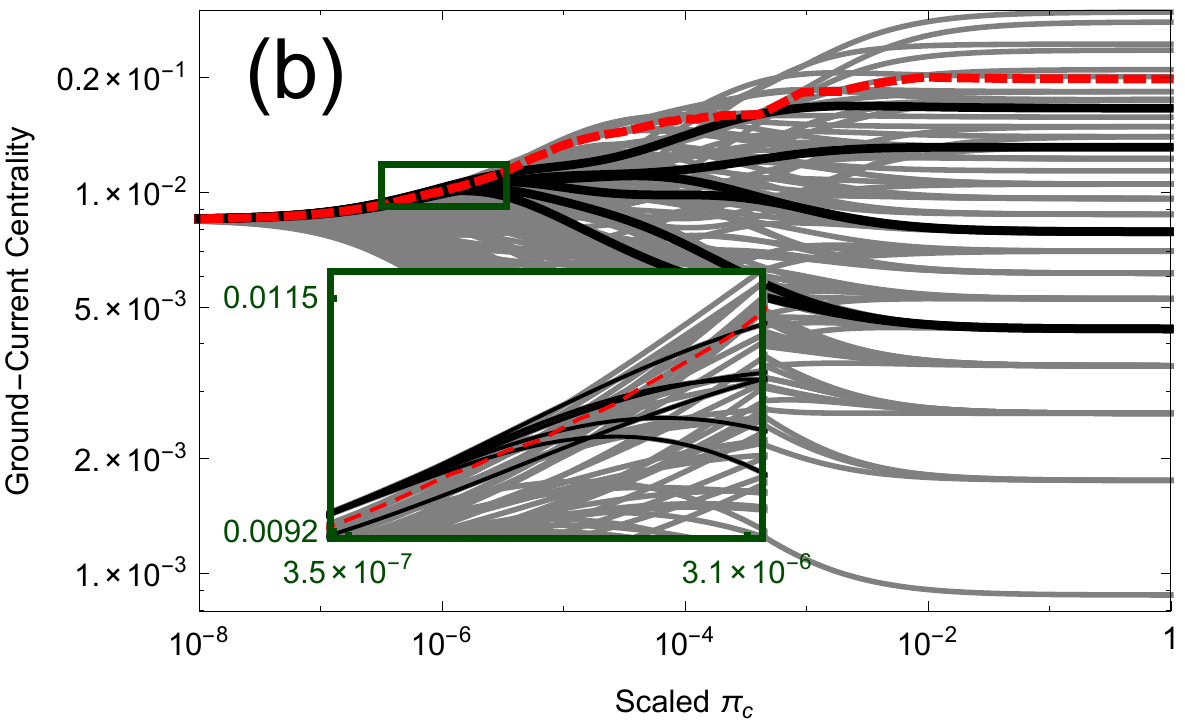}
\caption[(a) Exogenous communicability centrality on the vole network. (b) Exogenous ground-current centrality on the vole network]{\label{fig:vole_comm} { (a) Exogenous communicability centrality on the vole network \cite{nr-aaai15,voles}. (b) Exogenous ground-current centrality on the vole network}.  For details, see the text and the caption to Fig.~\ref{fig:unweightedFlorida_comm}.}

\end{figure}

We quantify the preference of a centrality X for bottlenecks by the ratio $f_X$: the number of nodes that are highly ranked in both $X$ and betweenness  divided by the number of nodes that are highly ranked in betweenness. Here, ``high'' means ranked in the top 5\%. This measurement is illustrated for the exogenous communicability  and  the exogenous ground-current centralities in Figs.~\ref{fig:unweightedFlorida_comm} and \ref{fig:vole_comm}. The solid curves in the figures indicate the centrality values of the nodes in the corresponding networks (respectively, the unweighted version of the Florida power grid depicted in Fig.~\ref{fig:reach_intermediate}, and the trapping network of voles depicted in Fig.~\ref{fig:voles}). The thick black curves correspond to nodes that lie in the top 5\% of betweenness rank. The dotted red curve indicates the cutoff for high centrality: all the values above this curve lie in the top 5\% of communicability centrality in  part (a) or ground-current centrality in part (b). The centrality's sensitivity to bottlenecks is  measured as the fraction $f$ of thick black curves that lie above the dotted red curve. (In these, and the following, figures, we use a scaled form of $\para$ that is constrained to lie between zero and one. See Appendix \ref{app:scaledparams} for  details.)

Figures \ref{fig:unweightedFlorida_comm} and \ref{fig:vole_comm} also illustrate the unique properties of the vole network. The low reach (high parameter) region in these plots display the networks' degree centralities. Unlike the weighted Florida power grid network, the vole network has no high-betweenness nodes in the upper ranks of degree centrality. This absence of high-betweenness nodes persists across most of the parameter range, for both the communicability and the ground-current centrality. At very high reach, the ground-current centrality assigns close to equal importance to every node. However, the high-betweenness  nodes rise in relative rank. As a result, the fraction $f_\mathrm{GCC}$ quickly rises from zero at low $\para_C$.

\begin{figure}[h]
\hspace{2cm}\includegraphics[scale=0.8, trim={0cm 0cm 0cm 0cm},clip]{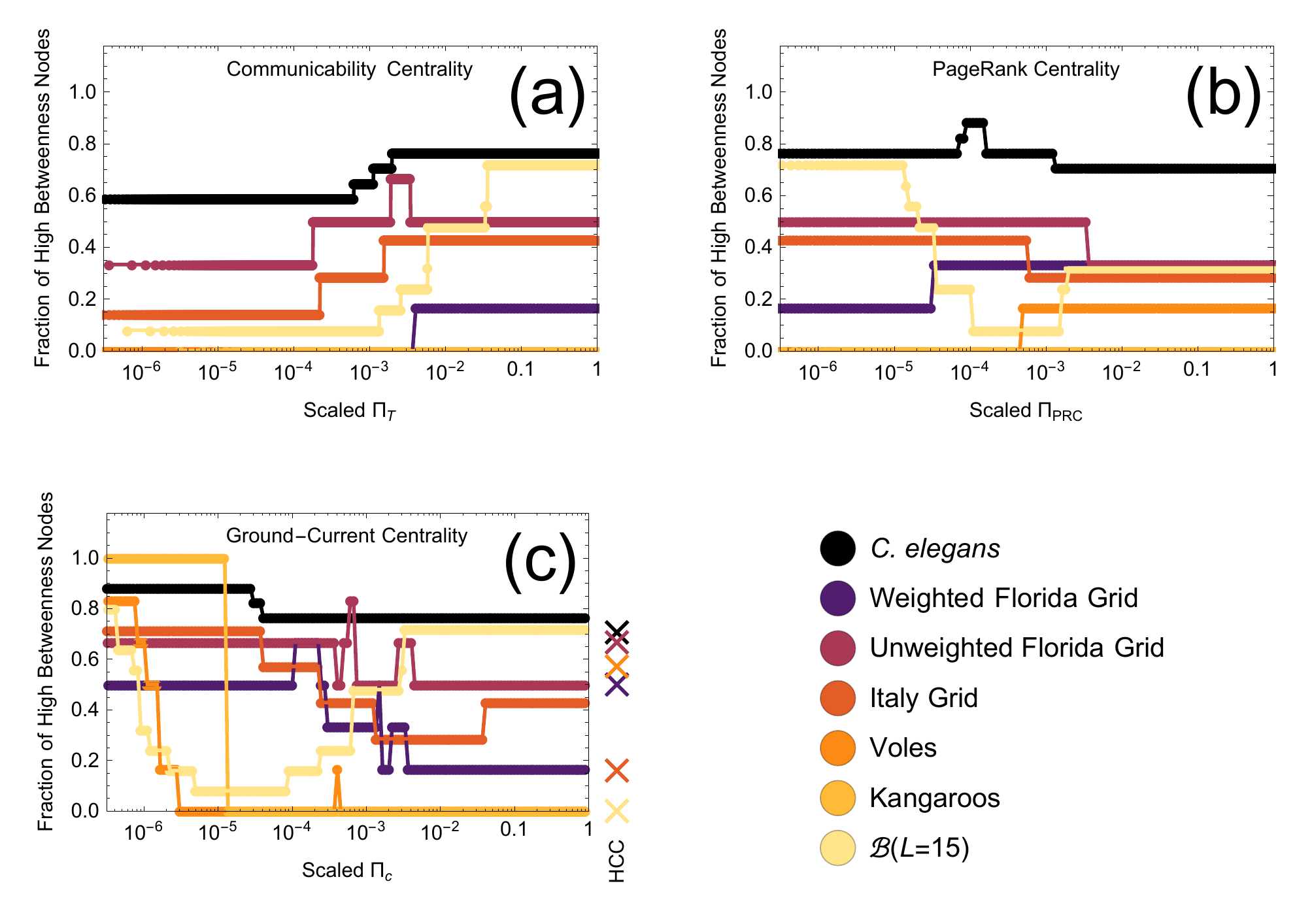}
\caption{\label{fig:comm_all_nets} Fractions of high betweenness nodes among nodes with high (a) exogenous communicability, (b) exogenous PageRank Centrality, and (c) exogenous ground-current centrality.  The fraction in (a) is equal to zero at all parameter values for both the vole network and the kangaroo network. The fraction in (b) is equal to zero at all parameter values for  the kangaroo network. The crosses in (c) depict the fractions  of high betweenness nodes among nodes with high harmonic closeness centrality (HCC). These are always less than or equal to the highest fractions obtained by the ground-current centrality. }
\end{figure}

The values of $f$ are reported in Fig. \ref{fig:comm_all_nets}. 
In part (a), the communicability centrality is not sensitive to bottlenecks:  for all but one of the networks under consideration (the unweighted Florida grid), $f_\mathrm{COM}$ is maximized at large $\para_T$, where $c^\mathrm{COM}$ is equivalent to the degree centrality. Note that $f_\mathrm{COM}$ is zero for the vole network at all parameter values. This is because the high betweenness nodes do not have the highest degrees and are not in the most highly connected regions of the network. Part (b) shows that the PageRank centrality is also not sensitive to bottlenecks. In 3 out of 7 example networks, $f_\mathrm{PRC}$ is maximized at  high reach (low $\para_\mathrm{PRC}$), which is equivalent to the degree centrality. In the other 4 cases, the amount of variation in $f_\mathrm{PRC}$ is small.  

This is in sharp contrast to $f_\mathrm{GCC}$, illustrated in Fig.~\ref{fig:comm_all_nets}(c). In every example network, the highest value of $f_\mathrm{GCC}$ is achieved at the lowest $\para_c$ (highest reach), and these maxima are significantly larger than the values at large $\para_C$, which is equivalent to degree centrality. Notably, $f_\mathrm{GCC}$ is very high for the vole and kangaroo networks, which had $f_\mathrm{COM}=0$ for all $\para_T$. The low-$\para_c$ values of $f_\mathrm{GCC}$ are also greater than or equal to the values $f_\mathrm{HCC}$ for the harmonic closeness centrality, as indicated by crosses in the figure. 

In summary, the ground-current centrality at high reach captures features of the betweenness centrality, assigning high ranks to bottleneck nodes. 

\subsubsection{Localization}
\label{sec:local}
Centrality localization \cite{martin2014localization,pradhan2020principal}  describes the situation when a small number of nodes account for a large fraction of the total centrality. (This can be viewed as a generalization of Freeman's  centralization metric \cite{freeman1978centrality}.) As shown in Fig.~\ref{fig:cayley_results}, the Communicability, Katz, and PageRank centralities exhibit virtually no localization on closed Cayley trees, since the centrality values of all nodes are nearly equal. In \cite{martin2014localization}, the amount of localization of a {\it square-normalized } centrality $c$ is measured with the {\it inverse participation ratio } (IPR):
\begin{equation}
\mathrm{IPR}(c) = \sum_i c_i^4.
\end{equation}
The minimum IPR value for a network of size $N$ is $1/N$, and occurs in the trivial case where all centrality values are identical. The largest value of $ \mathrm{IPR}(\widetilde{c}^\mathrm{COM}) N$ for the closed Cayley tree ($k=3,n=7$), across all possible parameters, is approximately $1.004$.  The fact that this is close to 1 confirms that localization is absent to the extent that the centrality is nearly trivial. The ground-current centrality is still highly unlocalized, but farther from the trivial limit:  $\mathrm{IPR}(\widetilde{c}^\mathrm{GCC})N  \approx 2.243$.

While the communicability centrality exhibits  little localization (is nearly trivial) in the case of regular networks, in many cases it exhibits so much localization that most nodes have centralities that are nearly zero.  In \cite{martin2014localization}, it is shown that networks with prominent hub nodes ({\it i.e.}, nodes directly connected to a large number of other nodes) lead to highly localized eigenvector centrality, which is the high-reach limit of communicability centrality. Among the networks studied by the authors is the electrical circuit network 838 from the ISCAS 89 benchmark set \cite{milo2004superfamilies}. The maximum IPR value for any network is 1, and occurs when all nodes but one have zero centrality. The eigenvector centrality for the circuit network  has relatively high  localization: $ \mathrm{IPR}\approx .179$, corresponding to very little centrality assigned  to nodes other than the hub node and its neighbors. Thus we see that in cases of both high and low localization, the centrality is not informative about most of the nodes in the network.  

Hub networks are not the only network architecture that leads to strongly-localized eigenvector centralities. For example, the vole network eigenvector centrality leads to $\mathrm{IPR}\approx .218$. Here, the localization is due to nodes with high {\it weighted} degree that do not have high {\it unweighted } degree, and so are not hubs in the usual sense. See Fig.~\ref{fig:voles} for an illustration. Here, the top 5\% of nodes in eigenvector centrality rank account for about 87\% of the total centrality. 

Another metric of localization is the Gini coefficient, frequently used by economists to quantify wealth or income inequality \cite{gastwirth1972estimation}. The simplest definition is the following weighted average of centrality differences:
\begin{equation}
\label{eq:gini}
\mbox{Gini coefficient }(c) =\frac{\sum_i^N\sum_j^N \left| c_i - c_j \right|}{2 (N-1) \sum_i^N c_i}.
\end{equation}
An advantage of the Gini coefficient over the IPR is that the latter is constrained between 0 (trivially unlocalized) and 1 (maximally localized) for all networks. We report similar results with both metrics, though the Gini may be easier to interpret. For example, the Gini coefficient for the eigenvector centralities of the circuit and vole networks are approximately .780 and .939, respectively, which indicates significant localization.

\begin{figure}[h]
\includegraphics[scale=.52, trim={0cm 0cm 0cm 0cm},clip]{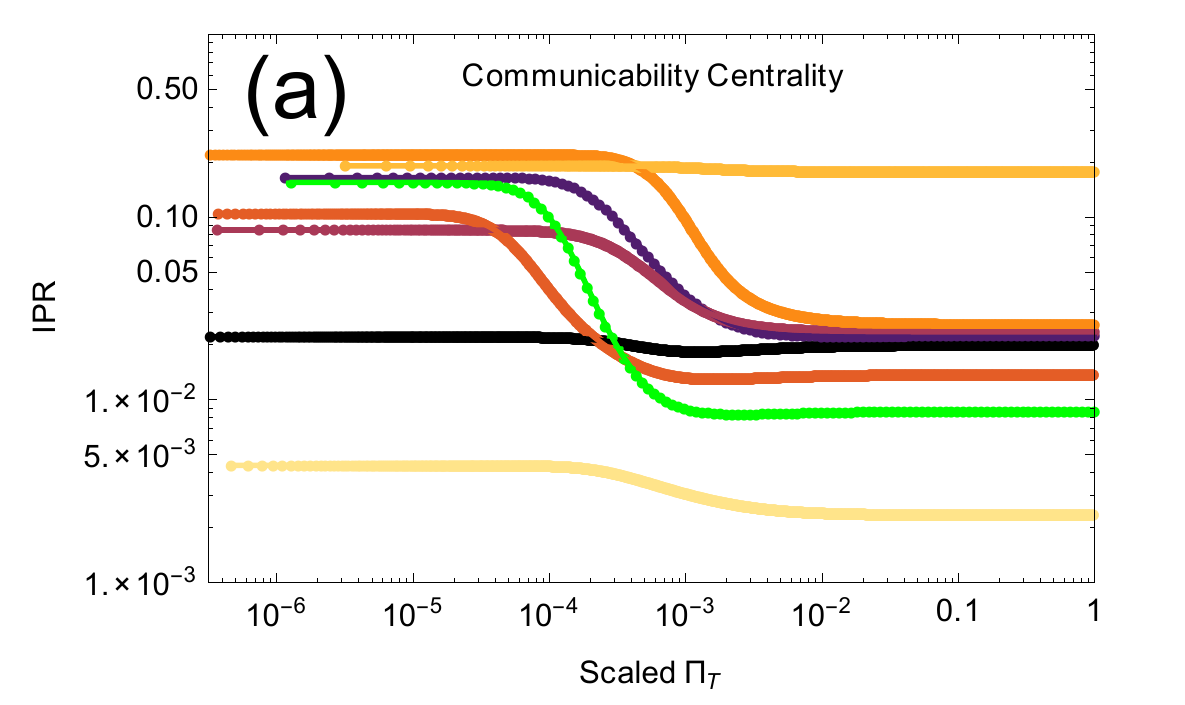}
\includegraphics[scale=.73, trim={0cm 0cm 0cm 0cm},clip]{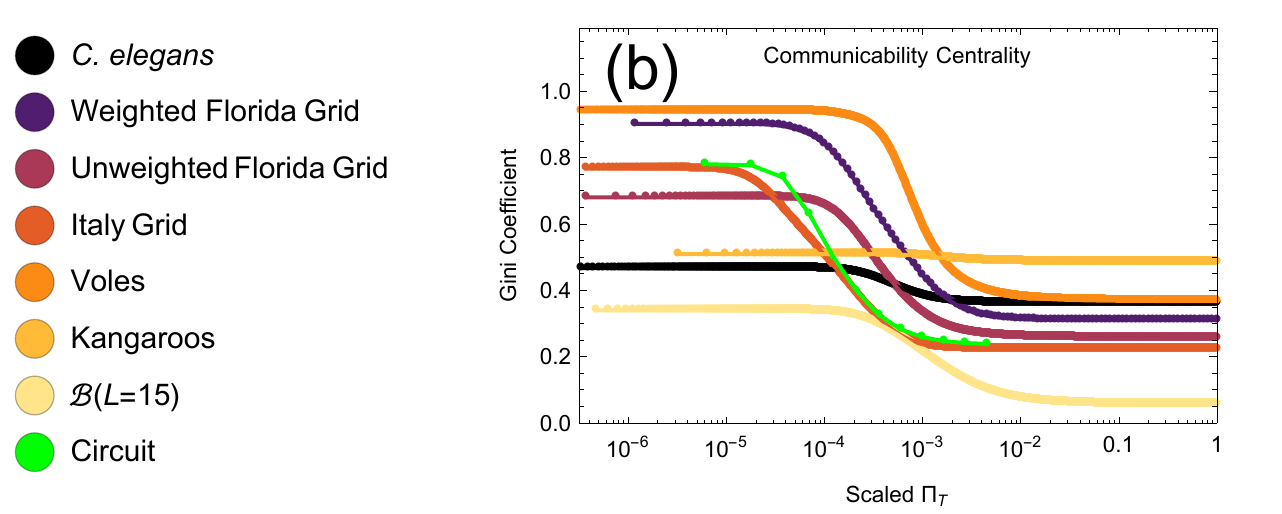}
\caption{\label{fig:IPR_comm_all_nets} (a) The $\mathrm{IPR}$ of $\widetilde{c}^\mathrm{COM}$ and (b) the  Gini coefficient of $\widetilde{c}^\mathrm{COM}$ for our example networks. The IPR is plotted on a log scale. The network labeled ``circuit'' is the electrical circuit network  838 from the ISCAS 89 benchmark set. The  low reach (scaled $\para \approx 1$) results are equivalent to those of the degree centrality. The high-reach results (scaled $\para \approx 0$) results are equivalent to those of the eigenvector centrality.}
\end{figure}

So far we have only considered the eigenvector centrality, which is the high-reach limit of the communicability centrality. The IPR and Gini coefficient values for all parameter values of the exogenous communicability centrality, as applied to all the considered example networks, are reported in Fig.~\ref{fig:IPR_comm_all_nets}(a) and (b),  respectively. The localization almost always increases with increasing reach, and in several cases it reaches values indicating a significant degree of localization. At high reach, the vole network scores higher than the circuit network on both localization measures. The Italian power grid network scores higher than the circuit network on the Gini coefficient. This result is reasonable: the top 5\% of nodes in eigenvector centrality rank account for approximately 44\% of all centrality, indicating the presence of localization.  In general, as can be seen in Fig.~\ref{fig:IPR_comm_all_nets}(b) the communicability centrality cannot produce unlocalized results, except in the case of regular networks as discussed in Sec.~\ref{sec:reg}, or in the case of nearly-regular networks such as $\mathcal{B}(L=15)$.

\begin{figure}[h]
\includegraphics[scale=.546, trim={0cm 0cm 0cm 0cm},clip]{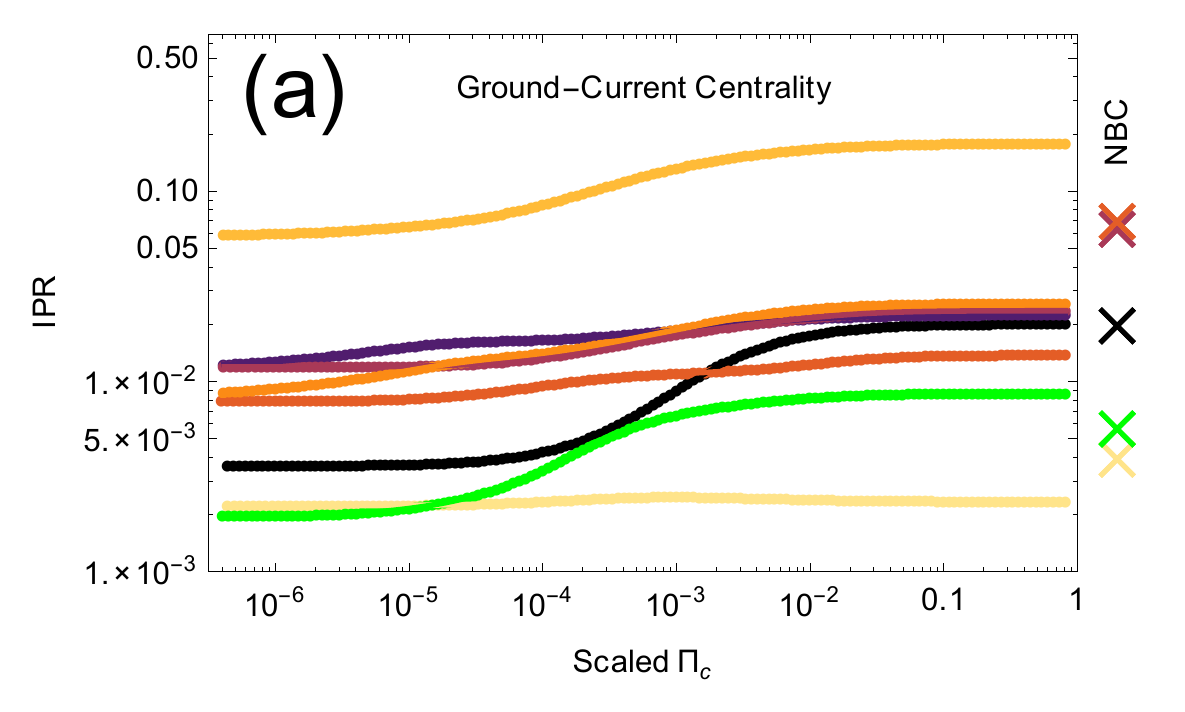}
\includegraphics[scale=.704, trim={0cm 0cm 0cm 0cm},clip]{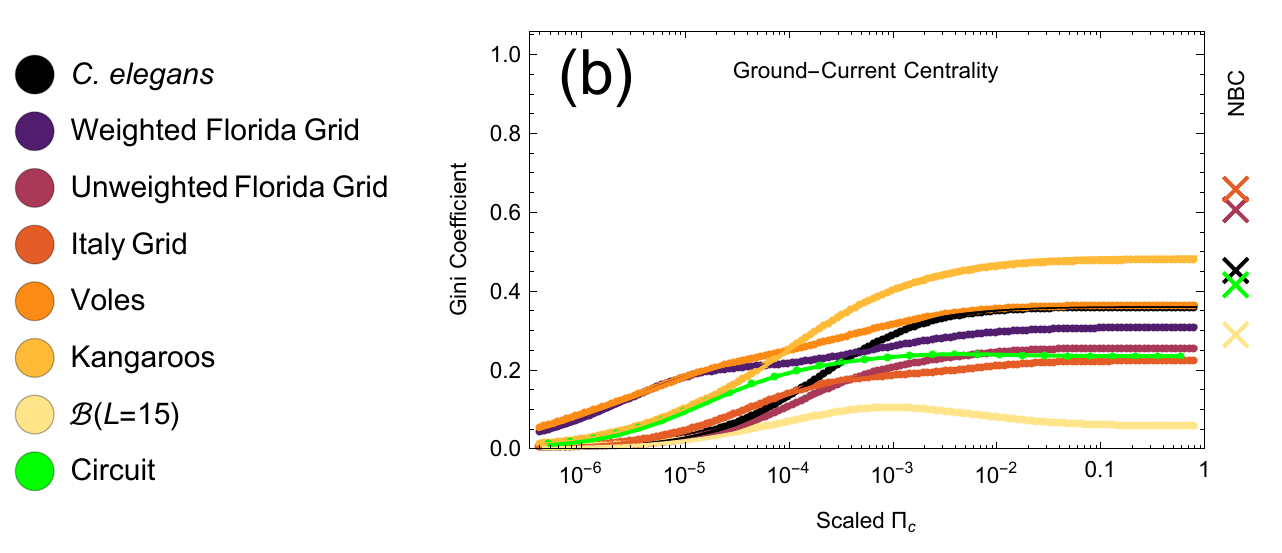}
\caption{\label{fig:IPR_gcc_all_nets} (a) The $\mathrm{IPR}$ of $\widetilde{c}^\mathrm{GCC}$ and (b)  the Gini coefficient of $\widetilde{c}^\mathrm{GCC}$ for our example networks. See the caption to Fig.~\ref{fig:IPR_comm_all_nets}. Crosses represent the values of the nonbacktracking centrality  (NBC) \cite{martin2014localization}, based on the Hashimoto  matrix \cite{hashimoto1989zeta}.  }
\end{figure}

The pattern is reversed with the ground-current centrality, which tends to produce unlocalized centrality values.
The IPR and Gini coefficient for the exogenous ground-current centrality are shown in Fig.~\ref{fig:IPR_gcc_all_nets}(a) and (b), respectively. Almost always, the localization values  decrease with increasing reach. At very high reach they invariably reach the minimum values ($N^{-1}$ for IPR, 0 for Gini), since the ground-current  centrality  always produces uniform centrality in the limit of high reach.  However, this occurs only at very high reach, meaning that the centrality is unlocalized, but not trivial. In general, the Gini coefficients are between .15 and .50 for much of the parameter range. For comparison, the range of Gini coefficients for income across  all nations is .24 to .63, according to the World Bank \cite{worldbank}. The crosses in the figure represent the IPR and Gini values for the nonbacktracking centrality (defined only for unweighted networks), which is presented in \cite{martin2014localization} as a nonlocalizing alternative to the eigenvector centrality. 

\section{Conclusion: acyclic, conserved-flow centralities} \label{sec:conc}

Network  centrality measures can be described as more or less appropriate only relative to the specific demands of a given application.  Here we have shown that the ground-current centrality is particularly well suited for purposes requiring low localization. However, there may be situations in which it would be desirable to pick out only some important nodes from a network. In this case, high localization would be desirable, and both the ground-current centrality and the nonbacktracking centrality from \cite{martin2014localization} would be inappropriate choices. 

A key aim of centrality research is to identify the properties that may render a centrality more or less useful in different situations. To aid in this task, we have expanded Borgatti's centrality typology \cite{borgatti2005centrality,borgatti2006graph} to  categorize the properties of {\it parametrized} centralities  (see Table \ref{tab:classification}). The expanded typology  includes our newly introduced reach-parametrized and grasp-parametrized categories. [From this perspective, the communicability centrality is a reach-parametrized centrality that increases localization with increasing reach (Fig.~\ref{fig:IPR_comm_all_nets}), while the ground-current centrality is a reach-parametrized centrality that decreases localization with increasing reach (Fig.~\ref{fig:IPR_gcc_all_nets}).] Along with the reach/grasp distinction, we categorize parametrized centralities as  to their Walk Position (radial vs. medial), as well as whether  they are based on acyclic and conserved flows.

The utility of the ground-current centrality stems from its unique position in this classification system. The ground-current centrality is the only radial reach-parametrized centrality based on acyclic, conserved flows (see Table \ref{tab:classification} and Sec.~\ref{sec:newcent}).  As a result, it closely matches intuitive aspects of the harmonic closeness and betweenness centrality orderings, unlike the PageRank, Katz, and communicability centralities. It is noteworthy that the closeness and betweenness are, respectively, the low-grasp limits of the conditional resistance-closeness and the conditional current-betweenness centralities, which are also based on acyclic, conserved flows. This behavior is demonstrated on a variety of networks, including line networks, star networks, regular networks, and networks with bottlenecks, as discussed in Sec.~\ref{sec:adv_reg}. The reason is that, with acyclic, conserved flows, influence cannot get trapped in any part of the network; as the reach is increased, the influence  must always flow toward as yet unvisited nodes  \footnote{In principle, an acyclic flow could reach a dead end and therefore fail to visit all nodes. This scenario cannot occur with the ground-current centrality, since it is based on electrical currents flowing to ground, which is connected to all nodes.}. We now consider how this manifests on the types of networks listed above.

In the line network (see Sec.~\ref{sec:lines}), the random walkers of the PageRank centrality ``bounce" off  the end nodes, so that walkers on  nodes near the periphery are less likely to leave the periphery than walkers near the center are likely to leave the center. This leads to a higher centrality for peripheral nodes.  (However, end nodes   have the lowest centrality of all, because all walkers on them have no choice but to leave.) This scenario cannot occur with acyclic centralities, because ``bouncing'' off the end node always creates cycles of length 2.

For cyclic  centralities on the closed Cayley tree (see Sec.~\ref{sec:reg}), influence that originates on the periphery is less likely to arrive at the center node than it is to stay on the periphery. This is because all nodes have the same degree, and so the influence is not biased toward the center. The same reasoning holds for any regular network that has a central location. In acyclic centralities like the ground-current centrality, all sufficiently high-reach (and thus long) paths must pass through the center. Thus, the ground-current  centrality provides a  sensitive, nonlocal measure of centrality for regular networks (see Fig.~\ref{fig:cayley_results}). We  propose that it may also be  the appropriate choice for {\it nearly}-regular networks, such as the Manhattan street grid, though further study is needed.

The weighted bottleneck network $\mathcal{B}$ (see the first part of Sec.~\ref{sec:bnecks}) behaves similarly to regular networks: cyclic-centrality influence originating in one of the sublattices is likely to stay there, since the nodes there have higher degrees than the bottleneck nodes. In acyclic centralities, all sufficiently long paths must pass through the bottleneck node. This reasoning also holds for real networks with bottlenecks (see the second part of Sec.~\ref{sec:bnecks}), where the ground-current centrality prefers high-betweenness nodes at high reach (low $\para_C$). For example, the high-betweenness nodes in the vole network (black nodes in Fig.~\ref{fig:voles}) do not have very high weighted or unweighted degree. At high reach (low $\para_T$), the highest communicability centrality (gray nodes) occurs in nodes with high weighted degree, near clusters of high unweighted degree. The influence is trapped in these parts of the network, just as it was in the sublattices of $\mathcal{B}$. In contrast, the acyclic ground-current centrality must pass influence through the high-betweenness nodes when the reach (and thus the path length) is sufficiently high; see Figs.~\ref{fig:unweightedFlorida_comm}-\ref{fig:comm_all_nets}. 

The cyclic nature of the communicability and eigenvector centralities  also contributes to their tendency toward strong localization on some networks (see Sec.~\ref{sec:local}). In \cite{martin2014localization}, the nonbacktracking centrality is used as a less localizing alternative. It is based on the Hashimoto matrix \cite{hashimoto1989zeta}, whose definition prevents influence from traveling in cycles of length 2. The ground-current centrality does not allow influence to travel in cycles of any length, and consequently tends to have even less localization than the nonbacktracking centrality, as seen in Fig.~\ref{fig:IPR_gcc_all_nets}.

In addition to being acyclic, the ground-current centrality is based on conserved, rather than duplicating, flows. (Though  cyclicity and duplication are generally independent dimensions of centrality type,  Table \ref{tab:classification} demonstrates that they coincide for the metrics considered here.) The reliance on duplicating flows leads  the  communicability (and Katz) rankings to deviate from those of the other centralities in the subdivided star network $\mathcal{S}$ (see  Sec.~\ref{sec:sstar}). As shown in Fig.~\ref{fig:exp_star}, communicability influence originating on the  central node $n_0$ of $\mathcal{S}_{\{1,2,5,10,18,30\}}$ flows primarily to $n_{30}$ at high reach (low $\para_T$).  This is paradoxical  because $n_{30}$ is the node at the highest unweighted distance from $n_{0}$. 
The situation is explained by the pattern of influence duplication within the communicability centrality, defined in Eq.~(\ref{eq:communicability}). There, each factor $\mathbf{A}^l$ corresponds to influence  traveling $l$ steps, duplicating at every node  in proportion to its weighted degree. Because nodes on the $n_{30}$ spoke have the highest weighted degrees in the network, most of the duplication occurs there. In fact, when the reach (and therefore $l$) is high, $\approx $ 99.4\% of the influence is created along the  $n_{30}$ spoke, even though its original source is $n_0$. As a result, $n_{30}$ receives the highest centrality. 

Thus, the high-degree regions of a network are doubly challenging for the communicability centrality and similar measures. Because of cyclicity, influence tends to get trapped in these areas and, because of duplication, even more influence is created there. These phenomena can lead to very high centrality localization \cite{martin2014localization}. However, these situations do not arise with the acyclic, conserved (nonduplicating) ground-current centrality.

 In summary, the unique features of the ground-current centrality arise from its position in the classification system of Table~\ref{tab:classification}, which encompasses parametrized measures of two types: reach and grasp. The ground-current centrality is the only acyclic,  conserved measure with parametrized reach. Furthermore, the other acyclic, conserved centralities have more complicated descriptions and formulas, since grasp parametrization requires more involved calculations \cite{gurfinkel2020absorbing}. Real-world processes on networks usually have limitations on both travel distance (reach) and the number of paths that can be traveled (grasp).  An appropriate choice of $\para$ is required to  apply parametrized centralities to study such processes. We are currently developing methods to quantify the levels of reach and grasp across different centrality measures.


\appendix

\section{Communicability betweenness and similar centralities}
\label{app:cmb}

The communicability betweenness centrality (CMB) \cite{estrada2009communicability} is described by
\begin{equation}\label{eq:combet}
\mathbf{M}^\mathrm{CMB}_{i j}(\para_T) = \sum_{s} \frac{ \exp(\mathbf{A}/\para_T)_{sj}- \exp(\mathbf{A}/\para_T- \mathbf{E}(i)/\para_T)_{sj}}{\exp(\mathbf{A}/\para_T)_{sj}}, \hspace{1em} i\neq s, i\neq j, s\neq j. 
\end{equation}
Here, $\mathbf{E}(i)$ is the $i$th row and column of $\mathbf{A}$, with zeroes elsewhere, so the numerator quantifies $i$'s contribution to the communicability $\mathbf{M}^\mathrm{COM}_{sj}$ of $s$ and $j$. In this context, $\para_T$---usually a reach parameter---acts as a grasp parameter. This works similarly to the conditional current in \cite{gurfinkel2020absorbing}: as the reach is decreased, the shortest path between $s$ and $j$ becomes dominant. While the numerator goes to zero, the denominator does as well, which allows for a finite contribution. 

This technique can be generalized to convert any radial reach-parametrized centrality into a medial grasp-parametrized centrality. The effect of the expression $\mathbf{A}-\mathbf{E}(i)$ is simply to remove node $i$, resulting in a modified network. From there, the fractional differences in centrality between the original and modified networks are calculated. For example, the resulting grasp-parametrized medial form of the Katz centrality would be:
\begin{equation*}\label{eq:combet}
\mathbf{M}_{i j}(\para_T) = \sum_{s} \frac{ \sum_l(\mathbf{A}^l/\para_T^l)_{sj}- \sum_l([\mathbf{A}- \mathbf{E}(i)]^l/\para_T^l)_{sj}}{\sum_l(\mathbf{A}^l/\para_T^l)_{sj}}, \hspace{1em} i\neq s, i\neq j, s\neq j
\end{equation*}

\section{The reduced Laplacian assigns $V=0$ to the removed node}
\label{app:reduced}
Resistor networks are described by the  system of equations $\ket{I}=\mathbf{L} \ket{V}$, where $\mathbf{L}$ is defined in terms of the elements of the conductance matrix $c_{ij}$. This system is underdetermined when solving for  $\ket{V}$ because of the gauge invariance of the scalar potential; this  fact is captured by the equation $\mathbf{L} \ket{1}=0$. Standard methods to solve this underdetermined system include (a) using the pseudoinverse of the Laplacian matrix $\mathbf{L}$ and (b) removing one node $g$ from the network, leaving  $(N-1)$-dimensional reduced vectors $\ket{V}^\mathrm{red}$ and $\ket{I}^\mathrm{red}$, and the $(N-1)\times(N-1)$-dimensional reduced matrix $\mathbf{L}^\mathrm{red}$. With the latter method, the resulting system is no longer underdetermined and can be solved with standard matrix inversion. Here we show that this forces the gauge such that the potential of the removed node is zero, hence $g$ for ``ground". This result is commonly quoted, but the explanation is almost always omitted and is included here for completeness.

Consider the description of the unreduced linear system in terms of the reduced one:
\begin{equation}
\label{eq:guage}
\begin{array}{cccc}
\mathbf{L}&\ket{V}&=&\ket{I}\\
\veq&\veq&&\veq\\
\left(\begin{array}{cc}
\sum_i c_{g i} & \hspace{1em}- \bra{c_g}\\
 - \ket{c_g}& \mathbf{L}^\mathrm{red}
\end{array}\right)
&
\left(\begin{array}{c}
V_g\\
\ket{V}^\mathrm{red}
\end{array}\right)
&
=
&
\left(\begin{array}{c}
I_g\\
\ket{I}^\mathrm{red}
\end{array}\right)
\end{array}.
\end{equation}
Here we have, without loss of generality, chosen $g$ to be the node in position one, and the vector $\ket{c_g}$ is defined to have $i$th element equal to $c_{gi}$. Note that, by the solution  of the reduced problem, $\ket{V}^\mathrm{red}= (\mathbf{L}^\mathrm{red})^{-1}\ket{I}^\mathrm{red}$. With this substitution, the second row of the multiplication in Eq.~(\ref{eq:guage}) results in
\begin{equation}
\ket{I}^\mathrm{red} =  -\ket{c_g} V_g + \ket{I}^\mathrm{red},
\end{equation}
which forces $V_g=0$, as claimed.

\section{Asymptotic forms of the ground-current centrality}
\label{app:gcc}

\newcommand{\avpic}{\langle \para_C \rangle}

In this appendix, we derive the limiting values of both variants of the ground-current centrality in the regimes of both high and low ground conductances. To demonstrate the robustness of our reasoning, we will not rely on the physical analogy with current flow; all calculations will follow solely from  the matrix formula Eq.~(\ref{eq:gccf}).

Here we consider an \em arbitrary \em vector $\ket{C}$ of ground conductances, and so rather than $\para_C$, we rely on the  the average ground conductance: $\avpic \stackrel{\mathrm{def}}{=} N^{-1}\sum_{i}C_i = N^{-1}C_\mathrm{tot}$. When all ground conductances are identical, as in Eq.~(\ref{gccfinal}), $\avpic$ reduces to $\para_C$.  When analyzing the limiting behavior of the centrality, we only consider cases in which all ground conductances are small or all are large, though there may be relative fluctuations around the average value $\avpic$.

\subsubsection{Precise calculation of the  low $\avpic$ limits.}
The general form of the ground-current centrality depends primarily on the elements of the matrix $\left[\mathbf{L}+\pmb{\mathrm{Diag}}(\ket{C}) \,\right]^{-1}$, and the low $\avpic$ limit of the centrality can be extracted from the low $\avpic$ limit of the matrix.  As $\avpic$ goes to zero, the $\left[\mathbf{L}+\pmb{\mathrm{Diag}}(\ket{C}) \,\right]^{-1}$ matrix ceases to converge because $\mathbf{L}$ is singular. The manner of the  divergence of this matrix can be specified precisely by using the eigendecomposition as follows.

We separate out the asymptotic portion of the matrix by writing it in terms of the  diagonal matrix $\mathbf{Q}$ with elements $\mathbf{Q}_{i i}= \sqrt{C_\mathrm{tot}/C_i}$. This ``quotient matrix" is convenient because it is invariant under a uniform scaling of the $C_i$, so does not depend on the value of $\avpic$. Furthermore, $\mathbf{Q}$ satisfies $\mathbf{Q} \;\pmb{\mathrm{Diag}}(\ket{C})\; \mathbf{Q} =  \pmb{\mathbb{I}}\, C_\mathrm{tot} = \pmb{\mathbb{I}}\, \avpic N $. We can then write
\begin{equation} 
\left[\mathbf{L}+\pmb{\mathrm{Diag}}(\ket{C}) \,\right]^{-1} = \mathbf{Q} \left[\mathbf{Q} \mathbf{L} \mathbf{Q}+ \pmb{\mathbb{I}}\, \avpic N\right]^{-1} \mathbf{Q}.
\end{equation} 

From the well-known fact that all symmetric graph Laplacians are positive semidefinite, we have that the matrix $\mathbf{Q L Q}$ has all eigenvalues $\lambda_i \ge 0$. In fact, there is only one eigenvalue equal to zero: $\lambda_0=0$, with corresponding normalized eigenvector  $\ket{v_0}$ such that $\ket{v_0}_i=\sqrt{C_i/C_\mathrm{tot}}$. This is because $\mathbf{Q}$ is invertible and the entire nullspace of $\mathbf{L}$ is spanned by $\ket{1}$, thus the nullspace of $\mathbf{Q}\mathbf{L}\mathbf{Q}$ is spanned by the vector $\ket{v}$ satisfying $\mathbf{Q}\ket{v}=\ket{1}$.

Therefore we have 
$$ \mathbf{Q}\mathbf{L}\mathbf{Q} =\ket{v_0} \, 0\, \bra{v_0} + \sum_{i=1}^{N-1}\ket{v_i}\lambda_i\bra{v_i},$$
where the eigenvectors $\ket{v}$ of $\mathbf{Q}\mathbf{L}\mathbf{Q} $ form an orthonormal basis.  With the addition of the identity matrix term, we are able to take the inverse:
\begin{equation*} 
\left[\mathbf{Q}\mathbf{L}\mathbf{Q} + \pmb{\mathbb{I}}\, \avpic N\right]^{-1} =\ket{v_0}\, \left(\avpic N\right)^{-1}\, \bra{v_0}\; +\; \sum_{i=1}^{N-1}\ket{v_i}\left(\lambda_i+\avpic N\right)^{-1}\bra{v_i}.
\end{equation*} 
As $\avpic$ approaches $0$, the first term dominates because all the $\lambda_i$ are greater than 0:

\begin{eqnarray} 
&\left[\mathbf{Q}\mathbf{L}\mathbf{Q} + \pmb{\mathbb{I}}\, \avpic N\right]^{-1} & \;\approx  \ket{v_0} \, \left(\avpic N\right)^{-1}\, \bra{v_0}\\
\left[\mathbf{L}+\pmb{\mathrm{Diag}}(\ket{C})\right]^{-1}=&\;\mathbf{Q}\left[\mathbf{Q}\mathbf{L}\mathbf{Q} + \pmb{\mathbb{I}}\, \avpic N\right]^{-1}\mathbf{Q}&\;\approx \ket{1} \left(\avpic N\right)^{-1} \bra{1},
\end{eqnarray} 
where Eq.~(\ref{app:gcc}3) comes from Eq.~(\ref{app:gcc}1) and the definition of $\mathbf{Q}$. Using this result in Eq.~(\ref{eq:gccf}),
we find that 
\begin{eqnarray}
c^\mathrm{}_i &=&  \avpic N = C_{tot} \\
\mathbf{M}^\mathrm{}_{i j} &=   & C_j
\end{eqnarray}
This is the same result that was obtained when reasoning about the physical properties of resistor networks in the low $\avpic$ limit. 

A seeming difficulty in the preceding is posed by the possibility of zero ground-conductance values, since the $\mathbf{Q}$ matrix will then have infinitely large entries.  However, since the contribution of the $\mathbf{Q}$ matrices cancels out in Eq.~(\ref{app:gcc}3),  we see that the results hold for arbitrarily small ground-conductance values. As a result, Eq.~(\ref{app:gcc}4) still holds for the exogenous ground-current centrality, with only the caveat being that diagonal elements of $\widetilde{\mathbf{M}}$ are 0 because self-influence is  disallowed. As a result, the exogenous form will  have $\tilde{c}_i=C_{tot} -C_i$.

\subsubsection{Precise calculation of the  high $\avpic$ limits.}
The high $\avpic$ limit of the ground-current centrality can be found transparently from the limiting form of Eq.~(\ref{eq:gccf}):
\begin{alignat}{4} \label{appmri}
c^\mathrm{}_i&\to\lim_{\avpic \to  \infty}&\; 1/ \left[\mathbf{L}+\pmb{\mathrm{Diag}}(\ket{C}) \,\right]_{i i}^{-1}& &=&\; C_i\nonumber \\
\mathbf{M}^\mathrm{}_{i j} &\to\lim_{\avpic \to  \infty}&c_i \left[\mathbf{L}+\pmb{\mathrm{Diag}}(\ket{C}) \,\right]_{i j}^{-1}& \;C_j \;\, &=&\;  \delta_{i j}\,  C_j.
\end{alignat}
 This is again in agreement with the behavior of physical resistor networks, as described in Section \ref{ssc:gccp}.

The above limiting procedure fails in the case of the exogenous ground-current centrality. Recall that for this measure, the diagonal elements $\widetilde{\mathbf{M}}^\mathrm{}_{ii}$ are set to zero by construction. While Eq.~(\ref{appmri}) shows that the diagonal component of the matrix $\left[\mathbf{L}+\pmb{\mathrm{Diag}}(\ket{C}) \,\right]^{-1}$ becomes dominant in the high $\avpic$ limit, we are now looking for the significantly smaller off-diagonal terms. These terms can be found by utilizing the well-known Woodbury matrix identity \cite{higham2002accuracy}: 
$$
\left[\pmb{\mathbb{A}} + \pmb{\mathbb{U}}\pmb{\mathbb{C}}\pmb{\mathbb{V}} \right]^{-1}
= \pmb{\mathbb{A}}^{-1}-\pmb{\mathbb{A}}^{-1}\pmb{\mathbb{U}}
\left[\pmb{\mathbb{C}}^{-1}+\pmb{\mathbb{V}}\pmb{\mathbb{A}}^{-1}\pmb{\mathbb{U}}\right]^{-1}
\pmb{\mathbb{V}}\pmb{\mathbb{A}}^{-1},
$$
where double-struck letters refer to arbitrary, but compatibly-sized matrices. Here, we take $\pmb{\mathbb{A}}= \pmb{\mathrm{Diag}}(\ket{C})$, $\pmb{\mathbb{U}}=\mathbf{L}$, and $\pmb{\mathbb{C}}=\pmb{\mathbb{V}}=\pmb{\mathbb{I}}$, for the identity matrix $\pmb{\mathbb{I}}$. Let us denote the inverse of $\pmb{\mathrm{Diag}}(\ket{C})$ as $\pmb{\mathbb{D}}$, where $\pmb{\mathbb{D}}_{ij}=\delta_{ij}C_j^{-1}$. In the high $\avpic$ limit, $\pmb{\mathbb{D}}$ will approach zero. Applying the formula, Eq.~(\ref{appmri}) becomes
\newcommand*{\vD}{\pmb{\mathbb{D}}}
\begin{equation}
{\mathbf{M}}^\mathrm{}_{ij}=
\frac
{ 
\Big(\vD-\vD \mathbf{L}\left[\pmb{\mathbb{I}}+\vD \mathbf{L} \right]^{-1} \vD
\Big)_{ij}
}
{
\Big(\vD-\vD \mathbf{L} \left[\pmb{\mathbb{I}}+\vD \mathbf{L} \right]^{-1} \vD
\Big)_{ii}
}C_j\,
\stackrel{\mathrm{large }\,\avpic}{\longrightarrow}
\frac
{ 
\Big(\vD-\vD \mathbf{L}  \vD
\Big)_{ij}
}
{
\Big(\vD-\vD \mathbf{L}  \vD
\Big)_{ii}
}C_j\,
,
\end{equation}
where we have kept only terms up to second order in $\vD$.

Here we can see that, in the high $\avpic$ limit, ${\mathbf{M}}^\mathrm{}_{ii}$ approaches the  value of $C_i$, which is diverging. This is an illustration of the dominance of the diagonal seen in Eq.~(\ref{appmri}). For the exogenous centrality ($\widetilde{\mathbf{M}}$), however, only the off-diagonal elements are needed; they are found by taking the second term in the numerator and the first term in the denominator of the preceding equation. This is because in the latter case we are free to throw away   $O(\vD^2)$ terms, but in the former the  $O(\vD^2)$ terms are all that remain off diagonal. The result, using the definition of the Laplacian matrix [with the diagonal weighted degree matrix $\pmb{\mathrm{Diag}}(\ket{k})$], is 
\begin{equation}
\widetilde{\mathbf{M}}^\mathrm{}_{ij\,:\,j \neq i}
\stackrel{\mathrm{large }\,\avpic}{\longrightarrow}
-
\frac
{ 
\Big(\vD \left[\pmb{\mathrm{Diag}}(\ket{k})-\mathbf{A}\right] \vD
\Big)_{ij\,:\,j \neq i}
}
{
\vD_{ii}
}C_j\,
=
\frac
{ 
\Big(\vD \mathbf{\mathbf{A}}\, \vD
\Big)_{ij}
}
{
C_i^{-1}
}C_j\,
=
 \frac{
C_j^{-1}\mathbf{A}_{ij}C_i^{-1}}
{C_i^{-1}}C_j
=\mathbf{A}_{ij}
\end{equation}

\noindent (Here, we have dropped the $j\neq i$ after the first equals sign because $\vD \mathbf{\mathbf{A}}\, \vD$ has zeroes on the diagonal.)

Thus, the exogenous ground-current centrality matrix reduces to the adjacency matrix in the limit of large $\avpic$. This behavior is what motivates the introduction of this variant of the ground-current centrality.

	Finally, we underscore that the asymptotic reasoning in this section only works when every element of $\ket{C}$ goes to infinity with $\avpic$; \em i.e.\em, $\ket{C}= \avpic \widetilde{\ket{C}}$, where every term in $\widetilde{\ket{C}}$ does not approach zero. Thus, $O(\vD^2)$ is equivalent to $O({\avpic}^{-2})$. 

\section{Scaled parameters}
\label{app:scaledparams}

The horizontal axis in many of the figures in Sec.~\ref{sec:adv_reg} uses a rescaled form of the parameters $\para_T$, $\para_\mathrm{PRC}$, and $\para_C$. This is done because parameter values for different networks are, in general, not comparable: {\it e.g.},  $\para_T=2.5$ means something very different for the kangaroo network (see Fig.~\ref{fig:grasp_demo}) than it does for the Florida power-grid network (Fig.~\ref{fig:reach_intermediate}). In the former, there is almost no variation in the centrality values at $\para_T\lessapprox \para_T^\mathrm{left}=8.25$, while in the latter, $\para_T\approx2.5$ is a region of dramatic variation (while stability is obtained at $\para_T\lessapprox\para_T^\mathrm{left}= 0.40657$). 
Specifically, the left boundary $\para^\mathrm{left}$ of the varying region is calculated  to be the largest parameter that satisfies 
\begin{equation} \label{eq:leftlimit}
\frac{\Delta c_i(\para) /c_i(\para) } {\Delta \para / \para} <.001 \hspace{3em} \forall_i , \forall \para<\para^\mathrm{left},
\end{equation}
with the right boundary $\para^\mathrm{right}$ being defined similarly. 
The parameter range between $\para^\mathrm{left}$ and $\para^\mathrm{right}$ accounts for the vast majority of variation in the centrality values. 

The dimensionless quantity $\frac{\Delta c_i(\para) /c_i(\para) } {\Delta \para / \para}$ is the discrete derivative  of the log-log centrality plot  [such as the one in Fig.~\ref{fig:exp_star}(b)].  Because the PageRank centrality fails to converge at $\para_\mathrm{PRC}<1$, it is appropriate to plot $\log c_\mathrm{PRC}$ against $\log (\para_\mathrm{PRC}-1)$. This replaces ${\Delta \para / \para}$ with ${\Delta \para / (\para-1)}$ in Eq.~(\ref{eq:leftlimit}).

To plot our results for several different networks on the same axes in Figs.~\ref{fig:comm_all_nets}-\ref{fig:IPR_gcc_all_nets}, we produce a ``scaled $\para$'' where all relevant parameter values are constrained between zero and one:
\begin{equation}\label{eq:scaled_pi}
\mbox{scaled $\para$} = (\para- \para^\mathrm{left})/(\para^\mathrm{right}-\para^\mathrm{left})
.
\end{equation}

\begin{acknowledgements}
We thank  G. M. Buend\'\i a, G. Brown, and three anonymous Referees for useful comments and critical reading of the manuscript. 

Supported in part by U.S.  National Science Foundation Grant No. DMR-1104829. 

Work at the University of Oslo was  supported by the Research Council of Norway 
through the Center of Excellence funding scheme, Project No. 262644.
\end{acknowledgements}

\bibliography{paper2,paper1}

\begin{thebibliography}{47}%
\makeatletter
\providecommand \@ifxundefined [1]{%
 \@ifx{#1\undefined}
}%
\providecommand \@ifnum [1]{%
 \ifnum #1\expandafter \@firstoftwo
 \else \expandafter \@secondoftwo
 \fi
}%
\providecommand \@ifx [1]{%
 \ifx #1\expandafter \@firstoftwo
 \else \expandafter \@secondoftwo
 \fi
}%
\providecommand \natexlab [1]{#1}%
\providecommand \enquote  [1]{``#1''}%
\providecommand \bibnamefont  [1]{#1}%
\providecommand \bibfnamefont [1]{#1}%
\providecommand \citenamefont [1]{#1}%
\providecommand \href@noop [0]{\@secondoftwo}%
\providecommand \href [0]{\begingroup \@sanitize@url \@href}%
\providecommand \@href[1]{\@@startlink{#1}\@@href}%
\providecommand \@@href[1]{\endgroup#1\@@endlink}%
\providecommand \@sanitize@url [0]{\catcode `\\12\catcode `\$12\catcode
  `\&12\catcode `\#12\catcode `\^12\catcode `\_12\catcode `\%12\relax}%
\providecommand \@@startlink[1]{}%
\providecommand \@@endlink[0]{}%
\providecommand \url  [0]{\begingroup\@sanitize@url \@url }%
\providecommand \@url [1]{\endgroup\@href {#1}{\urlprefix }}%
\providecommand \urlprefix  [0]{URL }%
\providecommand \Eprint [0]{\href }%
\providecommand \doibase [0]{http://dx.doi.org/}%
\providecommand \selectlanguage [0]{\@gobble}%
\providecommand \bibinfo  [0]{\@secondoftwo}%
\providecommand \bibfield  [0]{\@secondoftwo}%
\providecommand \translation [1]{[#1]}%
\providecommand \BibitemOpen [0]{}%
\providecommand \bibitemStop [0]{}%
\providecommand \bibitemNoStop [0]{.\EOS\space}%
\providecommand \EOS [0]{\spacefactor3000\relax}%
\providecommand \BibitemShut  [1]{\csname bibitem#1\endcsname}%
\let\auto@bib@innerbib\@empty
\bibitem [{\citenamefont {Page}\ \emph {et~al.}(1999)\citenamefont {Page},
  \citenamefont {Brin}, \citenamefont {Motwani},\ and\ \citenamefont
  {Winograd}}]{page1999pagerank}%
  \BibitemOpen
  \bibfield  {author} {\bibinfo {author} {\bibfnamefont {L.}~\bibnamefont
  {Page}}, \bibinfo {author} {\bibfnamefont {S.}~\bibnamefont {Brin}}, \bibinfo
  {author} {\bibfnamefont {R.}~\bibnamefont {Motwani}}, \ and\ \bibinfo
  {author} {\bibfnamefont {T.}~\bibnamefont {Winograd}},\ }\href@noop {} {\emph
  {\bibinfo {title} {The PageRank Citation Ranking: Bringing Order to the
  Web.}}},\ \bibinfo {type} {Tech. Rep.}\ (\bibinfo  {institution} {Stanford
  InfoLab},\ \bibinfo {year} {1999})\BibitemShut {NoStop}%
\bibitem [{\citenamefont {Joyce}\ \emph {et~al.}(2010)\citenamefont {Joyce},
  \citenamefont {Laurienti}, \citenamefont {Burdette},\ and\ \citenamefont
  {Hayasaka}}]{joyce2010new}%
  \BibitemOpen
  \bibfield  {author} {\bibinfo {author} {\bibfnamefont {K.~E.}\ \bibnamefont
  {Joyce}}, \bibinfo {author} {\bibfnamefont {P.~J.}\ \bibnamefont
  {Laurienti}}, \bibinfo {author} {\bibfnamefont {J.~H.}\ \bibnamefont
  {Burdette}}, \ and\ \bibinfo {author} {\bibfnamefont {S.}~\bibnamefont
  {Hayasaka}},\ }\bibfield  {title} {\enquote {\bibinfo {title} {A new measure
  of centrality for brain networks},}\ }\href@noop {} {\bibfield  {journal}
  {\bibinfo  {journal} {PloS One}\ }\textbf {\bibinfo {volume} {5}},\ \bibinfo
  {pages} {e12200} (\bibinfo {year} {2010})}\BibitemShut {NoStop}%
\bibitem [{\citenamefont {Gurfinkel}\ and\ \citenamefont
  {Rikvold}(2020)}]{gurfinkel2020absorbing}%
  \BibitemOpen
  \bibfield  {author} {\bibinfo {author} {\bibfnamefont {A.~J.}\ \bibnamefont
  {Gurfinkel}}\ and\ \bibinfo {author} {\bibfnamefont {P.~A.}\ \bibnamefont
  {Rikvold}},\ }\bibfield  {title} {\enquote {\bibinfo {title} {Absorbing
  random walks interpolating between centrality measures on complex
  networks},}\ }\href@noop {} {\bibfield  {journal} {\bibinfo  {journal} {Phys.
  Rev. E}\ }\textbf {\bibinfo {volume} {101}},\ \bibinfo {pages} {012302}
  (\bibinfo {year} {2020})}\BibitemShut {NoStop}%
\bibitem [{\citenamefont {Xu}\ \emph {et~al.}(2014)\citenamefont {Xu},
  \citenamefont {Gurfinkel},\ and\ \citenamefont
  {Rikvold}}]{xu2014architecture}%
  \BibitemOpen
  \bibfield  {author} {\bibinfo {author} {\bibfnamefont {Y.}~\bibnamefont
  {Xu}}, \bibinfo {author} {\bibfnamefont {A.~J.}\ \bibnamefont {Gurfinkel}}, \
  and\ \bibinfo {author} {\bibfnamefont {P.~A.}\ \bibnamefont {Rikvold}},\
  }\bibfield  {title} {\enquote {\bibinfo {title} {Architecture of the
  {F}lorida power grid as a complex network},}\ }\href@noop {} {\bibfield
  {journal} {\bibinfo  {journal} {Physica A}\ }\textbf {\bibinfo {volume}
  {401}},\ \bibinfo {pages} {130--140} (\bibinfo {year} {2014})}\BibitemShut
  {NoStop}%
\bibitem [{\citenamefont {Estrada}\ and\ \citenamefont
  {Hatano}(2008)}]{estrada2008communicability}%
  \BibitemOpen
  \bibfield  {author} {\bibinfo {author} {\bibfnamefont {E.}~\bibnamefont
  {Estrada}}\ and\ \bibinfo {author} {\bibfnamefont {N.}~\bibnamefont
  {Hatano}},\ }\bibfield  {title} {\enquote {\bibinfo {title} {Communicability
  in complex networks},}\ }\href@noop {} {\bibfield  {journal} {\bibinfo
  {journal} {Phys. Rev. E}\ }\textbf {\bibinfo {volume} {77}},\ \bibinfo
  {pages} {036111} (\bibinfo {year} {2008})}\BibitemShut {NoStop}%
\bibitem [{\citenamefont {Estrada}\ \emph {et~al.}(2009)\citenamefont
  {Estrada}, \citenamefont {Higham},\ and\ \citenamefont
  {Hatano}}]{estrada2009communicability}%
  \BibitemOpen
  \bibfield  {author} {\bibinfo {author} {\bibfnamefont {E.}~\bibnamefont
  {Estrada}}, \bibinfo {author} {\bibfnamefont {D.~J.}\ \bibnamefont {Higham}},
  \ and\ \bibinfo {author} {\bibfnamefont {N.}~\bibnamefont {Hatano}},\
  }\bibfield  {title} {\enquote {\bibinfo {title} {Communicability betweenness
  in complex networks},}\ }\href@noop {} {\bibfield  {journal} {\bibinfo
  {journal} {Physica A}\ }\textbf {\bibinfo {volume} {388}},\ \bibinfo {pages}
  {764--774} (\bibinfo {year} {2009})}\BibitemShut {NoStop}%
\bibitem [{\citenamefont {Katz}(1953)}]{katz1953new}%
  \BibitemOpen
  \bibfield  {author} {\bibinfo {author} {\bibfnamefont {L.}~\bibnamefont
  {Katz}},\ }\bibfield  {title} {\enquote {\bibinfo {title} {A new status index
  derived from sociometric analysis},}\ }\href@noop {} {\bibfield  {journal}
  {\bibinfo  {journal} {Psychometrika}\ }\textbf {\bibinfo {volume} {18}},\
  \bibinfo {pages} {39--43} (\bibinfo {year} {1953})}\BibitemShut {NoStop}%
\bibitem [{\citenamefont {Ghosh}\ and\ \citenamefont
  {Lerman}(2011)}]{ghosh2011parameterized}%
  \BibitemOpen
  \bibfield  {author} {\bibinfo {author} {\bibfnamefont {R.}~\bibnamefont
  {Ghosh}}\ and\ \bibinfo {author} {\bibfnamefont {K.}~\bibnamefont {Lerman}},\
  }\bibfield  {title} {\enquote {\bibinfo {title} {Parameterized centrality
  metric for network analysis},}\ }\href@noop {} {\bibfield  {journal}
  {\bibinfo  {journal} {Phys. Rev. E}\ }\textbf {\bibinfo {volume} {83}},\
  \bibinfo {pages} {066118} (\bibinfo {year} {2011})}\BibitemShut {NoStop}%
\bibitem [{\citenamefont {Borgatti}(2005)}]{borgatti2005centrality}%
  \BibitemOpen
  \bibfield  {author} {\bibinfo {author} {\bibfnamefont {S.~P.}\ \bibnamefont
  {Borgatti}},\ }\bibfield  {title} {\enquote {\bibinfo {title} {Centrality and
  network flow},}\ }\href@noop {} {\bibfield  {journal} {\bibinfo  {journal}
  {Soc. Networks}\ }\textbf {\bibinfo {volume} {27}},\ \bibinfo {pages}
  {55--71} (\bibinfo {year} {2005})}\BibitemShut {NoStop}%
\bibitem [{\citenamefont {Borgatti}\ and\ \citenamefont
  {Everett}(2006)}]{borgatti2006graph}%
  \BibitemOpen
  \bibfield  {author} {\bibinfo {author} {\bibfnamefont {S.~P.}\ \bibnamefont
  {Borgatti}}\ and\ \bibinfo {author} {\bibfnamefont {M.~G.}\ \bibnamefont
  {Everett}},\ }\bibfield  {title} {\enquote {\bibinfo {title} {A
  graph-theoretic perspective on centrality},}\ }\href@noop {} {\bibfield
  {journal} {\bibinfo  {journal} {Soc. Networks}\ }\textbf {\bibinfo {volume}
  {28}},\ \bibinfo {pages} {466--484} (\bibinfo {year} {2006})}\BibitemShut
  {NoStop}%
\bibitem [{\citenamefont {Dekker}(2005)}]{Dekker2005}%
  \BibitemOpen
  \bibfield  {author} {\bibinfo {author} {\bibfnamefont {A.}~\bibnamefont
  {Dekker}},\ }\bibfield  {title} {\enquote {\bibinfo {title} {Conceptual
  distance in social network analysis},}\ }\href@noop {} {\bibfield  {journal}
  {\bibinfo  {journal} {Journal of social structure}\ }\textbf {\bibinfo
  {volume} {6}} (\bibinfo {year} {2005})}\BibitemShut {NoStop}%
\bibitem [{\citenamefont {Rochat}(2009)}]{Rochat2009}%
  \BibitemOpen
  \bibfield  {author} {\bibinfo {author} {\bibfnamefont {Y.}~\bibnamefont
  {Rochat}},\ }\href
  {https://infoscience.epfl.ch/record/200525/files/%5bEN%5dASNA09.pdf} {\emph
  {\bibinfo {title} {Closeness centrality extended to unconnected graphs: The
  harmonic centrality index}}},\ \bibinfo {type} {Tech. Rep.}\ (\bibinfo
  {institution} {Institute of Applied Mathematics, University of Lausanne},\
  \bibinfo {year} {2009})\BibitemShut {NoStop}%
\bibitem [{\citenamefont {Newman}(2018)}]{NEWM10}%
  \BibitemOpen
  \bibfield  {author} {\bibinfo {author} {\bibfnamefont {M.~E.~J.}\
  \bibnamefont {Newman}},\ }\href@noop {} {\emph {\bibinfo {title} {Networks:
  an Introduction}}}\ (\bibinfo  {publisher} {Oxford University Press},\
  \bibinfo {address} {Oxford, UK},\ \bibinfo {year} {2018})\BibitemShut
  {NoStop}%
\bibitem [{\citenamefont {Martin}\ \emph {et~al.}(2014)\citenamefont {Martin},
  \citenamefont {Zhang},\ and\ \citenamefont
  {Newman}}]{martin2014localization}%
  \BibitemOpen
  \bibfield  {author} {\bibinfo {author} {\bibfnamefont {T.}~\bibnamefont
  {Martin}}, \bibinfo {author} {\bibfnamefont {X.}~\bibnamefont {Zhang}}, \
  and\ \bibinfo {author} {\bibfnamefont {M.~E.~J.}\ \bibnamefont {Newman}},\
  }\bibfield  {title} {\enquote {\bibinfo {title} {Localization and centrality
  in networks},}\ }\href@noop {} {\bibfield  {journal} {\bibinfo  {journal}
  {Phys. Rev. E}\ }\textbf {\bibinfo {volume} {90}},\ \bibinfo {pages} {052808}
  (\bibinfo {year} {2014})}\BibitemShut {NoStop}%
\bibitem [{\citenamefont {Arnaudon}\ \emph {et~al.}(2020)\citenamefont
  {Arnaudon}, \citenamefont {Peach},\ and\ \citenamefont
  {Barahona}}]{arnaudon2020scale}%
  \BibitemOpen
  \bibfield  {author} {\bibinfo {author} {\bibfnamefont {A.}~\bibnamefont
  {Arnaudon}}, \bibinfo {author} {\bibfnamefont {R.~L.}\ \bibnamefont {Peach}},
  \ and\ \bibinfo {author} {\bibfnamefont {M.}~\bibnamefont {Barahona}},\
  }\bibfield  {title} {\enquote {\bibinfo {title} {Scale-dependent measure of
  network centrality from diffusion dynamics},}\ }\href@noop {} {\bibfield
  {journal} {\bibinfo  {journal} {Phys. Rev. Research}\ }\textbf {\bibinfo
  {volume} {2}},\ \bibinfo {pages} {033104} (\bibinfo {year}
  {2020})}\BibitemShut {NoStop}%
\bibitem [{Note1()}]{Note1}%
  \BibitemOpen
  \bibinfo {note} {In \cite {gurfinkel2020absorbing}, we chose to present
  {\protect \it unnormalized } centrality results to better track centrality
  behavior across a range of parameter values.}\BibitemShut {Stop}%
\bibitem [{not()}]{noteoninverseparam}%
  \BibitemOpen
  \href@noop {} {\ }\bibinfo {note} {Generally, the Katz centrality is defined
  as $\mathbf{M}= (\pmb{\mathbb{I}}- \para\mathbf{A})^{-1}$: the parameter is
  the inverse of our parameter $\para_\mathrm{KC}$. We also make a similar
  change to the definition of PageRank centrality. The definitions in this
  paper ensure that all the considered centrality parameters consistently
  restrict reach or grasp as they increase.}\BibitemShut {Stop}%
\bibitem [{ran()}]{random_walk_note}%
  \BibitemOpen
  \href@noop {} {\ }\bibinfo {note} {This is a standard random walk, where a
  walker on node $j$ has probability $A_{ij}/k_j$ of crossing to node $j$ in
  one step.}\BibitemShut {Stop}%
\bibitem [{\citenamefont {Benzi}\ and\ \citenamefont
  {Klymko}(2013)}]{benzi2013total}%
  \BibitemOpen
  \bibfield  {author} {\bibinfo {author} {\bibfnamefont {M.}~\bibnamefont
  {Benzi}}\ and\ \bibinfo {author} {\bibfnamefont {C.}~\bibnamefont {Klymko}},\
  }\bibfield  {title} {\enquote {\bibinfo {title} {Total communicability as a
  centrality measure},}\ }\href@noop {} {\bibfield  {journal} {\bibinfo
  {journal} {J. Complex Networks}\ }\textbf {\bibinfo {volume} {1}},\ \bibinfo
  {pages} {124--149} (\bibinfo {year} {2013})}\BibitemShut {NoStop}%
\bibitem [{\citenamefont {Gurfinkel}\ \emph {et~al.}(2015)\citenamefont
  {Gurfinkel}, \citenamefont {Silva},\ and\ \citenamefont
  {Rikvold}}]{gurfinkel2015centrality}%
  \BibitemOpen
  \bibfield  {author} {\bibinfo {author} {\bibfnamefont {A.~J.}\ \bibnamefont
  {Gurfinkel}}, \bibinfo {author} {\bibfnamefont {D.~A}\ \bibnamefont {Silva}},
  \ and\ \bibinfo {author} {\bibfnamefont {P.~A.}\ \bibnamefont {Rikvold}},\
  }\bibfield  {title} {\enquote {\bibinfo {title} {Centrality fingerprints for
  power grid network growth models},}\ }\href@noop {} {\bibfield  {journal}
  {\bibinfo  {journal} {Phys. Proc.}\ }\textbf {\bibinfo {volume} {68}},\
  \bibinfo {pages} {52--55} (\bibinfo {year} {2015})}\BibitemShut {NoStop}%
\bibitem [{kan(2017)}]{kangadata}%
  \BibitemOpen
  \href@noop {} {\enquote {\bibinfo {title} {Kangaroo network dataset},}\
  }\bibinfo {howpublished} {KONECT
  \url{http://konect.uni-koblenz.de/networks/moreno_kangaroo}} (\bibinfo {year}
  {2017})\BibitemShut {NoStop}%
\bibitem [{\citenamefont {Grant}(1973)}]{grant1973dominance}%
  \BibitemOpen
  \bibfield  {author} {\bibinfo {author} {\bibfnamefont {T.~R.}\ \bibnamefont
  {Grant}},\ }\bibfield  {title} {\enquote {\bibinfo {title} {Dominance and
  association among members of a captive and a free-ranging group of grey
  kangaroos ({M}acropus giganteus)},}\ }\href@noop {} {\bibfield  {journal}
  {\bibinfo  {journal} {Anim. Behav.}\ }\textbf {\bibinfo {volume} {21}},\
  \bibinfo {pages} {449--456} (\bibinfo {year} {1973})}\BibitemShut {NoStop}%
\bibitem [{\citenamefont {Newman}(2005)}]{newman2005measure}%
  \BibitemOpen
  \bibfield  {author} {\bibinfo {author} {\bibfnamefont {M.~E.~J.}\
  \bibnamefont {Newman}},\ }\bibfield  {title} {\enquote {\bibinfo {title} {A
  measure of betweenness centrality based on random walks},}\ }\href@noop {}
  {\bibfield  {journal} {\bibinfo  {journal} {Soc. Networks}\ }\textbf
  {\bibinfo {volume} {27}},\ \bibinfo {pages} {39--54} (\bibinfo {year}
  {2005})}\BibitemShut {NoStop}%
\bibitem [{\citenamefont {Stephenson}\ and\ \citenamefont
  {Zelen}(1989)}]{stephenson1989rethinking}%
  \BibitemOpen
  \bibfield  {author} {\bibinfo {author} {\bibfnamefont {K.}~\bibnamefont
  {Stephenson}}\ and\ \bibinfo {author} {\bibfnamefont {M.}~\bibnamefont
  {Zelen}},\ }\bibfield  {title} {\enquote {\bibinfo {title} {Rethinking
  centrality: Methods and examples},}\ }\href@noop {} {\bibfield  {journal}
  {\bibinfo  {journal} {Soc. Networks}\ }\textbf {\bibinfo {volume} {11}},\
  \bibinfo {pages} {1--37} (\bibinfo {year} {1989})}\BibitemShut {NoStop}%
\bibitem [{\citenamefont {Brandes}\ and\ \citenamefont
  {Fleischer}(2005)}]{brandes2005centrality}%
  \BibitemOpen
  \bibfield  {author} {\bibinfo {author} {\bibfnamefont {U.}~\bibnamefont
  {Brandes}}\ and\ \bibinfo {author} {\bibfnamefont {D.}~\bibnamefont
  {Fleischer}},\ }\bibfield  {title} {\enquote {\bibinfo {title} {Centrality
  measures based on current flow},}\ }in\ \href@noop {} {\emph {\bibinfo
  {booktitle} {Annual symposium on theoretical aspects of computer science}}}\
  (\bibinfo {organization} {Springer},\ \bibinfo {address}
  {Berlin-Heidelberg},\ \bibinfo {year} {2005})\ pp.\ \bibinfo {pages}
  {533--544}\BibitemShut {NoStop}%
\bibitem [{\citenamefont {Estrada}\ and\ \citenamefont
  {Rodriguez-Velazquez}(2005)}]{estrada2005subgraph}%
  \BibitemOpen
  \bibfield  {author} {\bibinfo {author} {\bibfnamefont {E.}~\bibnamefont
  {Estrada}}\ and\ \bibinfo {author} {\bibfnamefont {J.~A.}\ \bibnamefont
  {Rodriguez-Velazquez}},\ }\bibfield  {title} {\enquote {\bibinfo {title}
  {Subgraph centrality in complex networks},}\ }\href@noop {} {\bibfield
  {journal} {\bibinfo  {journal} {Phys. Rev. E}\ }\textbf {\bibinfo {volume}
  {71}},\ \bibinfo {pages} {056103} (\bibinfo {year} {2005})}\BibitemShut
  {NoStop}%
\bibitem [{\citenamefont {Avrachenkov}\ \emph {et~al.}(2013)\citenamefont
  {Avrachenkov}, \citenamefont {Litvak}, \citenamefont {Medyanikov},\ and\
  \citenamefont {Sokol}}]{avrachenkov2013alpha}%
  \BibitemOpen
  \bibfield  {author} {\bibinfo {author} {\bibfnamefont {K.}~\bibnamefont
  {Avrachenkov}}, \bibinfo {author} {\bibfnamefont {N.}~\bibnamefont {Litvak}},
  \bibinfo {author} {\bibfnamefont {V.}~\bibnamefont {Medyanikov}}, \ and\
  \bibinfo {author} {\bibfnamefont {M.}~\bibnamefont {Sokol}},\ }\bibfield
  {title} {\enquote {\bibinfo {title} {Alpha current flow betweenness
  centrality},}\ }in\ \href@noop {} {\emph {\bibinfo {booktitle} {International
  Workshop on Algorithms and Models for the Web-Graph}}}\ (\bibinfo
  {organization} {Springer},\ \bibinfo {address} {Berlin-Heidelberg},\ \bibinfo
  {year} {2013})\ pp.\ \bibinfo {pages} {106--117}\BibitemShut {NoStop}%
\bibitem [{\citenamefont {Doyle}\ and\ \citenamefont
  {Snell}(1984)}]{Doyle06randomwalks}%
  \BibitemOpen
  \bibfield  {author} {\bibinfo {author} {\bibfnamefont {P.~G.}\ \bibnamefont
  {Doyle}}\ and\ \bibinfo {author} {\bibfnamefont {J.~L.}\ \bibnamefont
  {Snell}},\ }\href@noop {} {\emph {\bibinfo {title} {Random Walks and Electric
  networks}}}\ (\bibinfo  {publisher} {Mathematical Association of America},\
  \bibinfo {year} {1984})\ \bibinfo {note} {\!\!. Since 2006, available under
  the GNU FDL at
  \url{https://math.dartmouth.edu/~doyle/docs/walks/walks.pdf}}\BibitemShut
  {NoStop}%
\bibitem [{Note2()}]{Note2}%
  \BibitemOpen
  \bibinfo {note} {The interpretation of non-zero adjacency matrix elements as
  conductances in a resistor network makes sense for {\protect \it
  affinity-weighted} networks \cite {newman2005measure}, as well as multigraphs
  where $\protect \mathbf {A}_{ij}$ stands for the number of parallel edges
  between $i$ and $j$. When an adjacency matrix element stands for, {\protect
  \it e.g.}, a distance rather than an affinity, it is more appropriate for it
  to be interpreted as a resistance. In either case, when $\protect \mathbf
  {A}_{ij}=0$, there is no edge between $i$ and $j$.}\BibitemShut {Stop}%
\bibitem [{\citenamefont {Dale}\ \emph {et~al.}(2009)\citenamefont {Dale},
  \citenamefont {Alqutami}, \citenamefont {Baldwin}, \citenamefont {Faruque},
  \citenamefont {Langston}, \citenamefont {McLaren}, \citenamefont {Meeker},
  \citenamefont {Steurer},\ and\ \citenamefont {Schoder}}]{dale}%
  \BibitemOpen
  \bibfield  {author} {\bibinfo {author} {\bibfnamefont {S.}~\bibnamefont
  {Dale}}, \bibinfo {author} {\bibfnamefont {T.}~\bibnamefont {Alqutami}},
  \bibinfo {author} {\bibfnamefont {T.}~\bibnamefont {Baldwin}}, \bibinfo
  {author} {\bibfnamefont {O.}~\bibnamefont {Faruque}}, \bibinfo {author}
  {\bibfnamefont {J.}~\bibnamefont {Langston}}, \bibinfo {author}
  {\bibfnamefont {P.}~\bibnamefont {McLaren}}, \bibinfo {author} {\bibfnamefont
  {P.}~\bibnamefont {Meeker}}, \bibinfo {author} {\bibfnamefont
  {M.}~\bibnamefont {Steurer}}, \ and\ \bibinfo {author} {\bibfnamefont
  {K.}~\bibnamefont {Schoder}},\ }\href@noop {} {\emph {\bibinfo {title}
  {Progress Report for the Institute for Energy Systems, Economics and
  Sustainability and the {F}lorida Energy Systems Consortium}}},\ \bibinfo
  {type} {Tech. Rep.}\ (\bibinfo  {institution} {Florida State University},\
  \bibinfo {address} {Tallahassee, FL},\ \bibinfo {year} {2009})\BibitemShut
  {NoStop}%
\bibitem [{\citenamefont {Brandes}(2001)}]{brandes2001faster}%
  \BibitemOpen
  \bibfield  {author} {\bibinfo {author} {\bibfnamefont {U.}~\bibnamefont
  {Brandes}},\ }\bibfield  {title} {\enquote {\bibinfo {title} {A faster
  algorithm for betweenness centrality},}\ }\href@noop {} {\bibfield  {journal}
  {\bibinfo  {journal} {J. Math. Sociol.}\ }\textbf {\bibinfo {volume} {25}},\
  \bibinfo {pages} {163--177} (\bibinfo {year} {2001})}\BibitemShut {NoStop}%
\bibitem [{\citenamefont {Choe}\ \emph {et~al.}(2004)\citenamefont {Choe},
  \citenamefont {McCormick},\ and\ \citenamefont
  {Koh}}]{Choe-2004-connectivity}%
  \BibitemOpen
  \bibfield  {author} {\bibinfo {author} {\bibfnamefont {Y.}~\bibnamefont
  {Choe}}, \bibinfo {author} {\bibfnamefont {B.~H.}\ \bibnamefont {McCormick}},
  \ and\ \bibinfo {author} {\bibfnamefont {W.}~\bibnamefont {Koh}},\ }\bibfield
   {title} {\enquote {\bibinfo {title} {Network connectivity analysis on the
  temporally augmented \em {C}. elegans \em web: A pilot study},}\ }\bibfield
  {booktitle} {\emph {\bibinfo {booktitle} {Society of Neuroscience
  Abstracts}},\ }\href@noop {} {\ \textbf {\bibinfo {volume} {30}} (\bibinfo
  {year} {2004})}\BibitemShut {NoStop}%
\bibitem [{\citenamefont {Abou~Hamad}\ \emph {et~al.}(2010)\citenamefont
  {Abou~Hamad}, \citenamefont {Israels}, \citenamefont {Rikvold},\ and\
  \citenamefont {Poroseva}}]{hama10}%
  \BibitemOpen
  \bibfield  {author} {\bibinfo {author} {\bibfnamefont {I.}~\bibnamefont
  {Abou~Hamad}}, \bibinfo {author} {\bibfnamefont {B.}~\bibnamefont {Israels}},
  \bibinfo {author} {\bibfnamefont {P.~A.}\ \bibnamefont {Rikvold}}, \ and\
  \bibinfo {author} {\bibfnamefont {S.~V.}\ \bibnamefont {Poroseva}},\
  }\bibfield  {title} {\enquote {\bibinfo {title} {Spectral matrix methods for
  partitioning power grids: applications to the {I}talian and {F}loridian
  high-voltage networks},}\ }\href@noop {} {\bibfield  {journal} {\bibinfo
  {journal} {Phys. Proc.}\ }\textbf {\bibinfo {volume} {4}},\ \bibinfo {pages}
  {125--129} (\bibinfo {year} {2010})}\BibitemShut {NoStop}%
\bibitem [{\citenamefont {Rossi}\ and\ \citenamefont {Ahmed}()}]{nr-aaai15}%
  \BibitemOpen
  \bibfield  {author} {\bibinfo {author} {\bibfnamefont {R.~A.}\ \bibnamefont
  {Rossi}}\ and\ \bibinfo {author} {\bibfnamefont {N.~K.}\ \bibnamefont
  {Ahmed}},\ }\href@noop {} {\enquote {\bibinfo {title} {The network data
  repository with interactive graph analytics and visualization},}\ }\bibinfo
  {howpublished} {Available at
  \url{http://networkrepository.com/mammalia-voles-bhp-trapping-24.php}
  (2015)}\BibitemShut {NoStop}%
\bibitem [{\citenamefont {Davis}\ \emph {et~al.}(2015)\citenamefont {Davis},
  \citenamefont {Abbasi}, \citenamefont {Shah}, \citenamefont {Telfer},\ and\
  \citenamefont {Begon}}]{voles}%
  \BibitemOpen
  \bibfield  {author} {\bibinfo {author} {\bibfnamefont {S.}~\bibnamefont
  {Davis}}, \bibinfo {author} {\bibfnamefont {B.}~\bibnamefont {Abbasi}},
  \bibinfo {author} {\bibfnamefont {S.}~\bibnamefont {Shah}}, \bibinfo {author}
  {\bibfnamefont {S.}~\bibnamefont {Telfer}}, \ and\ \bibinfo {author}
  {\bibfnamefont {M.}~\bibnamefont {Begon}},\ }\bibfield  {title} {\enquote
  {\bibinfo {title} {Spatial analyses of wildlife contact networks},}\
  }\href@noop {} {\bibfield  {journal} {\bibinfo  {journal} {J. Roy. Soc.
  Interface}\ }\textbf {\bibinfo {volume} {12}},\ \bibinfo {pages} {20141004}
  (\bibinfo {year} {2015})}\BibitemShut {NoStop}%
\bibitem [{\citenamefont {Milo}\ \emph {et~al.}(2004)\citenamefont {Milo},
  \citenamefont {Itzkovitz}, \citenamefont {Kashtan}, \citenamefont {Levitt},
  \citenamefont {Shen-Orr}, \citenamefont {Ayzenshtat}, \citenamefont
  {Sheffer},\ and\ \citenamefont {Alon}}]{milo2004superfamilies}%
  \BibitemOpen
  \bibfield  {author} {\bibinfo {author} {\bibfnamefont {R.}~\bibnamefont
  {Milo}}, \bibinfo {author} {\bibfnamefont {S.}~\bibnamefont {Itzkovitz}},
  \bibinfo {author} {\bibfnamefont {N.}~\bibnamefont {Kashtan}}, \bibinfo
  {author} {\bibfnamefont {R.}~\bibnamefont {Levitt}}, \bibinfo {author}
  {\bibfnamefont {S.}~\bibnamefont {Shen-Orr}}, \bibinfo {author}
  {\bibfnamefont {I.}~\bibnamefont {Ayzenshtat}}, \bibinfo {author}
  {\bibfnamefont {M.}~\bibnamefont {Sheffer}}, \ and\ \bibinfo {author}
  {\bibfnamefont {U.}~\bibnamefont {Alon}},\ }\bibfield  {title} {\enquote
  {\bibinfo {title} {Superfamilies of evolved and designed networks},}\
  }\href@noop {} {\bibfield  {journal} {\bibinfo  {journal} {Science}\ }\textbf
  {\bibinfo {volume} {303}},\ \bibinfo {pages} {1538--1542} (\bibinfo {year}
  {2004})}\BibitemShut {NoStop}%
\bibitem [{Note3()}]{Note3}%
  \BibitemOpen
  \bibinfo {note} {In the final calculation, the many nonperipheral nodes
  (unlabeled in Fig.~\ref {fig:sstar}) account for the majority of the
  contribution to $i^\protect \mathrm {p}$'s centrality. This means that
  peripheral nodes on ``long'' spokes will have larger centrality, just because
  they are near many nonperipheral nodes. As a result, the $c_{i^\protect
  \mathrm {p}}$ will have an ordering that places greater importance on nodes a
  farther unweighted distance from the hub node $n_0$. All of the centralities
  under discussion reproduce this expected $c_{i^\protect \mathrm {p}}$
  ordering. We focus on the matrix elements $\protect \mathaccentV
  {tilde}07E{\protect \mathbf {M}}_{n_0 i^\protect \mathrm {p}}$ rather than
  the final centrality $c_i$ because they are a direct measurement of the
  influence between two nodes, and as such are more sensitive to the chosen
  centrality method.}\BibitemShut {Stop}%
\bibitem [{\citenamefont {Watts}\ and\ \citenamefont
  {Strogatz}(1998)}]{watts1998collective}%
  \BibitemOpen
  \bibfield  {author} {\bibinfo {author} {\bibfnamefont {D.~J.}\ \bibnamefont
  {Watts}}\ and\ \bibinfo {author} {\bibfnamefont {S.~H.}\ \bibnamefont
  {Strogatz}},\ }\bibfield  {title} {\enquote {\bibinfo {title} {Collective
  dynamics of 'small-world' networks},}\ }\href@noop {} {\bibfield  {journal}
  {\bibinfo  {journal} {Nature}\ }\textbf {\bibinfo {volume} {393}},\ \bibinfo
  {pages} {440} (\bibinfo {year} {1998})}\BibitemShut {NoStop}%
\bibitem [{\citenamefont {Newman}(2002)}]{newman2002assortative}%
  \BibitemOpen
  \bibfield  {author} {\bibinfo {author} {\bibfnamefont {M.~E.~J.}\
  \bibnamefont {Newman}},\ }\bibfield  {title} {\enquote {\bibinfo {title}
  {Assortative mixing in networks},}\ }\href@noop {} {\bibfield  {journal}
  {\bibinfo  {journal} {Phys. Rev. Lett.}\ }\textbf {\bibinfo {volume} {89}},\
  \bibinfo {pages} {208701} (\bibinfo {year} {2002})}\BibitemShut {NoStop}%
\bibitem [{\citenamefont {Yan}\ \emph {et~al.}(2017)\citenamefont {Yan},
  \citenamefont {V{\'e}rtes}, \citenamefont {Towlson}, \citenamefont {Chew},
  \citenamefont {Walker}, \citenamefont {Schafer},\ and\ \citenamefont
  {Barab{\'a}si}}]{yan2017network}%
  \BibitemOpen
  \bibfield  {author} {\bibinfo {author} {\bibfnamefont {G.}~\bibnamefont
  {Yan}}, \bibinfo {author} {\bibfnamefont {P.~E.}\ \bibnamefont {V{\'e}rtes}},
  \bibinfo {author} {\bibfnamefont {E.~K.}\ \bibnamefont {Towlson}}, \bibinfo
  {author} {\bibfnamefont {Y.~L.}\ \bibnamefont {Chew}}, \bibinfo {author}
  {\bibfnamefont {D.~S.}\ \bibnamefont {Walker}}, \bibinfo {author}
  {\bibfnamefont {W.~R.}\ \bibnamefont {Schafer}}, \ and\ \bibinfo {author}
  {\bibfnamefont {A.-L.}\ \bibnamefont {Barab{\'a}si}},\ }\bibfield  {title}
  {\enquote {\bibinfo {title} {Network control principles predict neuron
  function in the \em {C}aenorhabditis elegans \em connectome},}\ }\href@noop
  {} {\bibfield  {journal} {\bibinfo  {journal} {Nature}\ }\textbf {\bibinfo
  {volume} {550}},\ \bibinfo {pages} {519--523} (\bibinfo {year}
  {2017})}\BibitemShut {NoStop}%
\bibitem [{\citenamefont {Pradhan}\ \emph {et~al.}(2020)\citenamefont
  {Pradhan}, \citenamefont {Angeliya},\ and\ \citenamefont
  {Jalan}}]{pradhan2020principal}%
  \BibitemOpen
  \bibfield  {author} {\bibinfo {author} {\bibfnamefont {P.}~\bibnamefont
  {Pradhan}}, \bibinfo {author} {\bibfnamefont {C.U.}\ \bibnamefont
  {Angeliya}}, \ and\ \bibinfo {author} {\bibfnamefont {S.}~\bibnamefont
  {Jalan}},\ }\bibfield  {title} {\enquote {\bibinfo {title} {Principal
  eigenvector localization and centrality in networks: revisited},}\
  }\href@noop {} {\bibfield  {journal} {\bibinfo  {journal} {Physica A}\
  }\textbf {\bibinfo {volume} {554}},\ \bibinfo {pages} {124169} (\bibinfo
  {year} {2020})}\BibitemShut {NoStop}%
\bibitem [{\citenamefont {Freeman}(1978)}]{freeman1978centrality}%
  \BibitemOpen
  \bibfield  {author} {\bibinfo {author} {\bibfnamefont {L.~C.}\ \bibnamefont
  {Freeman}},\ }\bibfield  {title} {\enquote {\bibinfo {title} {Centrality in
  social networks conceptual clarification},}\ }\href@noop {} {\bibfield
  {journal} {\bibinfo  {journal} {Soc. Networks}\ }\textbf {\bibinfo {volume}
  {1}},\ \bibinfo {pages} {215--239} (\bibinfo {year} {1978})}\BibitemShut
  {NoStop}%
\bibitem [{\citenamefont {Gastwirth}(1972)}]{gastwirth1972estimation}%
  \BibitemOpen
  \bibfield  {author} {\bibinfo {author} {\bibfnamefont {J.~L.}\ \bibnamefont
  {Gastwirth}},\ }\bibfield  {title} {\enquote {\bibinfo {title} {The
  estimation of the {L}orenz curve and {G}ini index},}\ }\href@noop {}
  {\bibfield  {journal} {\bibinfo  {journal} {The Review of Economics and
  Statistics}\ }\textbf {\bibinfo {volume} {54}},\ \bibinfo {pages} {306--316}
  (\bibinfo {year} {1972})}\BibitemShut {NoStop}%
\bibitem [{\citenamefont {{The World Bank}}()}]{worldbank}%
  \BibitemOpen
  \bibfield  {author} {\bibinfo {author} {\bibnamefont {{The World Bank}}},\
  }\href@noop {} {\enquote {\bibinfo {title} {Gini index ({W}orld {B}ank
  estimate)},}\ }\bibinfo {howpublished} {Available at
  \url{https://data.worldbank.org/indicator/SI.POV.GINI} (2020)}\BibitemShut
  {NoStop}%
\bibitem [{\citenamefont {Hashimoto}(1989)}]{hashimoto1989zeta}%
  \BibitemOpen
  \bibfield  {author} {\bibinfo {author} {\bibfnamefont {K.}~\bibnamefont
  {Hashimoto}},\ }\bibfield  {title} {\enquote {\bibinfo {title} {Zeta
  functions of finite graphs and representations of p-adic groups},}\ }in\
  \href@noop {} {\emph {\bibinfo {booktitle} {Automorphic forms and geometry of
  arithmetic varieties}}}\ (\bibinfo  {publisher} {Elsevier},\ \bibinfo {year}
  {1989})\ pp.\ \bibinfo {pages} {211--280}\BibitemShut {NoStop}%
\bibitem [{Note4()}]{Note4}%
  \BibitemOpen
  \bibinfo {note} {In principle, an acyclic flow could reach a dead end and
  therefore fail to visit all nodes. This scenario cannot occur with the
  ground-current centrality, since it is based on electrical currents flowing
  to ground, which is connected to all nodes.}\BibitemShut {Stop}%
\bibitem [{\citenamefont {Higham}(2002)}]{higham2002accuracy}%
  \BibitemOpen
  \bibfield  {author} {\bibinfo {author} {\bibfnamefont {N.~J.}\ \bibnamefont
  {Higham}},\ }\href@noop {} {\emph {\bibinfo {title} {Accuracy and stability
  of numerical algorithms}}},\ Vol.~\bibinfo {volume} {80}\ (\bibinfo
  {publisher} {SIAM},\ \bibinfo {year} {2002})\BibitemShut {NoStop}%
\end{thebibliography}%

\end{document}